\documentclass[12pt,usenames,dvipsnames,a4paper]{article}

\DeclareUnicodeCharacter{2062}{\ensuremath{}}

\usepackage{amsmath,amsfonts,amssymb}
\usepackage{appendix}
\usepackage{caption}
\usepackage{cite}
\usepackage{color}
\usepackage{datetime}
\usepackage{float}
\usepackage{graphicx}
\usepackage[colorlinks,citecolor = blue]{hyperref}
\usepackage{indentfirst}
\usepackage[numbers,square,comma,sort&compress]{natbib} 
\usepackage{subfig}
\numberwithin{equation}{section}
\usepackage{mathtools}
\usepackage[colorinlistoftodos]{todonotes}
\usepackage[compat=1.1.0]{tikz-feynman} 
\usepackage{ytableau}
\usepackage{tabu}

\def\a{\alpha}

\def\b{\beta}
\def\c{\chi}

\def\const{\mathrm{const}}

\def\d{\delta}
\def\D{\Delta}

\def\eps{\varepsilon}
\def\f{\frac}
\def\g{\gamma}

\def\G{\Gamma}

\def\l{\left}

\def\la{\langle}
\def\ra{\rangle}

\def\mc{\mathcal}

\def\m{\mu}
\def\n{\nu}

\def\nn{\nonumber}

\def\p{\partial}
\def\vp{\varphi}

\def\r{\right}

\def\t{\theta}

\def\vp{\varphi}

\def\x{\xi}

\def\z{\zeta}

\def\dst{\displaystyle}

\def\be{\begin{equation}}
\def\ee{\end{equation}}

\def\bea{\begin{eqnarray}}
\def\eea{\end{eqnarray}}

\def\ba{\begin{array}}
\def\ea{\end{array}}

\def\bc{\begin{center}}
\def\ec{\end{center}}

\def\bl{\begin{flushleft}}
\def\el{\end{flushleft}}

\def\br{\begin{flushright}}
\def\er{\end{flushright}}

\def\bi{\begin{itemize}}
\def\ei{\end{itemize}}

\def\bt{\begin{tabular}}
\def\et{\end{tabular}}

\newcommand{\REf}[1]{(\ref{#1})}

\newcommand{\btf}[1]{\begin{tikzpicture}[baseline=-3pt] \begin{feynman}[inline=(#1.base)]}
\def\etf{\end{feynman}\end{tikzpicture}}

\tikzfeynmanset{
  bigdot/.style = {
    /tikzfeynman/dot,
    /tikz/minimum size=4pt,
  },
  bigcross/.style = {
    /tikzfeynman/crossed dot,
    /tikz/minimum size=6pt,
  },
  warn luatex = false,
}

\newcommand{\gsim}{\gtrsim} 
\newcommand{\lsim}{\lesssim}

\begin{document}

\bc

{\Large \bf Identifying Large Charge Operators}

\vspace*{0.5cm}

{\textsc {Gil~Badel$^\dagger$, Alexander~Monin$^{\star,\dagger}$, Riccardo~Rattazzi$^\dagger$}}

\vspace*{0.5cm}
{\it $^\dagger$Institute of Physics, Theoretical Particle Physics Laboratory (LPTP), \\ 
\'Ecole Polytechnique F\'ed\'erale de Lausanne (EPFL), \\ 
CH-1015 Lausanne, Switzerland}

\hspace*{0.3cm}

{\it $^\star$Department of Physics and Astronomy, \\
University of South Carolina, \\
Columbia SC 29208, USA}

\vspace*{0.5cm}
\texttt{\small gil.badel@epfl.ch}  \\
\texttt{\small amonin@mailbox.sc.edu}  \\
\texttt{\small riccardo.rattazzi@epfl.ch} \\

\ec


\abstract{ The Large Charge sector of Conformal Field Theory (CFT) can generically be 
described through a semiclassical expansion around  a superfluid background. In this work, focussing on $U(1)$ invariant Wilson-Fisher fixed points, we study the spectrum of spinning large charge operators. For sufficiently low spin
these correspond to the phonon excitations of the superfluid state.
We discuss the organization of these states into conformal multiplets and the form of the corresponding composite operators
in the free field theory limit. The latter entails  a mapping, built order-by-order in the inverse charge $n^{-1}$, between the Fock space of vacuum fluctuations and the Fock space of  fluctuations around the superfluid state.
We discuss the limitations of the semiclassical method, and find that the phonon description breaks down for spins of order $n^{1/2}$ while the computation of observables is valid up to spins of order $n$.
Finally, we apply the semiclassical method to compute some conformal 3-point and 4-point functions, and analyze the conformal block decomposition of the latter with our  knowledge of the operator spectrum.
}

\tableofcontents

\section{Introduction}

Semiclassical methods represent one of the main tools to investigate non-perturbative phenomena in Quantum Field Theory (QFT). Vacuum decay \cite{Kobzarev:1974cp,Coleman:1977py}, instantons \cite{tHooft:1976snw}, topological defects and multiparticle production \cite{Rubakov:1995hq,Son:1995wz} are just a partial list of the different declinations of  this  methodology, but there recently  came up  an interesting addition \cite{Hellerman:2015nra}.
That is in the context of  conformal field theory (CFT), where it was shown that operators carrying a large conserved quantum number $Q$  admit a universal description in terms of a finite density superfluid state, with $1/Q$ controlling the  semiclassical expansion. In practice the superfluid state is described by an effective field theory (EFT) for the hydrodynamic excitations. In particular that implies that there exists  a non trivial correspondence between large charge operators and the hydrodynamic excitations in a superfluid.
While the original application focussed on the operator spectrum, in  \cite{Monin:2016jmo} it was later shown how the methodology straightforwardly extends to the computation of correlators.
That motivated exploring   large charge operators using instead the conformal bootstrap 
\cite{Jafferis:2017zna}. Perfect agreement was found, thus remarkably showing that the superfluid phase dynamics is somewhat encapsulated
in the bootstrap constraints at large $Q$.

The above results define robust and universal features of generic CFTs with conserved global symmetries. For specific CFTs that admit a definition within perturbation theory, through the large $N$- or  $\varepsilon$-expansions, the semiclassical approach is even more powerful \cite{Orlando:2019hte,Alvarez-Gaume:2019biu,Badel:2019oxl,Badel:2019khk}. In \cite{Badel:2019oxl,Badel:2019khk} focussing on the $U(1)$ symmetric Wilson-Fisher fixed point in $4-\varepsilon$ and $3-\varepsilon$, this was elucidated by considering  the properties of the simplest charge $n$ operator $\phi^n$. In particular, given the coupling $\lambda$, it was found it is the combination $\lambda n$ that controls the convergence of the standard Fenynman diagram approach: only for $\lambda n\ll1 $ is perturbation theory applicable. On the other hand, the large charge semiclassical approach, applies as long as $n\gg 1$, for any $\lambda n$. Remarkably there then exists a non-trivial overlap for the application of the two methods, which was exploited in \cite{Badel:2019oxl} both to validate the semiclassical computation and to boost the available finite order Feynman loop computations. Amusingly the parameter $\lambda n$ shares some features with the 't Hooft coupling in AdS/CFT \cite{Maldacena:1997re}. In particular, 
$\lambda n\gg 1$ corresponds to the regime where all the modes beside the hydrodynamic ones are gapped and can be integrated out, very much like in AdS/CFT the  large 't Hooft coupling allows to integrate out  the string modes to obtain the supergravity limit.
Another interesting aspect of Wilson-Fisher models is that, at least for $\lambda n\ll 1$, the operator spectrum can be explicitly constructed both in terms of fields and derivatives and in terms of hydrodynamic modes around the semiclassical saddle.
This clearly invites to see how the hydrodynamic Fock space structure emerges in the ordinary construction based on the elementary fields and their derivatives.
Otherwise stated, the semiclassical approach delivers the operators spectrum, but it does so somewhat formally, without  telling concretely   what these operators look like. It is the main goal of this paper to investigate this issue, as we now explain in more detail.


%
%
%


\subsection{$\varepsilon$-expansion at large charge\label{sec:epsilonIntro}}

In this paper we will  mostly consider weakly coupled CFTs, focusing on either $U(1)$ invariant $\phi^4$ theory in $d= 4-\eps$
\be
\mc L_4 = \p \bar \phi \p \phi + \f{\lambda}{4} (\bar \phi \phi)^2
\label{eq:phi4}
\ee
at the Wilson-Fisher fixed point \cite{Kleinert:2001ax}
\begin{equation} \label{eq:phi4FixedPoint}
  \frac{\lambda_*}{(4\pi)^2} = \frac{\varepsilon}{5} + \frac{3}{25}\varepsilon^2 + \mathcal O (\varepsilon^3),
\end{equation}
or its $\phi^6$ cousin in $d=3-\eps$ dimensions
\be
\mc L_6 = \p \bar \phi \p \phi + \f{\lambda^2}{36} (\bar \phi \phi)^3
\label{eq:phi6}
\ee
at the Wilson-Fisher fixed point \cite{Pisarski:1982vz}
\begin{equation} \label{eq:phi6FixedPoint}
 \frac{\lambda_*^2}{ (4\pi)^2 } =  \frac{3}{7} \varepsilon + \mathcal O(\varepsilon^2) .
\end{equation}

As made  evident by equations \REf{eq:phi4} -- \REf{eq:phi6FixedPoint} both theories are in the perturbative regime provided $\eps \ll 1$. 
As long as the coupling $\lambda$ is the only relevant 
parameter, observables can be reliably computed -- putting aside the usual asymptotic nature of perturbative series -- through Feynman diagrams. Things are however different when  other parameters enter the game. In particular, the correlators of operators with sufficiently large charge $n$, more precisely satisfying  $\lambda n \gsim 1$, are not computable via standard perturbation theory. We believe  the same is true for operators with large spin.

This situation is rather generic when large quantum numbers are present. In practice, standard perturbation theory breaks down because large combinatoric factors render the effective expansion parameter large (see~\cite{Voloshin:1994yp,Rubakov:1995hq,Libanov:1997nt} for a review in the context of multi-particle scattering). Nevertheless, it is generally believed that, as long as the coupling is small,  observables involving large quantum numbers can still be computed by an alternative perturbative expansion, performed
around a non-trivial saddle.

Identifying this saddle presents a difficult problem in general. However, for the case at hand, CFT correlators (or any observable for that matter) involving large charge operators, can be computed in a double scaling limit
\be
n \gg 1, ~~ \lambda \ll 1, ~~ \lambda n =\textrm{fixed},
\label{eq:doubleScaling}
\ee
by finding the saddle explicitly and expanding around it. Computations are simplified due to the enhanced symmetry of the problem. In particular, the operator-state correspondence allows to map the theory on the cylinder, where the saddle point corresponds to a homogeneous superfluid state with spontaneously broken $U(1)$ group whose properties can be systematically computed. For instance, the scaling dimension of $\phi ^n$, which is the lightest operator in the sector of  charge $n$, is given by the superfluid ground state energy. In the saddle point approximation it is written as a power series in $\lambda$ with coefficients that are themselves functions of $\lambda n$ 
\be
\D_{\phi^n} = \f{1}{\lambda} \D_{-1} (\lambda n) + \D_{0} (\lambda n)+ \lambda \D_{1} (\lambda n) + \dots
\label{eq:phinDimension}
\ee
A more detailed discussion of this result can be found in~\cite{Badel:2019oxl,Badel:2019khk}, where the first two coefficients in \REf{eq:phinDimension} are also explicitly computed. Notice that, taking $\lambda n$ as a fixed parameter, the expansion in $\lambda$ is equivalent to an expansion in $1/n$. \footnote{ Notice that here and in what follows, when we write $\lambda$ we indeed mean $\lambda_*$, as away from the fixed point the notion of scaling dimension is ill-defined beyond the lowest order.}

Similarly, one finds that (see Section~\ref{sec:saddle} for a recap) the excitations of the superfluid are given by phonons of spin $\ell$ and energies
\be
\omega^2_{A,B} (\ell)= J_\ell+\Omega^2 \mp \sqrt{4 J_\ell \m^2 +\Omega^4},
\label{eq:phiSpectrum}
\ee
where
\be \label{eq:defJl}
J_\ell = \ell (\ell+d-2),
\ee
is the $SO(d)$ Casimir, and moreover
\be
\label{eq:Omega_phi} 
\Omega^2 = \l \{
\ba{ll}
3 \mu_4^2 - m^2 , &\text{ for } (\bar \phi \phi)^2, \\
2 \mu_6^2 - m^2 , &\text{ for } (\bar \phi \phi)^3,
\ea
\r.
\ee
with $m = \frac{d}{2}-1$ and
\begin{align}
\text{ for } (\bar \phi \phi)^2: \mu_4(\lambda n,d) & = \frac{(d-2)}{2}\frac{\left( 3^{1/3}+ \left[ \frac{9 \lambda n \Gamma(d/2)}{2 \pi^{d/2} (d-2)^3} -  \sqrt{ \left(\frac{9 \lambda n \Gamma(d/2)}{2 \pi^{d/2} (d-2)^3}\right)^2-3 }\right]^{2/3} \right)  }
       {3^{2/3}  \left[ \frac{9 \lambda n \Gamma(d/2)}{2 \pi^{d/2} (d-2)^3} -  \sqrt{ \left(\frac{9 \lambda n \Gamma(d/2)}{2 \pi^{d/2} (d-2)^3}\right)^2-3 }\right]^{1/3} }  , \\
\text{ for } (\bar \phi \phi)^3: \mu_6(\lambda n,d) & = \frac{(d-2)}{2} \frac{\sqrt{1+\sqrt{1+ \frac{\lambda^2 n^2 \Gamma(d/2)^2}{3 \pi^d (d-2)^4}  }}}{\sqrt{2}} .
\end{align}
The Fock space of phonon excitations corresponds to the space of operators with charge $n$, whose  spectrum  of scaling dimensions at next to leading order (NLO) is then given by \footnote{As $\D_{\phi^n}$ is $O(n)$ and the  $\omega_{A,B}(\ell)$ are $O(1)$, the tree level frequencies are sufficient to compute the dimension $\D\l (\{k^A\},\{k^B\} \r )$ at NLO. On the other hand, in order to compute the splittings at NLO, one would need to perform  a full 1-loop computation.}
\be
\D\l (\{k^A\},\{k^B\} \r )=\D_{\phi^n} +\sum_{\ell = 1}^\infty k^A_{\ell}\omega_A(\ell)+\sum_{\ell'=0}^\infty k^B_{\ell'}\omega_B(\ell'),
\label{eq:primaryspectrum}
\ee
with $k^A_\ell$ and $k^B_\ell$ non-negative integers. The above result applies for states with a finite number of phonons and finite spin as $n\to \infty$. For large enough total spin, one expects a non-homogeneous configuration to dominate the path integral (see for instance \cite{vortices1}).

We call $A$- and $B$-type the phonons with energy $\omega_A$ and  $\omega_B$ respectively.
Notice that primary operators correspond to states with $k_1^A=0$, and that descendants are obtained by adding 
spin-1 $A$-type phonons. Compatibly with that, and with the accuracy of \eqref{eq:primaryspectrum}, one indeed has
\begin{equation} \label{eq:omega-1}
	\omega_A(1) = 1+O(\varepsilon)\,.
\end{equation}


\subsection{Motivation and goals \label{sec:motivation}}

The approach outlined above provides the spectrum of the operators but it does not say anything about their explicit form
in terms of elementary fields  and derivatives.  Establishing such form is one of the goals of this paper. Notice though that the  explicit form of composite operators  depends on the renormalization procedure and that, moreover,  for large enough $\lambda n$ we do not possess such a procedure. We will thus content ourselves with the construction of the operators
 in the free field theory limit $\lambda\to 0$ and with their correspondance to superfluid excitations.  

\begin{figure}[h]
  \centering
  \includegraphics[width=7cm]{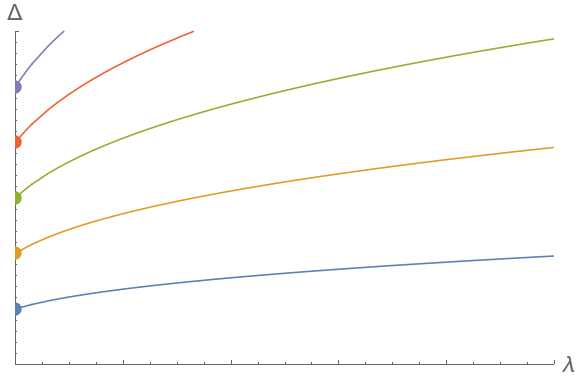}
  \caption{\label{fig:norms} Scaling dimension of families of operators as a function of $\lambda$. Each family can be labeled with the corresponding ``seed'' operator in free theory (at $\lambda = 0$).}
\end{figure}

As  we shall see,  the tree level result is already structurally informative. Indeed, the properties of the operator spectrum vary continuously with $\lambda$ (in truth with $\varepsilon$): by varying 
 $\lambda$ we obtain operator  {\it families} $\mathcal O^{(n,\ell,{\mathbb \alpha})}_\lambda(x)$, with $\mathbb \alpha$ a discrete  label characterizing the phonon composition (the $k^A$ and $k^B$ mentioned in the previous section). As qualitatively depicted in Fig.~\ref{fig:norms}, the dimensions $\Delta$, and OPE coefficients, of the $\mathcal O^{(n,\ell,{\mathbb \alpha})}_\lambda(x)$ are continuous functions of $\lambda$.  Our tree level construction will thus
correspond to  the starting point at $\lambda=0$ of each trajectory.  Such endpoints, however, fully  characterize the families non-perturbatively, even if indirectly.

%
%

This paper is organized as follows.
In section \ref{sec:vac_fluctuations}, we discuss the classification of operators with charge $n$  in 3D free field theory quantized around  $\phi=0$. We explain how to identify primary operators, and provide a systematic construction for a sub-class of them.
Section \ref{sec:superfluid_fluctuations} starts with a recap of the semiclassical expansion around the superfluid field configuration, followed by the construction of the mapping between superfluid Fock states and operators. We also discuss the identification of primary operators in this picture.
In section \ref{sec:large_spin}, we discuss the breakdown of the homogeneous superfluid description as the spin gets large.

In this paper, besides analyzing the operator spectrum, we apply the saddle point expansion to the computation of  correlators
between large charge operators and neutral operators  of the form $ \mc O =(\bar \phi \phi)^k$, with $k$ finite.
 In particular,  in Section~\ref{sec:3ptfunction}, we compute the 3-point functions $\la \bar \phi^n \mc O \phi^n \ra$ at next-to-leading (NLO) order in the $1/n$ expansion.
In Section~\ref{sec:4ptfunction} we  compute at NLO the  4-point functions $\la \bar \phi^n \mc O \mc O\phi^n \ra$ and 
study at the same order the $\mc O \times \phi^n$ operator product expansion (OPE).

\section{Operators corresponding to vacuum fluctuations in free theory\label{sec:vac_fluctuations}}

Our first goal is to classify the families of large charge operators by focussing on their representatives in the free limit, as sketched in figure \ref{fig:norms}.
As explained in section \ref{sec:motivation}, the first step is the classification of  the operators of free field theory  in terms of conformal multiplets.
This amounts to  identifying the conformal primaries.

In CFT every local operator corresponds to a state and vice versa (operator-state correspondence). In particular primary states, i.e. those annihilated by the  special conformal generators,  correspond to primary operators. The goal of this section is to set up the methodology for identifying these  states. To make things  explicit 
 we will fully construct a subclass of the operators.

Working in radial quantization we will now, in turn, construct the Fock space of vacuum fluctuations, derive the state-operator correspondence and write the conformal group generators.
We will  then write down in closed form a subset of primary states, 
also  showing by a combinatoric argument that it forms a complete basis
of the subspace of primary operators with a number of derivatives smaller than the charge.

\subsection{Fock space of vacuum fluctuations}

Let us consider a free  complex scalar field in $d=3$ Euclidean dimensions
\be
\mc L = \p \bar \phi \p \phi.
\label{eq:FreeLagrangian}
\ee
As usual in CFT, it is beneficial to put the theory \REf{eq:FreeLagrangian} on the cylinder $\mathbb R \times \mathbb S^2$ by redefining the coordinates
\be
\label{eq:PlaneCylinderCoords}
  x^\mu = r n^\mu, ~~  r = e^\tau, ~~ \vec n = (\sin \theta \cos\varphi, \sin \theta \sin \varphi, \cos\theta)
\ee
and the field
\be
\label{eq:PlaneCylinderField}
  \hat \phi (\tau,\t,\vp) = e^{\tau/2} \phi (x).
\ee



As a result we have the following action on the cylinder
\be
S = \int d\tau d\Omega_2 \l [ g^{\m \n} \p_\m \hat { \bar \phi} \p_\n \hat \phi + \f{1}{4}\hat {\bar\phi} \hat  \phi \r ], ~~ g_{\m \n} =\mathrm{diag}(1,1,\sin^2\t) .
\ee
Time translations on the cylinder are generated by the corresponding Hamiltonian $H$ in the following way
\begin{equation}
\label{HeisenbergCylinder}
  \hat \phi(\tau,\theta,\phi) = e^{H \tau} \hat \phi(0,\theta,\phi) e^{-H\tau},
\end{equation}
and are related to dilatations on the plane, which are generated by $D$
\begin{equation}
\label{HeisenbergPlane}
e^{D \lambda} \phi(x) e^{- D \lambda} = e^{\lambda/2 }\phi(e^\lambda x) .
\end{equation}
This implies
\begin{equation}
  H =  D,
\end{equation}
so that operator dimensions  are in one to one correspondence with energy levels  on the cylinder. 

Hermitian conjugation in radial quantization of the parent Euclidean field theory implies $\hat {\bar \phi}(0, \t,\vp)=\hat \phi(0, \t,\vp)^\dagger $,
which at arbitary $\tau$ on the cylinder and arbitrary $x$ on the plane implies respectively
\be
\hat {\bar \phi}(\tau, \t,\vp)=\hat \phi(-\tau, \t,\vp)^\dagger \qquad{\mathrm {and}}\qquad \bar\phi(x)=|x|^{-1}\phi(x^{-1})^\dagger\,.
\label{HermiteanConjugation}
\ee

Quantization proceeds by expanding the fields in  spherical harmonics $Y_{\ell m}$
\be
\hat \phi(\tau, \t,\vp)=\sum_{\ell=0}^\infty \sum_{m=-\ell}^\ell \f{1}{\sqrt{2\omega_\ell}} \l [ a^\dagger_{\ell m} e^{\omega_\ell \tau} Y^*_{\ell m}(\t,\vp) + b_{\ell m} e^{-\omega_\ell \tau} Y_{\ell m}(\t,\vp) \r ],
\label{eq:aaCylinderFieldComplex}
\ee
and\footnote{ Notice $\hat {\bar \phi}(\tau, \t,\vp)=\hat \phi(-\tau, \t,\vp)^\dagger $  in accordance with (\ref{HermiteanConjugation}).}
\be \label{eq:aaCylinderFieldComplexConj}
\hat {\bar \phi}(\tau, \t,\vp)=\sum_{\ell=0}^\infty \sum_{m=-\ell}^\ell \f{1}{\sqrt{2\omega_\ell}} \l [ b^\dagger_{\ell m} e^{\omega_\ell \tau} Y^*_{\ell m}(\t,\vp) + a_{\ell m} e^{-\omega_\ell \tau} Y_{\ell m}(\t,\vp)\r ],
\ee
with energies
\be
\omega_\ell = \ell+\f{1}{2}.
\ee
The corresponding canonically conjugated momenta are given by\footnote{There appears an  ``$i$'' in front of the time derivatives  because we work in Euclidean time.}
\be
p_{\hat\phi}(\tau, \t,\vp)= i \partial_\tau \hat{\bar\phi} =i\sum_{\ell=0}^\infty \sum_{m=-\ell}^\ell\sqrt{\f{\omega_\ell}{2}} \l [ b^\dagger_{\ell m} e^{\omega_\ell \tau} Y^*_{\ell m}(\t,\vp) - a_{\ell m} e^{-\omega_\ell \tau} Y_{\ell m}(\t,\vp)\r ],
\label{eq:aaCylinderMomentumComplex}
\ee
and
\be
p_{\hat{\bar\phi}}(\tau, \t,\vp)=i\partial_\tau \hat\phi =i\sum_{\ell=0}^\infty \sum_{m=-\ell}^\ell \sqrt{\f{\omega_\ell}{2}}\l [ a^\dagger_{\ell m} e^{\omega_\ell \tau} Y^*_{\ell m}(\t,\vp) - b_{\ell m} e^{-\omega_\ell \tau} Y_{\ell m}(\t,\vp) \r ].
\label{eq:aaCylinderMomentumComplexConj}
\ee
Creation and annihilation operators, satisfying the usual commutation relations
\be
[ a_{\ell m},a^\dagger_{\ell'm'}] = [ b_{\ell m},b^\dagger_{\ell'm'}] = \d_{\ell \ell'}\d_{mm'},
\ee
allow us to build the Hilbert space. Defining the vacuum state $| 0 \ra$ as 
\be
a_{\ell m} | 0 \ra =b_{\ell m} | 0 \ra =0, \qquad \forall\, \ell, m
\ee
states featuring a string of creation operators acting on the vacuum
\be
\prod_{i= 1}^{n_a} a^\dagger_{\ell_i m_i} \prod_{j=1}^{n_b} b^\dagger_{\ell'_j m'_j} | 0 \ra 
\label{eq:FockSpaceStates}
\ee
provide a basis of the Hilbert space, and give it 
the standard Fock space structure. The $U(1)$ charge of these states 
is determined 
 by the charge operator
\begin{equation} \label{eq:Qaabb}
  Q = \sum_{\ell=0}^\infty \sum_{m=-\ell}^\ell \left( a_{\ell m}^\dagger a_{\ell m} - b_{\ell m}^\dagger b_{\ell m} \right) .
\end{equation}


\subsection{Operator-state correspondence \label{sec:oppStateCorr}}

Combining (\ref{eq:aaCylinderFieldComplex}) with (\ref{eq:aaCylinderMomentumComplexConj}) and using the orthonormality of the $Y_{\ell m}$  (\ref{eq:orthonormalityHarmonics}) we get
\be
a_{\ell m}^\dagger = \f{e^{-\omega_\ell \tau}}{\sqrt{2 \omega _\ell}} \int d\Omega_2 \, Y_{\ell m} \l ( \partial_\tau \hat\phi(\tau) +\omega_\ell \hat \phi (\tau) \r ),
\ee
which is valid at any finite $\tau$. Remembering the change of coordinates (\ref{eq:PlaneCylinderCoords}) and the relation between fields on the plane and on the cylinder \REf{eq:PlaneCylinderField}, this expression can be rewritten as 
\be
a_{\ell m}^\dagger  = \f{r^{-\ell }}{\sqrt{2 \omega_\ell}} \int d\Omega_2 \, Y_{\ell m} \big ( x^\m \p_\m \phi (x) + (\ell +1) \phi(x) \big) ,
\ee
where the integral is over the unit sphere and $x^\mu = r n^\mu$.

Acting on the vacuum and Taylor expanding around the origin\footnote{As can be seen in (\ref{eq:aaCylinderFieldComplex}), the field is singular at the origin, $r\to 0$ or $\tau\to -\infty$ due to negative-frequency $b_{\ell m}$ modes. However, in $\phi(x) | 0 \ra$ 
the singular terms drop and Taylor expansion is legitimate.} we get
\begin{equation}
  \phi(x)|0\rangle = \sum_{\ell'=0}^\infty \frac{1}{\ell'!} x^{\{\mu_{1}}\cdots x^{\mu_{\ell'}\}} \partial_{\mu_1}\cdots \partial_{\mu_{\ell'}} \phi(0)|0\rangle\,,
\end{equation}
where by $\{\dots\}$ we indicate the  traceless symmetric combination, which arises because  of the equation of motion $\partial^2 \phi(x) = 0$.
Noting that
\be
\int d\Omega_2 \, Y_{\ell m} x^{\{\mu_{1}}\cdots x^{\mu_{\ell'}\}} = 0,~~ \ell'\neq \ell,
\ee
the expansion results in
\be
a_{\ell m}^\dagger | 0 \ra = \f{\sqrt{2\ell+1}}{\ell!} \int\! d\Omega_2 \, Y_{\ell m} n^{\m_1} \cdots n^{\m_\ell} \p_{\m_1}\cdots \p_{\m_\ell} \phi(0) | 0 \ra.
\label{eq:aVsDerivatives}
\ee
The integral can be easily done, yielding
\begin{equation} \label{eq:stateOperatorMapA}
  a_{\ell m}^\dagger |0\rangle = \mathcal Y_{\ell m}^{\mu_1\dots\mu_\ell} \partial_{\mu_1} \cdots \partial_{\mu_\ell} \phi(0)|0\rangle,
\end{equation}
with (see Appendix \ref{sec:appYlmCoeff})
\begin{equation} \label{eq:defineYlmCoeff}
  \mathcal Y_{\ell m}^{\mu_1\dots\mu_\ell} = \f{\sqrt{2\ell+1}}{\ell!} \int d\Omega_2 \, Y_{\ell m} n^{\m_1} \cdots n^{\m_\ell} .
\end{equation}  
Repeating the same steps starting with (\ref{eq:aaCylinderFieldComplexConj}) and (\ref{eq:aaCylinderMomentumComplex}), we get similarly
\begin{equation} \label{eq:stateOperatorMapB}
  b_{\ell m}^\dagger |0\rangle = \mathcal Y_{\ell m}^{\mu_1\dots\mu_\ell} \partial_{\mu_1} \cdots \partial_{\mu_\ell} \bar\phi(0)|0\rangle .
\end{equation}
In particular, we obtain
\begin{equation}
  \mathcal Y_{\ell\ell}^{\mu_1 \dots \mu_\ell} = \frac{(-1)^\ell 2^{\frac{\ell}{2}+1}\sqrt{\pi}}{\sqrt{(2\ell)!}} \delta^{\mu_1}_- \cdots \delta^{\mu_\ell}_- ,
\end{equation}
\begin{equation}
  a_{\ell\ell}^\dagger |0\rangle = \frac{(-1)^\ell 2^{\frac{\ell}{2}+1}\sqrt{\pi}}{\sqrt{(2\ell)!}} \big( \partial_- \big)^{\ell} \phi(0) |0\rangle , \label{eq:a_llFields}
\end{equation}
where the following change of variables was performed
\be \label{eq:x+-0}
x_\pm=\f{x_1\pm i x_2}{\sqrt{2}}, ~~ x_0=x_3.
\ee
This generalizes to multi-particle Fock states (\ref{eq:FockSpaceStates}). For example\footnote{In field products acting on the vacum  the singular terms at the origin are eliminated by   normal-ordering. In the rest of the paper,  normal ordering will always be intended and we will drop the ``$:$'' symbol.},
\begin{equation}
  a_{\ell_1 m_1}^\dagger b_{\ell_2 m_2}^\dagger |0\rangle = \mathcal Y_{\ell_1 m_1}^{\mu_1 \dots \mu_{\ell_1}} \mathcal Y_{\ell_2 m_2}^{\nu_1 \dots \nu_{\ell_2}} : \partial_{\mu_1} \!\cdots \partial_{\mu_{\ell_1}}\! \phi(0)\, \partial_{\nu_1}\!\cdots \partial_{\nu_{\ell_2}}\! \bar\phi(0) : |0\rangle .
\end{equation}
Hermitian conjugation of (\ref{eq:aVsDerivatives}), together with (\ref{HermiteanConjugation}), implies
\begin{equation}
  \langle 0 | a_{\ell m} = \frac{\sqrt{2\ell+1}}{\ell!} \int \! d\Omega_2 \, Y_{\ell m}^* n^{\m_1} \cdots n^{\m_\ell} 
                       \lim_{x\to\infty} \langle 0 |  \partial_{\mu_1}^{(1/x)}\cdots\partial_{\mu_\ell}^{(1/x)} \big( |x| \bar\phi(x) \big)  , 
\end{equation}
where we defined
\begin{equation}
  \partial_\mu^{(1/x)} =  \left( x^2 \delta_{\mu\nu} - 2 x^\mu x^\nu \right) \partial_\nu^{(x)} .
\end{equation}
Notice, this time the field is evaluated at infinity, because hermitian conjugation in radial quantization involves a space inversion.


\subsection{Conformal generators}

In order to proceed with the classification and construction of the operators we first need the explicit expression of the 
 conformal group generators in terms of the ladder operators. We provide them in this section.
 
In  $d=3$, defining
\be
J_i = \f{1}{2} \eps_{ijk}J_{jk}, ~~ J_\pm = J_1\pm iJ_2
\ee
and
\be
P_\pm = \f{1}{\sqrt{2}} \l ( P_1\pm i P_2 \r ), ~~ K_\pm = \f{1}{\sqrt{2}} \l ( K_1\pm i K_2 \r ), ~~ P_0\equiv P_3, ~~ K_0=K_3,
\ee
such that 
\be
P_\pm^\dagger = K_\mp, ~~ P_0^\dagger = K_0,
\ee
 the commutation relations take the form (with $X_\bullet = P_\bullet, K_\bullet$)
\bea
\label{eq:CRin3dSpinBasis}
&&[J_3,J_\pm] = \pm J_\pm, ~~ [J_+,J_-] = 2 J_3,  \nn \\ 
&&[J_3, X_\pm]=\pm X_\pm, ~~ [J_3, X_0]=0, \nn \\
&&[J_+, X_+]=0, ~~            [J_+, X_0]=-\sqrt{2}X_+, ~~ [J_+, X_-]=\sqrt{2}X_0, \\
&&[J_-, X_+]=-\sqrt{2}X_0, ~~ [J_-, X_0]=\sqrt{2}X_-,  ~~ [J_-, X_-]=0 \nn \\
&&[D, K_i]=-K_i, ~~ [D, P_i]=P_i, \nn \\
&&[K_-, P_+]=2 \l ( D+ J_3 \r ), ~~ [K_+, P_-]=2 \l ( D- J_3 \r ), ~~ [K_0, P_0] = 2D \nn \\
&&[K_0, P_+]=-\sqrt{2} J_+, ~~ [K_+, P_0]=\sqrt{2} J_+, ~~ [K_-, P_0]=-\sqrt{2} J_-, ~~ [K_0, P_-]=\sqrt{2} J_-. \nn
\eea

The generators, as computed from the Noether currents of the theory, read  \bea
\label{eq:Daabb}
D & = & \sum_{\ell=0}^\infty \sum_{m=-\ell}^\ell \omega_\ell \l( a^\dagger_{\ell m}a_{\ell m}+b^\dagger_{\ell m}b_{\ell m}\r), \\
\label{eq:J3aabb}
J_3 & = & \sum_{\ell =0}^\infty \sum_{m=-\ell}^\ell m \l( a^\dagger_{\ell m}a_{\ell m}+b^\dagger_{\ell m}b_{\ell m}\r ), \\
\label{eq:P0aabb}
P_0 & = & \sum_{\ell =0}^\infty \sum_{m=-\ell}^\ell \sqrt{(\ell+1)^2-m^2} \l( a^\dagger_{\ell+1,m}a_{\ell m}+b^\dagger_{\ell+1,m}b_{\ell m}\r ), \\
\label{eq:K0aabb}
K_0 & = & \sum_{\ell =0}^\infty \sum_{m=-\ell}^\ell \sqrt{(\ell+1)^2-m^2} \l( a^\dagger_{\ell m}a_{\ell+1,m}+b^\dagger_{\ell m}b_{\ell+1,m}\r ),
\eea
\bea
\label{eq:J+aabb}
J_+ & = & \sum_{\ell=0}^\infty \sum_{m=-\ell}^\ell \sqrt{\ell(\ell+1)-m(m+1)}\l( a^\dagger_{\ell,m+1}a_{\ell m}+b^\dagger_{\ell,m+1}b_{\ell m}\r ), \\
& = & 
\sum_{\ell=0}^\infty \sum_{m=-\ell}^\ell \sqrt{\ell(\ell+1)-m(m-1)} \l( a^\dagger_{\ell m}a_{\ell,m-1}+ b^\dagger_{\ell m}b_{\ell,m-1}\r ), \nn \\
\label{eq:J-aabb}
J_- & = & \sum_{\ell=0}^\infty \sum_{m=-\ell}^\ell \sqrt{\ell(\ell+1)-m(m+1)} \l ( a^\dagger_{\ell m}a_{\ell,m+1}+b^\dagger_{\ell m}b_{\ell,m+1}\r ) \\
& = &
\sum_{\ell=0}^\infty \sum_{m=-\ell}^\ell \sqrt{\ell(\ell+1)-m(m-1)} \l ( a^\dagger_{\ell,m-1}a_{\ell m}+b^\dagger_{\ell,m-1}b_{\ell m} \r ), \nn
\eea
\bea
\label{eq:P+aabb}
P_+ & = & -\sum_{\ell=0}^\infty \sum_{m=-\ell}^\ell \sqrt{\f{(\ell+m+1)(\ell+m+2)}{2}} \l ( a^\dagger_{\ell+1,m+1}a_{\ell m}+b^\dagger_{\ell+1,m+1}b_{\ell m}\r ), \\
\label{eq:K-aabb}
K_- & = & -\sum_{\ell=0}^\infty \sum_{m=-\ell}^\ell \sqrt{\f{(\ell+m+1)(\ell+m+2)}{2}} \l ( a^\dagger_{\ell m}a_{\ell+1,m+1}+b^\dagger_{\ell m}b_{\ell+1,m+1}\r ), \\
\label{eq:P-aabb}
P_- & = & \sum_{\ell=0}^\infty \sum_{m=-\ell}^\ell \sqrt{\f{(\ell-m+1)(\ell-m+2)}{2}} \l ( a^\dagger_{\ell+1,m-1}a_{\ell m} + b^\dagger_{\ell+1,m-1}b_{\ell m} \r ), \\
\label{eq:K+aabb}
K_+ & = & \sum_{\ell=0}^\infty \sum_{m=-\ell}^\ell \sqrt{\f{(\ell-m+1)(\ell-m+2)}{2}} \l ( a^\dagger_{\ell m}a_{\ell+1,m-1}+ b^\dagger_{\ell m}b_{\ell+1,m-1} \r ).
\eea

\subsection{Primary states and operators}

Besides the quantum numbers associated with the conformal group, states can be  characterized by their charge and by their parity.
Charge is quickly dealt with. Any state of the form (\ref{eq:FockSpaceStates}) is an eigenstate of the charge operator $Q$.

Consider now parity.
At fixed charge $n$ and spin $\ell$, states divide into two {\it polarity} classes: polar states with parity $(-1)^\ell$ and axial states with parity $(-1)^{\ell+1}$. The two classes can schematically be written as 
 \bea
{\mathrm {polar}}\qquad P=(-1)^\ell\,\,\,\,\,\,\quad &\Rightarrow&\quad \partial^{\ell+2k} \phi^{n_a}\bar\phi^{n_b} \delta^k  \qquad\quad \,\, k\geq 0\label{polar}\\
{\mathrm {axial}}\,\qquad P=(-1)^{\ell+1} \quad &\Rightarrow &\quad \partial^{\ell+2k+1} \phi^{n_a}\bar\phi^{n_b} \epsilon \delta^k  \qquad  k\geq 0 \label{axial}
\eea
where $\partial$, $\delta$ and $\epsilon$ represent respectively a spacetime derivative $\partial_i$, the Kronecker delta $\delta_{ij}$ and the Levi-Civita tensor $\epsilon_{ijk}$. The $\delta$'s and the $\epsilon$ are all contracted with a pair of derivatives, while  the remaining $\ell$ indices are symmetrized and trace-subtracted. 

As the $U(1)$ charge $Q$ commutes with the conformal group, the conformal multiplets have definite charge. On the other hand,  by considering that $\partial \to -\partial $ under parity and the standard rule for adding angular momenta, one is easily convinced that the descendants of an operator with given polarity (polar or axial) can have either polarity. One can nonetheless label a conformal multiplet by the polarity of its primary state.

In this paper, we will provide a systematic construction of all primaries whose number of derivatives is bounded by the charge $n$ (see \cite{deMelloKoch:2018klm,deMelloKoch:2017dgi,Henning:2019mcv} for a different but less explicit procedure to construct primaries of given spin and charge). In a first time we describe this construction, before proving via a combinatorics argument that our procedure indeed generates all such primaries. This will enable us to concretely illustrate the emergence of the superfluid Fock space structure within the operator spectrum at large charge.




\subsubsection{Construction of primaries\label{sec:vacuum_primaries_construction}}

States of the form
\be
a^\dagger_{\ell_1,m_1} \dots a^\dagger_{\ell_{n_a},m_{n_a}} b^\dagger_{j_1,m_1} \dots b^\dagger_{j_{n_b},m_{n_b}} | 0 \ra,
\ee
can be decomposed into irreducible representations of $SO(3)$
\be
\ell_1\otimes \dots \otimes \ell_{n_a} \otimes j_1 \dots \otimes j_{n_b} = \l ( \ell_1+ \dots+\ell_{n_a} + j_1 \dots + j_{n_b}\r ) \oplus \dots
\label{eq:spinDecomposition}
\ee
Let us first consider the states with the highest total spin $\ell=\ell_1+\dots + j_{n_b}$ in the tensor product \REf{eq:spinDecomposition}, indicating them by
\be
| n; \ell,m \ra
\ee
where $n=n_a-n_b$ and $m$ are respectively the $Q$ and $J_3$ eigenvalues.
The highest weight state from which all the multiplet is constructed by repeatedly acting with $J_-$
is 
\begin{equation} \label{eq:highestWeightState}
	|n ; \ell, \ell \rangle = a^\dagger_{\ell_1,\ell_1} \dots a^\dagger_{\ell_{n_a},\ell_{n_a}} b^\dagger_{j_1,j_1} \dots b^\dagger_{j_{n_b},j_{n_b}} | 0 \ra.
\end{equation}
By eqs.~\REf{eq:stateOperatorMapA}, \REf{eq:stateOperatorMapB} and by the discussion at the beginning of this section, the corresponding operators are polar and have the schematic form $\phi^{n_a}\bar \phi^{n_b}\partial^\ell$. 
In the basis (\ref{eq:x+-0}), the operator corresponding to (\ref{eq:highestWeightState}) involves only $\partial_-$ derivatives, as is made clear by (\ref{eq:a_llFields}). We can now search for combinations of states of the form  \REf{eq:highestWeightState} that correspond to primaries.

Let us first consider states involving creation operators of only one sort, say $a^\dagger$. 
A first obvious example is the state of charge $n$ with lowest dimension, which is given by
\begin{equation} \label{eq:chargenspin0}
  | n \rangle = \frac{1}{\sqrt{n!}} (a_{00}^\dagger)^n |0\rangle = \frac{(4\pi)^{n/2}}{\sqrt{n!}} \phi^n(0) |0\rangle
\end{equation}
This state has spin $0$ and is a primary as it is annihilated by the $K_i$'s. 

To find  a spin-$\ell$ primary we start with the ansatz
\be
| n; \ell,\ell \ra^{(0)}_A= (a_{00}^\dagger)^{n-1} a_{\ell\ell}^\dagger | 0 \ra.
\ee
Acting on it with $K_-$ we get\footnote{ As can be derived from \REf{eq:K0aabb}, \REf{eq:K+aabb}, $[K_0,a_{\ell\ell}^\dagger] = [K_+,a_{\ell\ell}^\dagger] = 0$ for all $\ell$, so the state is annihilated by  both $K_0$ and $K_+$. Moreover \REf{eq:K-aabb} yields
\begin{equation}
  [K_-,a_{\ell\ell}^\dagger] = -\sqrt{\frac{(2\ell)(2\ell-1)}{2} } a_{\ell-1,\ell-1}^\dagger .
\end{equation}
}
\be
K_- | n; \ell,\ell \ra^{(0)}_A = -\sqrt{\f{2\ell(2\ell-1)}{2}} (a_{00}^\dagger)^{n-1} a_{\ell-1,\ell-1}^\dagger | 0 \ra, ~~ \ell\neq 1.
\ee
In order to cancel this contribution we modify the vector
\be
| n; \ell, \ell \ra^{(1)}_A= (a_{00}^\dagger)^{n-1} a_{\ell\ell}^\dagger | 0 \ra-\sqrt{\f{2\ell(2\ell-1)}{2}} (a_{00}^\dagger)^{n-2} a_{\ell-1,\ell-1}^\dagger a_{1,1}^\dagger | 0 \ra.
\label{2ndstep}
\ee
Acting with $K_-$ on the new state we find
\be
K_- | n; \ell,\ell \ra^{(1)}_A  = \sqrt{\f{2\ell(2\ell-1)}{2}} \sqrt{\f{(2\ell-2)(2\ell-3)}{2}}(a_{00}^\dagger)^{n-2} a_{\ell-2,\ell-2}^\dagger  a_{1,1}^\dagger | 0 \ra\, .
\ee
Again, to cancel this contribution we add an extra term to (\ref{2ndstep}) and we continue 
 further until we finally arrive at an exact primary
\be
| n; \ell,\ell \ra_A= \a_0\sum_{k=0}^{\ell}\g_{k,\ell} (a_{00}^\dagger)^{n-k-1} ( a_{1,1}^\dagger )^{k} a_{\ell-k, \ell-k}^\dagger | 0 \ra,
\label{eq:one-phononStateA}
\ee
with
\be
\g_{k,\ell}= \f{(-1)^k}{k!}\sqrt{\f{(2\ell)!}{2^k(2\ell-2k)!}},
\ee
with  the overall coefficient $\a_0$  fixed by the normalization condition $\parallel | n; \ell,\ell \ra_A\parallel^2=1$
\be
\label{eq:PrimaryNormalization}
\a_0^2\l[ \sum_{k=0}^{\ell-2}\g^2_{k,\ell} (n-k-1)! k! + \l( \g_{\ell-1,\ell}+\g_{\ell,\ell} \r )^2 (n-\ell)! \ell! \r ] =1.
\ee
As can be verified, this construction works if $1< \ell \leq n$.
Explicit constructions of these states and of the corresponding operators for  $\ell=0,1,2,3$ can be found in Appendix~\ref{app:spin23}.
By using \REf{eq:Daabb} one  can also check  that the energy of this state, or equivalently the dimension of the corresponding operator, is given by
\be
\D_A(n,\ell) = \f{n}{2}+\ell,
\ee
as expected in free theory.

Similarly we can consider states that involve one creation operator $b^\dagger$. Repeating the construction it is straightforward to construct,  for $\ell\leq n+1$,  a primary 
\be
| n; \ell,\ell \ra_B= \b_0\sum_{k=0}^{\ell}\g_{k,\ell} (a_{00}^\dagger)^{n-k+1} ( a_{1,1}^\dagger )^{k} b_{\ell-k, \ell-k}^\dagger | 0 \ra,
\label{eq:one-phononStateB}
\ee
with
\be
\b_0^2\sum_{k=0}^{\ell}\g^2_{k,\ell} (n-k-1)! k! =1.
\ee

These special cases can be combined to generate more primaries. Indeed, one can define spin $\ell$ multiplets of operators $\left\{\mathcal A_{\ell,m}^\dagger \right\}$, $\left\{\mathcal B_{\ell,m}^\dagger\right\}$ with $m=-\ell,\dots,\ell$ whose highest weight elements are
\begin{eqnarray}\label{eq:define_C_atomic}
	\mathcal A_{\ell,\ell}^\dagger &=& \sum_{k=0}^{\ell}\g_{k,\ell} (a_{00}^\dagger)^{\ell-k-1} ( a_{1,1}^\dagger )^{k} a_{\ell-k, \ell-k}^\dagger, \quad\quad \ell \geq 2 
\\
  \mathcal B_{\ell,\ell}^\dagger &=& \sum_{k=0}^{\ell}\g_{k,\ell} (a_{00}^\dagger)^{\ell-k+1} ( a_{1,1}^\dagger )^{k} b_{\ell-k, \ell-k}^\dagger,  \quad\quad \ell \geq 0 .  
\end{eqnarray}
$\mathcal A_{\ell,m}^\dagger$ and $\mathcal B_{\ell,m}^\dagger $ are polar   primaries, because they commute with all $K_i$, and they have charge $\ell$.
Notice, that $\mathcal A_{0,0}$ is not defined and $\mathcal A_{1,m}=0$, while $\mathcal B_{0,0}^\dagger=a_{00}^\dagger b_{00}^\dagger$. The primary states we constructed  are then given by
\begin{eqnarray}
	|n; \ell,\ell\rangle_A &=& \alpha_0 (a_{00}^\dagger)^{n-\ell} \mathcal A_{\ell,\ell}^\dagger |0\rangle ,
\\
	|n; \ell,\ell\rangle_B &=& \beta_0 (a_{00}^\dagger)^{n-\ell} \mathcal B_{\ell,\ell}^\dagger |0\rangle .
\end{eqnarray}
Since the $\mathcal A_{\ell,m}^\dagger$'s, $\mathcal B_{\ell,m}^\dagger$'s,  as well as $a_{00}^\dagger$, are all primaries, any product of them is a primary as well. This lets us generate primaries of various spins and charges by acting on the vacuum with these operators
\begin{equation} \label{eq:CDproduct}
  \left(a_{00}^\dagger\right)^{n-\sum_\alpha \ell_\alpha - \sum_\beta \tilde{\ell}_\beta}\prod_{\alpha} \mathcal A_{\ell_\alpha,m_\alpha}^\dagger \prod_\beta \mathcal B_{\tilde \ell_\beta,\tilde m_\beta}^\dagger | 0 \rangle ,
\end{equation}
where the number of derivatives of the corresponding operator is $P\equiv \sum_\alpha \ell_\alpha + \sum_\beta \tilde{\ell}_\beta$. Notice this state is an eigenstate of $J^2$ only for maximal spin states ($m_\alpha = \ell_\alpha, \tilde m_\beta = \tilde \ell_\beta$ or $m_\alpha = -\ell_\alpha, \tilde m_\beta =-\tilde \ell_\beta$). Otherwise, one must take linear combinations of these terms to build spin multiplets. 
By inspecting their definition, one can be convinced  that $\mathcal A_{\ell,\ell}^\dagger$ and $\mathcal B_{\ell,\ell}^\dagger$ (and hence the corresponding  spin multiplets) can not be written as  products of $\mathcal A^\dagger$'s and $\mathcal B^\dagger$'s with lower-spin -- they are in a sense ``irreducible''. Thus the above representation of a primary is unique.

Indeed, as it turns out, the above representation generates all the primaries with number of derivatives bounded by $n$. In the next section we will offer a combinatoric proof of that. Before moving to that, and to ease the counting, it is convenient to note  that the dimensionality of the space generated by (\ref{eq:CDproduct}) is the same as that of space generated by  (\ref{eq:FockSpaceStates}) barring the spin 1 ladder operators $a^\dagger_{1,m}$. This can be seen by picking only the $k=0$ terms in the series (\ref{eq:define_C_atomic}) for the $\mathcal A$'s  and $\mathcal B$'s. This remark will be used in the next section to prove that (\ref{eq:CDproduct}) provide a complete basis for primaries.
%



\subsubsection{Combinatorics: counting primaries}

As a warmup, we will first consider different subclasses of operators for which we can provide explicit expressions  for the number of primaries. After having done that, we will prove that the set (\ref{eq:CDproduct}) is indeed a complete basis for the primaries.

\paragraph{No $\bar\phi$, spin $\ell\leq n$, number of derivatives equal to $\ell$}\mbox{}\\
\noindent{Consider} the polar operators with $k=0$ and no $\bar\phi$ fields in eq.~(\ref{polar}). They correspond to symmetric traceless  tensors  with schematic form  $\partial^\ell \phi^n$. Using coordinates (\ref{eq:x+-0}),  the highest weight elements of the corresponding  $SO(3)$ multiplets have the  schematic form $\partial_-^\ell \phi^n$.  The counting is now straightforward:
there are as many operators as there are inequivalent ways of distributing $\ell$ derivatives $\p_{-}$ among $n$ fields $\phi$. That is given by the number of partitions of $\ell$  into at most $n$ integers, which we denote by $p(\ell,n)$. In the case $\ell\leq n$, the partition cannot contain more than $n$ elements, and so $p(\ell,n)$ reduces to the number $p(\ell)$ of unconstrained partitions of $\ell$.
 
For example, for $\ell=5$ we get the following partitions.
\be
5: ~~(5),~(4,1),~(3,2),~(3,1,1),~(2,2,1),~(2,1,1,1),~(1,1,1,1,1),\label{l=5}
\ee
Thus, there are 
\be
p(5,n) = 7
\ee
operators with spin $\ell=5$ and charge $n \geq 5$, while for charge $n=3$ there are only
\be
p(5,3)=5
\ee
operators in total, counted by the first five partitions in (\ref{l=5}).

We can now count primary operators. Obviously, at spin $\ell$, primaries  will be in one to one correspondence with operators that  cannot be obtained by acting with derivatives on all operators with spin $\ell-1$. Therefore the number of primaries is given by
\be \label{eq:primaryCount}
\mathrm{Prim}(\ell,n)\equiv p(\ell,n)-p(\ell-1,n).
\ee

For $\ell\leq n$ this number has a simple interpretation. Namely, it corresponds to the number of partitions of $\ell$, except those that can be obtained from partitions of $\ell-1$ by adding $1$, in other words partitions of $\ell$ containing $1$ should be eliminated \footnote{{That is because when acting with a derivative on an operator involving less than $n$ derivatives, among many terms,  there will always arise one involving a single derivative on $\phi$.}}. As an example, for $\ell=5$ and $\ell =4$ we have
respectively 
\be
\ba{cccccccccc}
5: && (5) & (4,1) & (3,2) & (3,1,1) & (2,2,1) & (2,1,1,1) & (1,1,1,1,1) \\
4: && & (4) & &(3,1) & (2,2) & (2,1,1) & (1,1,1,1).
\ea
\ee
Clearly, the $\ell=5$ primaries are counted by the partitions without $1$, so that for $n\geq 5$ 
\be
\mathrm{Prim}(5,n) = 2\,.
\ee
In the previous subsection we  found that any string  $(a_{00}^\dagger)^{n-\ell}\Pi_\alpha \mathcal A_{\ell_\alpha\ell_\alpha}^\dagger$ forms a primary  with total spin $\ell = \sum_\alpha \ell_\alpha \leq n$. Since there is no $\mathcal A_{1,1}$, it is clear that these primary states correspond to partitions of $\ell$ without 1's. Our counting argument then shows these are all the primaries of our class (polar with $k=0$ and no $\bar \phi$'s).

\paragraph{No $\bar\phi$, arbitrary spin $\ell$, number of derivatives equal to $\ell$}\mbox{}\\
\noindent{For} arbitrary $\ell$, the number of primaries (\ref{eq:primaryCount}) is given by the number of partitions of $\ell$ with each part bigger than 1 and not larger than $n$, i.e. by the number of solutions of the equation
\be
\sum_i \ell_i = \ell, ~~ 1< \ell_i \leq n.
\label{eq:SpinRestrictions}
\ee 
That can be proven as follows. Every partition $t$ can be associated with a Young tableau. For instance, the  partition $t=(4,3,2)$ of $9$  corresponds to
\be
t=\ytableaushort{~~~~,~~~,~~}
\ee
A conjugated Young tableau $t^*$ is defined by interchanging columns and rows, meaning that for the example above $t^*=(3,3,2,1)$
\be
t^*=\ytableaushort{~~~,~~~,~~,~}
\ee
This map obviously establishes an equality between the number $p(\ell,n)$ of partitions of $\ell$ into at most $n$ parts -- i.e. the number of Young tableaux with at most $n$ rows -- and the number of partitions $p^*(\ell,n)$ with parts bounded by $n$ -- i.e. the number of Young tableaux with at most $n$ columns. Therefore, the number of primaries can also be written as
\be \label{eq:count_higher_spin_primaries}
\mathrm{Prim}(\ell,n)=p(\ell,n)-p(\ell-1,n) = p^*(\ell,n)-p^*(\ell-1,n).
\ee
As before, we observe that every tableau counted by $p^*(\ell-1,n)$ can be promoted to a tableau counted by  $p^*(\ell,n)$ by adding a row with just one box,
\be
\ytableaushort{~~~,~~~,~~} \to \ytableaushort{~~~,~~~,~~,~}*[*(gray)]{0,0,0,1}
\ee
therefore, as claimed the number of primaries is given by the number $p^*(\ell, n, 2)$ of Young tableaux with each row bounded by $2\leq\ell_i\leq n$ (see Appendix \ref{app:counting_primaries} for examples).
Clearly, this is equal to the number of products of operators $\mathcal A_{\ell_\alpha,\ell_\alpha}^\dagger$ defined in (\ref{eq:define_C_atomic}) such that $2\leq \ell_\alpha \leq n$ and $\sum_\alpha {\ell_\alpha} = \ell$. 
Notice that, while the counting is still valid, the construction does not work for $\ell > n$, since the $\mathcal A_{\ell,\ell}^\dagger$ operators have charge equal to spin, and thus cannot be used to generate operators with spin higher than the charge.

\paragraph{$\phi$ and $\bar\phi$, arbitrary spin $\ell$, number of derivatives equal to $\ell$}\mbox{}\\
Consider now polar operators with $k=0$ but involving $\bar\phi$ fields. In this case the highest weight elements have  the schematic form $\partial_-^\ell \phi^{n_a} \bar\phi^{n_b}$.
To count the number of such operators, one can first distribute the derivatives as $\partial_-^{\ell-\ell'}\phi^{n_a} \times \partial_-^{\ell'}\bar\phi^{n_b}$, and compute the total number of operators as
\begin{equation}
  \sum_{\ell'=0}^\ell p(\ell-\ell',n_a) p(\ell',n_b) .
\end{equation}
This implies the number of primaries is given by
\begin{align}\label{eq:counting_highest_with_b}
  \mathrm{Prim}(\ell,n_a,n_b) &  = \sum_{\ell'=0}^\ell p(\ell-\ell',n_a) p(\ell',n_b) - \sum_{\ell'=0}^{\ell-1} p(\ell-1-\ell',n_a) p(\ell',n_b) \nonumber \\
  & = \sum_{\ell'=0}^{\ell-1} p^*(\ell-\ell',n_a,2) p^*(\ell',n_b) ,
\end{align}
where we have used the equalities deduced above from Young tableaux. This number is easy to interpret as the number of products of the form
\begin{equation} \label{eq:CDproductMax}
 (a_{00}^\dagger)^{n-\ell} \prod_{\alpha} \mathcal A_{\ell_\alpha,\ell_\alpha}^\dagger \prod_\beta \mathcal B_{\tilde \ell_\beta,\tilde \ell_\beta}^\dagger
\end{equation}
with $2\leq \ell_\alpha \leq n$, $0\leq \tilde \ell_\beta \leq n$ and $\sum_\alpha \ell_\alpha + \sum_\beta \tilde \ell_\beta = \ell$.
Thus, these products of operators are all the highest-weight polar primaries with $k=0$ and $\ell \leq n$.
Again, the counting (\ref{eq:counting_highest_with_b}) is valid for $\ell > n$, but the explicit construction does not apply in this regime.

\paragraph{All operators with number of derivatives bounded by $n$}\mbox{}\\
We finally consider operators made of both $\phi$ and $\bar\phi$ fields and a number of derivatives $P\leq n$, with eventually contracted indices.
We will not provide an explicit formula for the number of primaries in the general case, but will show that primaries are in one-to-one correspondence with operators of the form (\ref{eq:CDproduct}). The following argument is valid in any dimension.

A basis of the linear space of charge-$n$ operators is obtained by considering the set of monomials of the schematic form $\partial_{\mu_1}\dots \partial_{\mu_P}\phi^{n_a}\bar\phi^{n_b}$ with $n_a-n_b=n$ and with the $P$ derivatives distributed on the fields in all possible ways (removing the operators which are made redundant by the equations of motion $\partial^2 \phi = \partial^2\bar\phi=0$). We focus on a finite-dimensional subspace $H_{n_a,n_b,P}$ of fixed $n_a$, $n_b$ and $P$. The counting argument that we will provide works for each of those subspaces individually, and thus extends to the full space of operators. For each subspace, we construct a different basis, in which part of the elements are manifestly descendant states. The remaining elements of the basis span a subspace of same dimensionality as the subspace of explicitly known primary operators. This means that we have successfully identified complete basis of primaries and descendants.

The construction is the following. 
Monomials in the basis can be organized by factoring out all powers of $\phi$ carrying either 0 or 1  derivative \begin{equation} \label{eq:orderedBasis}
	\mathbb B_{n_a,n_b,P} = \left\{ \phi^q (\partial_{\mu_1} \phi) (\partial_{\mu_2}\phi) \cdots (\partial_{\mu_p} \phi)\, O^{(P-p)}_{n_a-p-q,n_b}\ , p \leq P \right\} ,
\end{equation}
with $O^{(p)}_{n, m}$  any monomial involving $n$  $\phi$'s, $m$ $\bar\phi$'s and $p$ derivatives, such that each $\phi$ is derived at least twice.
Notice that all $\partial_{\mu_i}$ factors commute with each other, hence without loss of generality we can assume they are ordered $\mu_1 \leq \mu_2 \leq \dots \leq \mu_p$.
For any $p\leq P$ we will also consider  the sub-basis of operators where $p$ $\phi$'s have a single derivative
\begin{equation} \label{eq:orderedBasisFixedp}
	\mathbb B_{n_a,n_b,P}^p = \left\{ \phi^q (\partial_{\mu_1} \phi) (\partial_{\mu_2}\phi) \cdots (\partial_{\mu_p} \phi)\, O^{(P-p)}_{n_a-p-q,n_b}\ \right\} .
\end{equation}
Now, for $p\geq 1$, we can rewrite the elements of  $\mathbb B_{n_a,n_b,P}^p$ as (for simplicity we write $O$ instead of $O^{(P-p)}_{n_a-p-q,n_b}$)
\begin{equation}
\begin{split}
	\phi^q (\partial_{\mu_1} \phi) (\partial_{\mu_2}\phi) & \cdots (\partial_{\mu_p} \phi)\, O \\
	=  \frac{1}{q+1} \Bigg( & \partial_{\mu_1} \Big( \phi^{q+1} (\partial_{\mu_2}\phi) \cdots (\partial_{\mu_p} \phi)\, O \Big) \\
	& - \sum_{k=2}^p \phi^{q+1} (\partial_{\mu_2} \phi) \cdots (\partial_{\mu_{k-1}}\phi) (\partial_{\mu_{k+1}}\phi) \cdots (\partial_{\mu_p}\phi) \partial_{\mu_1}\partial_{\mu_k} \phi\, O \\
	& - \phi^{q+1} (\partial_{\mu_2} \phi) \cdots ( \partial_{\mu_p}\phi) \partial_{\mu_1}O \Bigg) .
\end{split}
\end{equation}
The term in the first line of the right-hand side is obviously a descendant operator, while the two other lines contain monomials belonging to $\mathbb B_{n_a,n_b,P}^{p-2}$ and $\mathbb B_{n_a,n_b,P}^{p-1}$. 
This process can be repeated, rewriting the operators of the two last lines in the same way, as linear combinations of descendants and members of the lower sub-bases. The process can be iterated until the right hand side is written as a linear combination of descendants and monomials in $\mathbb B_{n_a,n_b,P}^0$. The latter involve no single-derivative $\phi$ fields and cannot be further rewritten.  Our result implies that  the subspace generated by $\mathbb B_{n_a,n_b,P}^p$ has the same primary content as the subspace generated by $\mathbb B_{n_a,n_b,P}^0$. Indeed, as this holds for any $p$, the very  space generated by $\mathbb B_{n_a,n_b,P}$ has the same primary content as the subspace generated by $\mathbb B_{n_a,n_b,P}^0$. We therefore conclude that the subspace of primaries within $H_{n_a,n_b,P}$ is $\leq$ than the number of elements in $\mathbb B_{n_a,n_b,P}^0$.

Our proof can now be completed by comparing the elements in $\mathbb B_{n_a,n_b,P}^0$ to the linearly independent primary states provided in (\ref{eq:CDproduct}). The latter, as we already remarked, are in a one-to-one correspondence with the set (\ref{eq:FockSpaceStates}), barring states involving $a^\dagger_{\ell=1,m}$.
By the operator-state correspondence the traceless symmetric derivatives $\partial_{\mu_1}\dots
\partial_{\mu_r} \phi$ and $\partial_{\mu_1}\dots
\partial_{\mu_s} \bar \phi$ match  respectively $a^\dagger_{r,m_r}$ and $b^\dagger_{s,m_s}$. It is then manifest that the elements in $\mathbb B_{n_a,n_b,P}^0$ and in (\ref{eq:FockSpaceStates}) are in a one-to-one correspondence. 
In particular  the exclusion of $\partial_i \phi$ factors in 
$\mathbb B_{n_a,n_b,P}^0$ crucially matches the exclusion of $a_{1,m}^\dagger$ in  (\ref{eq:FockSpaceStates}), which is mandated  in turn to  match  the building blocks  (\ref{eq:CDproduct}). As the cardinality of the basis $\mathbb B_{n_a,n_b,P}^0$ sets an upper bound to then dimension of the sub-space of primaries, it must be that the states  (\ref{eq:CDproduct}) with the   same $n_a, n_b$ and $P$  are a complete basis for the corresponding space of primaries.

\section{Fock space of superfluid fluctuations\label{sec:superfluid_fluctuations}}

We now turn to the semiclassical description, which applies, as we will see, for sufficiently large charge, regardless of the coupling. The goal of this section is to associate primary operators  to fluctuations around the non-trivial saddle.

\subsection{Fluctuations around a non-trivial saddle \label{sec:saddle}}

A detailed presentation of the large charge semiclassical method can be found in~\cite{Badel:2019oxl,Badel:2019khk}. Here we will outline the main ideas, providing formulae for further reference. 
We present the method in the context of an interacting Wilson-Fisher fixed point, but it applies 
also in free theory: all formulae  can be safely taken for  $\lambda=0$ and $\lambda n = 0$.

\subsubsection{Saddle point}

The method outlined in this section is suitable for computing correlators of the form
\be
\la \bar \phi^n(x_f) \mc O_N(x_N) \dots \mc O_1(x_1)  \phi ^n(x_i) \ra,
\label{eq:N+2Correlators} 
\ee
First, it proves useful to map the theory to  the cylinder $\mathbb R^d \to \mathbb R \times \mathbb S^{d-1}$ (generalizing (\ref{eq:PlaneCylinderCoords}) to to $d$~dimensions). For a generic primary operator, (\ref{eq:PlaneCylinderField}) generalizes to 
\be
\hat {\mc O} (\tau,\vec n) = e^{\D_{\mc O} \tau}\mc O(x),
\label{eq:PlaneCylinderOperators}
\ee
with $\D_{\mc O}$  the scaling dimension of $\mc O$. The theory on the cylinder is equivalent to that on the plane only   at the Wilson-Fisher fixed point, where the theory is conformal. However, to compute corrections we have to work off-criticality, and set the coupling at its critical value only at the end of computations. The advantage of this mapping  is that time-translation, which is a symmetry on the cylinder even off-criticality, corresponds to dilation on the plane, which is not a symmetry off-criticality. The additional symmetry of the non-critical theory on the cylinder makes it easier to find a saddle point. Explicit solutions for the saddle on the plane are known only in $d=3$ and $d=4$ for \REf{eq:phi6} and \REf{eq:phi4} respectively~\cite{Cuomo:2021ygt}.

In terms of the variables on the cylinder the correlator (\ref{eq:N+2Correlators}) has the form
\begin{equation}
 \langle \hat{\bar\phi}^n(\tau_f) \hat{\mathcal O}_N(\tau_N) \dots \hat{\mathcal O}_1(\tau_1) \hat{\phi}^n(\tau_i)\rangle 
 e^{-\Delta_{\phi^n}\tau_f}  e^{-\Delta_N\tau_N} \dots e^{-\Delta_1\tau_1} e^{-\Delta_{\phi^n}\tau_i}\,,
\end{equation}
where for simplicity we did not indicate the dependence of the operators on the  angular coordinates. The operator-state correspondence (see section \ref{sec:oppStateCorr}) yields
\begin{equation}
  |n\rangle = \frac{(4 \pi)^{n/2}}{\sqrt{n!}} \lim_{\tau\to-\infty} e^{-\Delta_{\phi^n}\tau_i} \hat\phi^n(\tau_i) |0\rangle\,,
\end{equation}
and its conjugate
\begin{equation}
  \langle n| = \frac{(4 \pi)^{n/2}}{\sqrt{n!}} \lim_{\tau_f \to \infty} \langle 0 | e^{ \Delta_{\phi^n}\tau_f} \hat{\bar\phi}^n(\tau_f)\,.
\end{equation}
Eq.~(\ref{eq:N+2Correlators}) is thus related to cylinder correlators according to
\be
\lim_{x_f \to \infty} \frac{(4\pi)^n}{n!} x_f^{2\D_{\phi^n}} 
\la \bar \phi^n (x_f) \mc O_N (x_N)\dots \mc O_1(x_1)  \phi ^n(0) \ra = \la n | \hat {\mc O}_N (\tau_N)\dots \hat {\mc O}_1(\tau_1) | n \ra
\prod_{j=1}^N e^{-\D_j \tau_j}.
\label{eq:CorrelatorsPlaneCylinder}
\ee

Since  $|n\rangle$, corresponding to operator $\phi^n$, is the lowest dimension state of charge $n$,   for any  charge $n$ state $|\psi_n\rangle$ with non-zero overlap with $|n\ra$, we have
\begin{align}
  \lim_{\tau_i\to-\infty} e^{H \tau_i} | \psi_n \rangle & = \lim_{\tau_i\to-\infty} e^{\Delta_{\phi^n}\tau_i} |n\rangle \langle n |\psi_n \rangle  \\
  \lim_{\tau_f\to\infty} \langle \psi_n | e^{-H \tau_f} & = \lim_{\tau_f\to\infty} e^{-\Delta_{\phi^n}\tau_f} \langle \psi_n |n \rangle \langle n |,
\end{align}
where $H$ is the Hamiltonian on the cylinder. 

Therefore we can also write
\be \label{eq:CorrelatorCylinderPsi}
\la n | \hat {\mc O}_N (\tau_N)\dots \hat {\mc O}_1(\tau_1)| n \ra = \lim_{\substack{\tau_f \to \infty \\ \tau_i \to - \infty}} \f{\la \psi_n | e^{-H \tau_f} \hat {\mc O}_N (\tau_N)\dots \hat {\mc O}_1(\tau_1) e^{H \tau_i} | \psi_n \ra}{\la \psi_n | e^{-H (\tau_f - \tau_i)} | \psi_n \ra}.
\ee
The right hand side can be represented by a path integral. For that purpose, it is useful to  introduce polar coordinates for the fields
\be
\hat \phi = \f{\rho}{\sqrt{2}} e^{i \c},~~ \hat{\bar \phi} = \f{\rho}{\sqrt{2}}  e^{-i \c}\,,
\label{eq:polarCoordinatesFields}
\ee
and single out their zero modes on the sphere
\bea
\c &=& \c_0+\c_\perp, ~~  \int \c (\vec n) d \Omega_{d-1} = \c_0 \Omega_{d-1}, ~~ \int \c_\perp (\vec n) d \Omega_{d-1} = 0\,,\\
\rho&=&\rho_0+\rho_\perp,~~\int \rho(\vec n)d \Omega_{d-1} = \rho_0 \Omega_{d-1}, ~~\int \rho_\perp (\vec n) d \Omega_{d-1} = 0\,.
\eea
with $\Omega_{d-1} = \frac{2\pi^{d/2}}{\Gamma(d/2)}$ the volume of $\mathbb S^{d-1}$. A convenient choice for  the state~$| \psi_n \ra$ is then
\be \label{eq:decomposeHomogeneous}
\la \rho, \chi | \psi_n \ra = \d(\rho_0-f) \d(\rho_\perp)\d(\c_\perp)e^{i n \c _0},
\ee
with $f$ a constant whose value will be suitably decided below. As a result eq.~(\ref{eq:CorrelatorCylinderPsi}) can be recast as 
\be \label{eq:cylinderPI}
\la n | \hat {\mc O}_N \dots \hat {\mc O}_1| n \ra \underset{_{\substack{\tau_f \to \infty \\ \tau_i \to - \infty}}
}{=} 
\mc Z^{-1} \int d\c_i d\c_f 
e^{- \f{in(\c_f-\c_i)}{\Omega_{d-1}}} 
\int_{\substack{\rho(\tau_i)=f \\ \c(\tau_i)=\c_i}} ^{\substack{\rho(\tau_f)=f \\ \c(\tau_f)=\c_f}}
\mc D \rho \mc D \c  \, \hat{\mc O}_N \dots \hat{\mc O}_1 \, e^{-S[\rho,\c]},
\ee
with
\begin{equation} \label{eq:cylinderPIPartitionFunction}
  \mathcal Z = \int d\c_i d\c_f 
e^{- \f{in(\c_f-\c_i)}{\Omega_{d-1}}} 
\int_{\substack{\rho(\tau_i)=f \\ \c(\tau_i)=\c_i}} ^{\substack{\rho(\tau_f)=f \\ \c(\tau_f)=\c_f}}
\mc D \rho \mc D \c \, e^{-S[\rho,\c]} \,.
\end{equation}
and  where the action is given by
\be \label{eq:polarLagrangian}
S[\rho, \c] = \int d\tau d\Omega_{d-1} \l [ \f{1}{2} (\p \rho)^2 +\f{1}{2} \rho^2 (\p \c)^2 + \f{1}{2}m^2 \rho^2 +V_{int}(\rho) \r]
\ee
with $m=\f{d}{2}-1$ and 
\begin{equation} \label{eq:Vint}
  V_{int}(\rho) = \begin{cases} \frac{\lambda}{16} \rho^4 & \text{ for } (\bar \phi \phi)^2 , \\ 
				\frac{\lambda^2}{288} \rho^6 & \text{ for } (\bar \phi \phi)^3 .
                  \end{cases}
\end{equation}

The saddle point is fixed by two conditions, corresponding to the variation of the action
with respect to $\phi$ in the bulk and on the boundary. The latter, in view of eq.~(\ref{eq:decomposeHomogeneous}),  reduces to variation with respect to the zero modes of $\chi$, $\chi_i$ and $\chi_f$.
From the bulk we have
\bea
\label{eq:chiEOM}
\p_\m \l ( \sqrt{g} g^{\m \n}\rho^2\p_{\n} \c\r ) &=& 0,\\
-\partial^2\rho+\rho\left[(\partial\chi)^2+m^2\right]+\partial_\rho V_{int}(\rho)&=&0
\label{eq:rhoEOM}
\eea
with $g_{\m \n}$ the metric on the cylinder. The first equation, corresponding to variation with respect to $\chi$,   coincides
with $U(1)$ current conservation. The variation at the boundaries gives instead 
\be
(\rho^2 \dot \c) (\tau_i) =(\rho^2 \dot \c) (\tau_f) = - \f{in}{\Omega_{d-1}},
\label{eq:boundaryEOM}
\ee
which fixes the charge to be $n$ and spatially homogeneous at the boundaries. Equations \REf{eq:chiEOM}, \REf{eq:rhoEOM}, \REf{eq:boundaryEOM} along with the constraint \REf{eq:decomposeHomogeneous} have the simple solution
\be
\rho_S (\tau) = f, ~~ \c_S (\tau)= -i \m (\tau-\tau_i)+\c_i,
\label{eq:SaddleSolution}
\ee
with $\m$ and $f$ satisfying
\bea \label{eq:muRelationf}
\m^2-m^2 &=& \f{1}{f} \f{\p V_{int}(f)}{\p f}\,,
\\
f^2 \m &=& \f{n}{\Omega_{d-1}}.
\label{eq:ChargeRelationfnmu}
\eea
A few comments are in order. The last two equations determine  the ``suitable'' value of $f$,  we alluded to below its definition in  \REf{eq:decomposeHomogeneous}. It is only for this specific choice of $f$ in \REf{eq:decomposeHomogeneous} that
the saddle point equations have a solution with a simple linear time dependence. Other choices would give solutions with a more complicated behaviour near the boundaries, but for $\tau_f-\tau_i\to \infty$ the result for \REf{eq:cylinderPI}
would be the same. Notice that as the solution is invariant under the combination $H-\mu Q$ of time translations and charge rotations, $\mu$ should be interpreted as the chemical potential. Finally notice that, while $\chi_f-\chi_i= -i\mu (\tau_f-\tau_i)$ is fixed by \REf{eq:SaddleSolution}, the zero mode $\c_i$ is not: integrating over it guarantees that correlators respect charge conservation.

Eqs. \REf{eq:muRelationf} and  \REf{eq:ChargeRelationfnmu}, for the two choices in \REf{eq:Vint}
imply
\begin{align}
\text{ for } (\bar \phi \phi)^2: \mu_4(\lambda n,d) & = \frac{(d-2)}{2}\frac{\left( 3^{1/3}+ \left[ \frac{9 \lambda n \Gamma(d/2)}{2 \pi^{d/2} (d-2)^3} -  \sqrt{ \left(\frac{9 \lambda n \Gamma(d/2)}{2 \pi^{d/2} (d-2)^3}\right)^2-3 }\right]^{2/3} \right)  }
       {3^{2/3}  \left[ \frac{9 \lambda n \Gamma(d/2)}{2 \pi^{d/2} (d-2)^3} -  \sqrt{ \left(\frac{9 \lambda n \Gamma(d/2)}{2 \pi^{d/2} (d-2)^3}\right)^2-3 }\right]^{1/3} }  \label{eq:mu_d_phi4} , \\
\text{ for } (\bar \phi \phi)^3: \mu_6(\lambda n,d) & = \frac{(d-2)}{2} \frac{\sqrt{1+\sqrt{1+ \frac{\lambda^2 n^2 \Gamma(d/2)^2}{3 \pi^d (d-2)^4}  }}}{\sqrt{2}} .\label{eq:mu_d_phi6}
\end{align}

Expanding around the saddle  we can  systematically compute any observable as a power series in $\lambda$ with coefficients that are themselves functions of $\lambda n$. For instance, given 
\be
\lim_{\substack{\tau_f \to \infty \\ \tau_i \to - \infty}} \la \psi_n | e^{-H (\tau_f - \tau_i)} | \psi_n \ra =  e^{-\D_{\phi^n} (\tau_f - \tau_i)} | \la n | \psi_n \ra|^2,
\ee
and its  path integral representation (\ref{eq:cylinderPIPartitionFunction}), the  evaluation of the action on the saddle point 
immediately gives  the scaling dimension of $\phi^n$ at leading order shown in \REf{eq:phinDimension}.


\subsubsection{Fluctuations \label{sec:SpectrumFluctuations}}

Expanding the fields (\ref{eq:cylinderPI}) around the saddle
\be \label{eq:cylinderFluctuations}
\rho=\rho_S+r, ~~ \c=\c_S+\f{\pi}{f}.
\ee
we can now write
\be
\la n | \hat {\mc O}_N \dots \hat {\mc O}_1| n \ra =
\f{\dst \int d\c_i \int \mc D r \mc D \pi  \, \hat{\mc O}_N \dots \hat{\mc O}_1 e^{-\hat S[r,\pi]}}
{\dst 2\pi \int \mc D r \mc D \pi  \, e^{-\hat S[r,\pi]}} ,
\label{eq:aroundSaddle}
\ee
where the action for the fluctuations is given by
\be
\hat S[r,\pi] = \int d\tau d\Omega_{d-1} \l ( \mc L_2+\mc L _{int} \r ),
\label{eq:FluctuationsLagrangian}
\ee
with
\be
\mc L_2=\f{1}{2} (\p r)^2+\f{1}{2} (\p \pi)^2-2i \m r \dot\pi +\f{1}{2}\l [ V^{''}_{int}(f) - (\m^2-m^2) \r ] r^2,
\label{eq:quadraticLagrangian}
\ee
and
\be
\mc L_{int}= \f{1}{f} \l [ r (\p\pi)^2 - i \m r^2 \dot \pi \r ] +\f{r^2 (\p\pi)^2}{2 f^2} + \l[ V_{int}(f+r) - \left( V_{int}(f) + V^{'}_{int}(f) r + \frac{1}{2} V^{''}_{int}(f) r^2 \right) \right].
\ee
Notice that the $\hat{\mc O}_i$ are local functions of $\rho$ and $\chi$. At the leading order the correlator is  then simply given by the product of the $\hat{\mc O}_i$ computed on  the saddle.

The canonically conjugated momenta\footnote{As in (\ref{eq:aaCylinderMomentumComplex}), the presence of ``$i$'' in front of time derivatives is because we work in Euclidean time.} forming pairs $(r,P)$ and $(\pi, \Pi)$ are
\bea
P = i \dot r, ~~ \Pi = i \dot \pi \l( 1+\f{r}{f} \r )^2+2\m r \l( 1+\f{r}{2f} \r ).
\label{eq:momentaSaddle}
\eea

These variables can be expanded  in harmonic modes as 
\begin{equation}\label{eq:harmonicComponents}
  \begin{pmatrix}
    r(\tau,\vec n) \\ \pi(\tau,\vec n)
    \end{pmatrix}
   = \sum_{\ell=0}^\infty \sum_{\vec m} \begin{pmatrix} r_{\ell \vec m}(\tau) \\ \pi_{\ell\vec m}(\tau) \end{pmatrix} Y_{\ell\vec m}(\vec n)
   \ , \quad
   \begin{pmatrix}
    P(\tau,\vec n) \\ \Pi(\tau,\vec n)
  \end{pmatrix}
   = \sum_{\ell=0}^\infty \sum_{\vec m} \begin{pmatrix} P_{\ell\vec m}(\tau) \\ \Pi_{\ell\vec m}(\tau) \end{pmatrix} Y^*_{\ell\vec m}(\vec n)\,,
\end{equation}
where $Y_{\ell\vec m}(\vec n)$ are the spherical harmonics in $d-1$ dimensions\footnote{$\vec m$ is a multi-index taking \begin{equation} N_{\ell,d} = (2\ell+d-2) \frac{(\ell+d-3)!}{(d-2)! \ell!} \end{equation} different values.} satisfying
\begin{equation}
	\Delta_{\mathbb S^{d-1}} Y_{\ell\vec m} (\vec n) = - J_\ell Y_{\ell\vec m}(\vec n) ,
\end{equation}
where $\D_{\mathbb S^{d-1}}$ is the Laplacian on the sphere $\mathbb S^{d-1}$ and where the eigenvalue $J_\ell$ was given  in \REf{eq:defJl}. The $Y_{\ell\vec m}(\vec n)$ also satisfy the normalization and completeness conditions
\be \label{eq:orthonormalityHarmonics}
\int Y_{\ell\vec m}(\vec n) Y^*_{\ell'\vec m'}(\vec n) d\Omega_{d-1} = \d_{\ell\ell'} \d_{\vec m\vec m'},
\ee
and
\be \label{eq:completenessHarmonics}
\sum_{\ell=0}^\infty \sum_{\vec m} Y_{\ell\vec m}(\vec n) Y^*_{\ell\vec m}(\vec n') = \d^{(\mathbb S^{d-1})} (\vec n-\vec n').
\ee
Notice in particular that $Y_{0\vec 0}=1/\sqrt {\Omega_{d-1}}$.
The harmonic modes are canonical variables satisfying  equal-time commutation relations
\begin{equation} \label{eq:commutationHarmonicComponents}
\begin{split}
  [r(\tau,\vec n),P(\tau,\vec n')] & = i \delta(\vec n-\vec n') \Leftrightarrow [r_{\ell\vec m}(\tau),P_{\ell'\vec m'}(\tau)] = i \delta_{\ell\ell'}\delta_{\vec m\vec m'} ,\\
  [\pi(\tau,\vec n),\Pi(\tau,\vec n')] & = i \delta(\vec n-\vec n') \Leftrightarrow [\pi_{\ell\vec m}(\tau),\Pi_{\ell'\vec m'}(\tau)] = i \delta_{\ell\ell'}\delta_{\vec m\vec m'} ,
\end{split}
\end{equation}
with the other commutators vanishing. 

\subsubsection{Linearized fluctuations}

In section \ref{sec:twoFock} we will need the modes  evolving according to the full lagrangian. To set the basis of perturbation theory and to compute the energy spectrum at lowest order we must however consider the modes of the quadratic Lagrangian \REf{eq:quadraticLagrangian}
\be
\mc L_2=\f{1}{2} (\p r)^2+\f{1}{2} (\p \pi)^2-2i \m r \dot\pi +\f{1}{2}M^2r^2,
\ee
with
\be
M^2 = V^{''}_{int}(f) - f^{-1}V'_{int} (f)= V^{''}_{int}(f) - (\m^2-m^2)\,.
\ee
At this order the canonical momenta are
\bea
\tilde P = i \dot r, ~~ \tilde \Pi = i \dot \pi +2\mu r .
\label{eq:momentalinearized}
\eea
The quantized fields (and the spectrum) are  obtained by considering    the linearized equations of motion 
\be
\l(
\ba{cc}
\p_\tau^2+\D_{\mathbb S^{d-1}} -M^2 & 2i \m \p_\tau \\
-2i \m \p_\tau & \p_\tau^2+\D_{\mathbb S^{d-1}}
\ea
\r)
\l (
\ba{c}
r \\
\pi
\ea
\r)=0,
\label{eq:EOMFluctuationsMatrix}
\ee
and by  finding the complete set of harmonic mode solutions of the form
\begin{equation}
	\begin{pmatrix} r_{\ell\vec m}(\tau) \\ \pi_{\ell\vec m}(\tau) \end{pmatrix} Y_{\ell\vec m}(\vec n) = \begin{pmatrix} C_1 \\ C_2 \end{pmatrix} e^{-\omega \tau} Y_{\ell\vec m}(\vec n)\,.
\end{equation}
For each $\ell$ we find two solutions
\begin{equation}
\begin{split}
\omega_{A}^2(\ell)=J_\ell+\f{V^{''}_{int}(f)+3\m^2+m^2}{2} - \sqrt{\l( \f{V^{''}_{int}(f)+3\m^2+m^2}{2}\r )^2+4 \m^2 J_\ell}\, , \\
\omega_{B}^2(\ell)=J_\ell+\f{V^{''}_{int}(f)+3\m^2+m^2}{2} + \sqrt{\l( \f{V^{''}_{int}(f)+3\m^2+m^2}{2}\r )^2+4 \m^2 J_\ell}\, ,
\label{eq:phiSpectrumV}
\end{split}
\end{equation}
with the corresponding coefficients $C_{1,2}^{A,B}(\ell)$, whose expression we do not need to display.
The $\omega_{A,B}^2(\ell)$ determine the energy spectrum shown in   \REf{eq:phiSpectrum}.
Expanding the fields $(\pi,r)$ in the complete set of solutions and imposing canonical  commutation relation with the conjugated momenta \REf{eq:momentalinearized}, we find
%
\bea 
\label{eq:rpiDYExpansion}
\l (
\ba{c}
r \\
\pi
\ea
\r)
& = &
\l (
\ba{c}
\dst \f{2\m}{\omega_B^2(0)} \, p_\pi \\
\dst \hat \pi -i p_\pi \tau \l ( 1- \f{4\m^2}{\omega_B^2(0)}\r ) 
\ea
\r)
Y_{0\vec 0} \\
&&
+
\sum_{\ell=1}^\infty \sum_{\vec m}
\sqrt{\f{\omega_A(\ell)}{2 \l[ \omega_B^2(\ell)-\omega_A^2(\ell) \r ]}}
\l [
\l (
\ba{c}
\dst \sqrt{\f{J_\ell}{\omega_A^2(\ell)}-1} \\
i \dst \sqrt{\f{\omega^2_+(\ell)}{J_\ell}-1}
\ea
\r)
A_{\ell\vec m}Y_{\ell\vec m} e^{- \omega_A(\ell) \tau}
+
h.c.
\r ]
\nn \\
&&
+
\sum_{\ell=0}^\infty \sum_{\vec m}
\sqrt{\f{\omega_B(\ell)}{2 \l[ \omega_B^2(\ell)-\omega_A^2(\ell) \r ]}}
\l [ 
\l (
\ba{c}
\dst \sqrt{1-\f{J_\ell}{\omega_B^2(\ell)}} \\
-i \dst \sqrt{1-\f{\omega_A^2(\ell)}{J_\ell}}
\ea
\r)
B_{\ell\vec m}Y_{\ell\vec m} e^{-\omega_B(\ell)\tau}
+
h.c.
\r ], \nn
\eea
where operators $(A_{\ell\vec m},A_{\ell\vec m}^\dagger)$, $(B_{\ell\vec m},B_{\ell\vec m}^\dagger)$ and $(\hat \pi,p_\pi)$ are canonically conjugated pairs:
\be
[A_{\ell\vec m},A^\dagger_{\ell'\vec m'}] = \d_{\ell\ell'} \d_{\vec m\vec m'}, ~~ [B_{\ell\vec m},B^\dagger_{\ell'\vec m'}] = \d_{\ell\ell'} \d_{\vec m\vec m'}, ~~ [\hat \pi, p_\pi] = i,
\label{eq:commutationAABBPpipi}
\ee
with all  other commutators vanishing. In the last sum's $\ell=0$ term, one has to use the limit
\begin{equation}
  \lim_{\ell\to 0} \frac{\omega^2_A(\ell)}{J_\ell} = 1- \frac{4\mu^2}{\omega_B^2(0)} .
\end{equation}
Notice that the $A_{\ell\vec m}$ are defined for $\ell\geq 1$ and have frequency $\omega_A(\ell)$, while the $B_{\ell\vec m}$ are defined for $\ell\geq 0$ and have frequency $\omega_B(\ell)$. The role of the $\ell =0$ mode 
in the $A$ sector is played by $\hat \pi$.

Several  features  of \REf{eq:phiSpectrumV} are worth remarking. The first is
\be
\omega_A(0)=0\,.
\ee
This is the manifestation of a Goldstone boson associated with   $U(1)$ symmetry  breaking around the saddle. The $U(1)$ acts as a constant shift of  $\pi$, while $\rho$ is invariant. Therefore $A_{\ell\vec m}$ and $B_{\ell\vec m}$ are all neutral while  $\hat \pi$ transforms by a constant shift. Notice that the conjugated momentum  $p_\pi$ precisely generates these transformations.
Indeed, applying Noether's theorem to the quadratic Lagrangian (\ref{eq:FluctuationsLagrangian}) and comparing the result to the generator $Q$  of $\chi$ shifts in (\ref{eq:polarLagrangian}),  we find
\begin{equation} \label{eq:pPiExact}
  p_\pi = (Q-n)\frac {Y_{0 \vec 0}}{f}.
\end{equation}
Up to a factor, the zero mode $\hat \pi$ is the phase $\chi_i$ that exactly parametrizes the family of solutions at the full non-linear level.
It therefore makes sense to treat this mode fully non-linearly, singling it out
when expressing $\phi$ in terms of the harmonic modes 
  \begin{equation}
  \label{eq:minchia}
	\pi(\tau, \vec n) = \hat \pi Y_{0\vec 0} + \tilde \pi(\tau, \vec n) ,
  \end{equation}
and factoring it out from $\phi$\footnote{Here we have also absorbed the $i\mu \tau_i$ in \REf{eq:SaddleSolution} into $\hat \pi /f$ or equivalently set $\tau_i=0$.}
\begin{equation} \label{eq:fieldPihatFactored}
  \hat \phi(\tau, \vec n) = \frac{f+r}{\sqrt{2}} e^{\mu \tau} e^{i \frac{\hat\pi Y_{00}}{f}} e^{i \frac{\tilde \pi}{f}} \simeq  \frac{f+r}{\sqrt{2}} e^{\mu \tau} e^{i \frac{\hat\pi Y_{0\vec 0}}{f}} (1+i \frac{\tilde \pi}{f}).
\end{equation}
As dictated by the commutation relations and by the definition of the modes, the factor $e^{i \frac{\hat\pi Y_{00}}{f}}$ has charge 1 while the fields $r, \tilde \pi$ are neutral, which is consistent with the transformation property of  $\hat \phi$. Thus $\hat \pi$ is a cyclic coordinate with periodicity $2\pi f/Y_{0\vec 0}$
and  the canonical pair $(\hat \pi,p_\pi)$ does not correspond to a harmonic oscillator with an associated Fock space.  The  Hamiltonian for this pair is 
\be 
H_{\hat \pi} = \f{p_\pi^2}{2} \l [ 1 - \l ( \f{4\m^2}{\omega_B^2(0)} \r )^2 \r ]\,.
\ee

The second important feature is that the $B$-mode is  gapped
\be
\omega^2_B(\ell )\geq\omega^2_B(0)=V^{''}_{int}(f)+3\m^2+m^2>0\,.
\ee
For large $\m$, or equivalently large $\lambda n$ (see \REf{eq:mu_d_phi4} and \REf{eq:mu_d_phi6}), we can then  integrate this mode out and derive an effective field theory description for  the  Goldstone mode~\cite{Hellerman:2015nra,Monin:2016jmo}, which consists of the $A_{\ell\vec m}$ and $\hat \pi$.

The third property is that, for the classically scale invariant cases, $(\bar\phi \phi)^2$ in $d=4$ and $(\bar\phi \phi)^3$  in $d=3$, we have  
\be
\omega_A(1)=1.
\label{eq:spectrumDescendant}
\ee
This equation is associated with the fact that $A_{1\vec m}$ and $A_{1\vec m}^\dagger$ 
are respectively the $K_{\vec {m}}$ and $P_{\vec {m}}$ generators (see for instance \REf{eq:LOgeneratorK}). As such they have scaling dimension $-1$ and $1$. Acting with $A_{1\vec m}^\dagger$ on a state therefore produces a descendant.
Finally, we have that in the free limit, $\lambda =0$, the two modes become
\be
\omega_A(\ell) = \ell, ~~ \omega_B(\ell) = \ell+d-2,
\label{eq:SpectrumLimit0}
\ee
and at finite coupling  their asymptotic behavior is given by
\be
\omega_A(\ell) \underset{\ell\to \infty}{=} \omega_B(\ell)\underset{\ell\to \infty}{=} \ell.
\label{eq:SpectrumLargeL}
\ee
Excitations  around the charge $n$ ground state $|n\ra$ are obtained by acting with 
the neutral modes $A_{\ell\vec m}^\dagger$ and $B_{\ell\vec m}^\dagger$ 
\be \label{eq:FockStates2}
(A^\dagger_{\ell_1m_1})^{n^A_1} \dots (B^\dagger_{j_1 k_1})^{n^B_1} \dots | n \ra .
\ee
Taking now into account that states involving at least one $A_{1\vec m}^\dagger$ are descendants
 leads to the  spectrum of primary operators at leading order which was mentioned at the end of section \ref{sec:epsilonIntro}.
As we said the $\hat\pi,p_\pi$ pair does not produce a Fock space. Instead  $e^{ i \frac{\hat\pi Y_{00}}{f}}$ and $e^{-i \frac{\hat\pi Y_{00}}{f}}$ respectively raise and decrease the charge by one unit, 
thus mapping to the corresponding fixed charge Fock spaces.


\subsection{Relation between different Fock spaces in free theory\label{sec:twoFock}}

Free field theory can be successfully studied around both the trivial $\phi =0$ and the non-trivial \REf{eq:SaddleSolution} saddles. That allows to find explicitly the mapping between the two corresponding Fock spaces. We will do that focussing on the $d=3$ case.

The map between the two spaces corresponds to a canonical transformation resulting from equations \REf{eq:aaCylinderMomentumComplex}, \REf{eq:polarCoordinatesFields}, \REf{eq:cylinderFluctuations} and \REf{eq:momentaSaddle}
\be (\hat \phi, p_{\hat \phi})\,,\quad (\hat {\bar \phi}, p_{\hat {\bar \phi}})
\qquad\Rightarrow\qquad  (r,P)\,, \quad (\pi, \Pi) .
\ee

We use the decomposition of the fields in harmonic components \REf{eq:harmonicComponents}. In three dimensions $m \in \{ -\ell,-\ell+1,\dots, \ell\}$ is a simple index. The fields clearly satisfy
\begin{equation} \label{eq:hermiticityHarmonicComponents}
\begin{split}
  r_{\ell m}(\tau) = (-1)^m \big( r_{\ell,-m}(-\tau) \big)^\dagger , & \quad \pi_{\ell m}(\tau) = (-1)^m \big( \pi_{\ell,-m}(-\tau) \big)^\dagger , \\
  P_{\ell m}(\tau) = (-1)^m \big( P_{\ell,-m}(-\tau) \big)^\dagger , & \quad \Pi_{\ell m}(\tau) = (-1)^m \big( \Pi_{\ell,-m}(-\tau) \big)^\dagger .
\end{split}
\end{equation}

These components are written in terms of the zero mode and creation and annihilation operators yielding
\begin{equation}\label{eq:canonicalTransform0}
\begin{split}
  r_{00}(\tau) & = p_\pi + \frac{1}{\sqrt{2}} \left( B_{00}(\tau) + B_{00}^\dagger (-\tau) \right) ,\\
  \pi_{00}(\tau) & = \hat \pi + \frac{i}{\sqrt{2}} \left( B_{00}^\dagger(-\tau) - B_{00}(\tau) \right) ,\\
  P_{00}(\tau) & = \frac{i}{\sqrt{2}} \left( B_{00}^\dagger(-\tau) - B_{00}(\tau) \right) ,\\
  \Pi_{00}(\tau) & = p_\pi ,
\end{split}
\end{equation}
and (for $\ell > 0$)
\begin{equation}\label{eq:canonicalTransformLM}
\begin{split}
  r_{\ell m}(\tau) & = \frac{1}{2\sqrt{\omega_\ell}} \left[ A_{\ell m}(\tau)+(-1)^m A_{\ell,-m}^\dagger(-\tau)+B_{\ell m}(\tau)+(-1)^m B_{\ell,-m}^\dagger (-\tau) \right] ,\\
  \pi_{\ell m}(\tau) & = \frac{i}{2\sqrt{\omega_\ell}} \left[ A_{\ell m}(\tau)-(-1)^m A_{\ell,-m}^\dagger(-\tau)-B_{\ell m}(\tau)+(-1)^m B_{\ell,-m}^\dagger (-\tau) \right] ,\\
  P_{\ell m}(\tau) & = \frac{i}{2\sqrt{\omega_\ell}} \left[-(-1)^m \ell A_{\ell,-m}(\tau)+ \ell A_{\ell,m}^\dagger(-\tau)-(-1)^m (\ell+1) B_{\ell,-m}(\tau)+(\ell+1) B_{\ell,m}^\dagger (-\tau) \right] ,\\
  \Pi_{\ell m}(\tau) & = \frac{1}{2\sqrt{\omega_\ell}} \left[ (-1)^m (\ell+1) A_{\ell,-m}(\tau)+(\ell+1) A_{\ell,m}^\dagger(-\tau)- (-1)^m \ell B_{\ell,-m}(\tau) - \ell B_{\ell,m}^\dagger (-\tau) \right] . 
\end{split}
\end{equation}
Here we do not consider only the quadratic Hamiltonian for fluctuations around the saddle but take into account the exact solutions of the equations of motion. Thus operators $A_{\ell m}(\tau),B_{\ell m}(\tau)$ have complicated time dependence, not just a simple phase rotation. However, they satisfy the commutation relations \REf{eq:commutationHarmonicComponents}, (\ref{eq:commutationAABBPpipi}) and hermiticity \REf{eq:hermiticityHarmonicComponents} at all $\tau$. At $\tau=0$ they 
coincide with the $\tau$-independent creation-annihilation operators introduced in section \ref{sec:SpectrumFluctuations} for quadratic fluctuations.

Our goal is to express these operators in terms of the ladder operators of vacuum fluctuations. The form of \REf{eq:polarCoordinatesFields} makes the mapping non-linear, which makes it difficult to find a closed form solution. However at large $n$ the solution can be reliably expressed as a systematic expansion in inverse powers of $n$.

We will be studying fluctuations around the lowest energy state with charge $n$, for which $\la a_{00}^\dagger a_{00} \ra \sim n$. The large charge expansion can then be organized by assigning to operators a scaling with $n$\be
a_{00} \sim O(\sqrt{n}), ~~~~~~ a_{\ell\neq0,m}\sim b_{\ell m} \sim O(1).
\label{eq:aanScaling}
\ee
For instance, by singling out  $a_{00}^\dagger a_{00}$ in the expression for  $Q$ \REf{eq:Qaabb}
we can write
\begin{equation} \label{eq:a00a00replacement}
  a_{00}^\dagger a_{00} = n\left\{1 + \frac{1}{n} \left[ Q-n + b_{00}^\dagger b_{00} - \sum_{\ell=1}^\infty \sum_{m=-\ell}^\ell \left(a_{\ell m}^\dagger a_{\ell m} - b_{\ell m}^\dagger b_{\ell m}\right) \right] \right\} ,
\end{equation}
where the term in square brackets represents an $\mathcal O(n^0)$ perturbation. In what follows we treat the fields as classical variables, disregarding issues of ordering. Expressions for quantum operators can be restored, in principle, by finding an appropriate ordering such that the commutation relations are satisfied.

Our goal can be achieved through the following  steps:
\begin{enumerate}
  \item 
Remembering that for   free theory in $d=3$ we have  
    \begin{equation}
      \mu_3(0,3) = \frac{1}{2} , \quad f = \sqrt{\frac{n}{2\pi}}, \quad \omega_\ell = \ell+\frac{1}{2} ,
    \end{equation}
and combining equations \REf{eq:aaCylinderFieldComplex}, \REf{eq:polarCoordinatesFields}, \REf{eq:SaddleSolution} and \REf{eq:cylinderFluctuations} we can write
    \begin{equation}
      \frac{f+r}{\sqrt{2}} e^{\frac{i \pi}{f}} = \sum_{\ell=0}^\infty \sum_{m=-\ell}^\ell \frac{1}{\sqrt{2\omega_\ell}} \left( a_{\ell m}^\dagger e^{\ell \tau} Y_{\ell m}^*(\vec n) + b_{\ell m}e^{-(\ell+1) \tau} Y_{\ell m}(\vec n) \right) \equiv h(\tau, \vec n)\,.
    \end{equation}
It is also convenient to write
    \begin{align}
		r(\tau, \vec n) & = \sqrt{2 h(\tau, \vec n) h(-\tau,\vec n)^\dagger} -f , \\
		e^{\frac{i \pi(\tau,\vec n)}{f}} & = \frac{h(\tau, \vec n)}{\sqrt{ h(\tau,\vec n) h(-\tau,\vec n)^\dagger } } . \label{eq:eipi}
    \end{align}
   Notice that, here and later, we formally treat $a_{0,0}$ and $a_{0,0}^\dagger$ as invertible
as  we are working in a subspace with large charge. For example, we can write
    \begin{equation}
      \frac{1}{\sqrt{ h(\tau,\vec n) h(-\tau,\vec n)^\dagger } } = \frac{1}{\sqrt{ \frac{n}{2\pi} + s(\tau,\vec n)} } \approx \sqrt{\frac{2\pi}{n}} -\sqrt{\frac{2\pi^3}{n^3}} s(\tau,\vec n) + 3\sqrt{\frac{\pi^5}{2n^5}}(s(\tau,\vec n))^2 + \dots
    \end{equation}
    where we used $a_{00}^\dagger a_{00}=n+\dots$ and parametrized all subleading effects by $s(\tau,\vec n)$.
    
    
  \item Using the orthonormality of spherical harmonics \REf{eq:orthonormalityHarmonics}, we extract the harmonic components $r_{\ell m},\pi_{\ell m},P_{\ell m},\Pi_{\ell m}$ from \REf{eq:harmonicComponents}.
    
  \item We finally solve \REf{eq:canonicalTransform0} and \REf{eq:canonicalTransformLM}, for $A_{\ell m},B_{\ell m}$, $\hat \pi$ and $p_\pi$.
\end{enumerate}

\paragraph{Leading order}\hfill

At leading order in the $n^{-1}$ expansion, we get 
\be
p_\pi = 0, ~~
\exp \l[ i \f{\hat \pi }{\sqrt{2n}}  \r]=\f{a^\dagger_{00}}{\sqrt{n}}, ~~ 
B_{\ell m}(\tau) = \f{a_{00}b_{\ell m}}{\sqrt{n}}e^{-(\ell +1)\tau}, ~~
A_{\ell m} (\tau)  = \f{a^\dagger_{00}a_{\ell m}}{\sqrt{n}}e^{-\ell \tau}.
\label{eq:twoFockSpacesRel}
\ee
As explained, the zero-mode $\hat\pi$ is kept in the exponential. This also ensures that the expressions are polynomial (monomial at this order) in the vacuum ladder operators. One further justification of the exponential notation will appear when computing the propagator in Appendix \ref{sec:propagatorCylinder}.

The commutation relations have the  form 
\bea 
\l [ A_{\ell m}, A_{\ell 'm'}^\dagger \r ] & = & 
\f{1}{n} \l (a_{00}^\dagger a_{00} \, \d_{\ell\ell'} \d_{mm'}  -  a_{\ell m} a_{\ell'm'}^\dagger\r )=\d_{\ell\ell'} \d_{mm'}+O\l ( n^{-1}\r ), \\
\l [  B_{\ell m}, B_{\ell'm'}^\dagger \r ] & = & \f{1}{n} \l (a_{00} a_{00}^\dagger \, \d_{\ell\ell'} \d_{mm'}  + b_{\ell m} b_{\ell 'm'}^\dagger  \r )=\d_{\ell\ell'} \d_{mm'}+O\l ( n^{-1}\r ),
\eea
which are canonical at the required accuracy (see \REf{eq:commutationAABBPpipi}).

\paragraph{Next to leading order}\hfill

We find that $\exp\l[ i \f{\hat \pi }{\sqrt{2n}}  \r]$ is still given by  (\ref{eq:twoFockSpacesRel}) while  $p_\pi$ is given by its exact result (\ref{eq:pPiExact}).  The expressions for the other ladder operators are long.  Therefore, here we provide only that  for $A_{\ell m}$, since we will need it in the next section (that for $B_{\ell m}$ can be found in Appendix \ref{sec:appNLOFockStates})
\begin{equation} \label{eq:NLOFockA}
\begin{split} 
	A_{\ell m} =\ & \frac{a_{00}^\dagger a_{\ell m}}{\sqrt{n}} +
		\frac{1}{4(1+2\ell )n^{3/2}} \Big((1+4\ell ) \big( n b_{00} - b_{00}^\dagger (a_{00}^\dagger)^2 \big) a_{\ell m} -2 n b_{00}^\dagger b_{\ell m} \\
					&  \hspace{126pt} + (-1)^m \big( (-1+2\ell ) n b_{00}^\dagger + (1+2\ell ) b_{00} a_{00}^2\big) a_{\ell ,-m}^\dagger  \\
					&  \hspace{126pt} - (-1)^{m} \big( 2(1+\ell ) n b_{00} + 2 \ell  b_{00}^\dagger (a_{00}^\dagger)^2 \big)b_{\ell ,-m}^\dagger \Big) \\
			& + \!\!\!\! \sum_{\substack{\ell_1,\ell_2 > 0 \\ \text{all } m_1,m_2}}  \!\!\!\! \frac{(-1)^m\sqrt{\pi}C_{-m,m_1,m_2}^{\ell ,\ell_1,\ell_2}}{8\sqrt{2\omega_\ell \omega_{\ell_1}\omega_{\ell_2}}n^{3/2}}
			\Big( - (2+3\ell +\ell_1+\ell_2) (a_{00}^\dagger)^2 a_{\ell_1,m_1}a_{\ell_2,m_2}   \\[-20pt]
			&  \hspace{145pt} + 2 (1+\ell -\ell_1+3\ell _2) n b_{\ell_1,m_1} a_{\ell_2,m_2} \\
			& \hspace{145pt} + (\ell -\ell_1-\ell_2) a_{00}^2 b_{\ell_1,m_1}b_{\ell_2,m_2} \\
			& \hspace{145pt} + 2(-1)^{m_2} (2+\ell+3\ell_1+\ell_2) n a_{\ell_1,m_1}a_{\ell_2,-m_2}^\dagger \\
			& \hspace{145pt} - 2 (-1)^{m_1}(1+3\ell-\ell_1+\ell_2) (a_{00}^\dagger)^2 b_{\ell_1,-m_1}^\dagger a_{\ell_2,m_2} \\
			& \hspace{145pt} + 2 (-1)^{m_2} (1+\ell-\ell_1+\ell_2) a_{00}^2 b_{\ell_1,m_1} a_{\ell_2,-m_2}^\dagger \\
			& \hspace{145pt} - 2 (-1)^{m_2}(2-\ell+\ell_1+3\ell_2) n b_{\ell_1,m_1}b_{\ell_2,-m_2}^\dagger \\
			& \hspace{145pt} + (-1)^{m_1+m_2}(2+\ell+\ell_1+\ell_2)a_{00}^2 a_{\ell_1,-m_1}^\dagger a_{\ell_2,-m_2}^\dagger \\
			& \hspace{145pt} - 2 (-1)^{m_1+m_2} (1-\ell+3\ell_1-\ell_2) n b_{\ell_1,-m_1}^\dagger a_{\ell_2,-m_2}^\dagger \\
			& \hspace{145pt} - (-1)^{m_1+m_2}(3\ell-\ell_1-\ell_2)(a_{00}^\dagger)^2 b_{\ell_1,-m_1}^\dagger b_{\ell_2,-m_2}^\dagger
			\Big) \,.
\end{split}
\end{equation}
Here we introduced the Gaunt coefficients
\begin{equation} \label{eq:gaunt}
\begin{split}
  C_{m_1 m_2 m_3}^{\ell_1\ \, \ell_2\ \, \ell_3} & = \int  Y_{\ell_1 m_1} Y_{\ell_2 m_2} Y_{\ell_3 m_3} d\Omega_2 \\
    & = \sqrt{\frac{(2\ell_1+1)(2\ell_2+1)(2\ell_3+1)}{4\pi}} \begin{pmatrix} \ell_1 & \ell_2 & \ell_3 \\ 0 & 0 & 0 \end{pmatrix} \begin{pmatrix} \ell_1 & \ell_2 & \ell_3 \\ m_1 & m_2 & m_3 \end{pmatrix} ,
\end{split}
\end{equation}
given in terms of Wigner $3j$ symbols. These coefficients vanish unless the spins satisfy the triangle inequality
\begin{equation} \label{eq:triangle}
  | \ell_1-\ell_2 | \leq \ell_3 \leq \ell_1+\ell_2  ,
\end{equation}
meaning each spin has to be in the tensor product of the other two. Moreover Gaunt coefficients vanish unless $m_1+m_2+m_3=0$ and $\ell_1+\ell_2+\ell_3$ is even.

Some remarks on (\ref{eq:NLOFockA}) are in order. First, notice that the NLO corrections have  relative size  $n^{-1/2}$. Higher orders behave similarly, resulting in an expansion in powers of $n^{-1/2}$. However, when computing observables, the NLO  terms do not interfere with the leading order terms, resulting in an expansion in powers of  $1/n$, as expected in the semiclassical framework.
 This is exemplified in section \ref{sec:observable2_estimate}. 
 
 When considering  even higher orders, $A_{\ell m}$ will contain sums over $4,6,\dots$ spins with coefficients that, like \REf{eq:gaunt}, are integrals of products of respectively $5,7,\dots$ spherical harmonics. As these coefficients go like powers of  $\ell$ one would expect the parameter controlling the convergence of the expansion to go like  $\frac{\ell^\kappa}{n}$ for some $\kappa$. We will discuss this in detail in 
 section \ref{sec:large_spin}.
 

Finally, notice that  some of the NLO terms don't annihilate the state $|n\rangle$, so that $A_{\ell m}|n\rangle \not = 0$ . The reason is that $|n\rangle$ is the lowest energy state of charge $n$  for the full hamiltonian (the one associated with \REf{eq:FluctuationsLagrangian}), while  $A_{\ell m}$ and $B_{\ell m}$ are the ladder operators  for  the quadratic hamiltonian (associated with \REf{eq:quadraticLagrangian}). The vacuum $|\Omega\rangle$, which is annihilated by $A_{\ell m}$ and $B_{\ell m}$, coincides with $|n\rangle$  only at leading order, hence our result.

\subsection{Mapping superfluid excitations to operators \label{sec:FluidFluctuationsToOperators}}

With  the tools presented in the previous sections, we are now ready to identify operators and map them to  superfluid excitations. The latter,  as defined in \REf{eq:FockStates2},   can be expressed as a power series in
$n^{-1/2}$ of polynomials of $a_{\ell m}, a^\dagger_{\ell m}, b_{\ell m}, b^\dagger_{\ell m}$ acting on the free Fock vacuum $|0\rangle$. These, by the operator state correspondence, can in turn  be written in terms of operators involving $\bar\phi, \phi$ and their derivatives. 

To identify primary states, we must express the special conformal generators in terms of 
$A_{\ell m}, B_{\ell m}$ and $A_{\ell m}^\dagger, B_{\ell m}^\dagger$. This is done by inverting \REf{eq:NLOFockA} and the other formulae relating ladder operators in the two frames, plugging the result in (\ref{eq:K0aabb}), (\ref{eq:K-aabb}), and (\ref{eq:K+aabb}). For instance, at leading order, using (\ref{eq:twoFockSpacesRel}), we get
\be \label{eq:LOgeneratorK}
K_0 = \sqrt{n} A_{1,0} , ~~ K_- = -\sqrt{n} A_{1,-1}, ~~ K_+ = \sqrt{n} A_{1,1}.
\ee
Thus, as was already discussed, at leading order only strings of creation operators not containing $A_{1,m}^\dagger$ are primaries.
There is a clear parallel with the conclusion of section \ref{sec:vacuum_primaries_construction}. This is due to the fact that, at leading order, states generated by creation operators $A_{\ell m}^\dagger, B_{\ell m}^\dagger$ correspond to the states generated by $\mathcal A_{\ell,m}^\dagger, \mathcal B_{\ell,m}^\dagger$.

As a result, due to the following identities
\be
A^\dagger_{\ell m} | n \ra = \f{a_{00}a_{\ell m}^\dagger}{\sqrt{n}} \f{ (a_{00}^\dagger )^n}{\sqrt{n!}} | 0 \ra=
\f{a_{\ell m}^\dagger (a_{00}^\dagger )^{n-1}}{\sqrt{(n-1)!}} | 0 \ra = 
\f{(4\pi)^{\f{n-1}{2}}}{\sqrt{(n-1)!}}\mathcal Y_{\ell m}^{\mu_1\dots\mu_\ell} \phi^{n-1} \partial_{\mu_1} \cdots \partial_{\mu_\ell} \phi 
|0\rangle,
\ee
the state $A^\dagger_{\ell m} | n \ra$ corresponds at leading order  to an operator with  $\ell$ derivatives all acting on the same field
\be
\label{eq:LOSpinlOperator}
\phi^{n-1} \p_{\m_1\dots\m_\ell}\phi.
\ee


\section{How large is ``large spin'' ?\label{sec:large_spin}}

Quantization around the saddle offers a systematic computation  of observables for states with charge $n$ as a power series in $n^{-1}$. Clearly, as $n\to \infty$ the procedure works for states with finite spin $\ell$, for the ground state $| n \ra$ in particular. In this section we will study the convergence of the expansion when both $\ell$ and $n$ become large.
%

\subsection{Matrix elements for excited states\label{sec:observable1_estimate}}

On general grounds we expect the expansion to be controlled by the ratio $\ell^\kappa/n$ for some $\kappa$. One way
 to find out what  $\kappa$ is, would be to perform NLO computations around the non-trivial saddle. However, we'll make use of the fact that we know  a class of primary states in free theory in exact form  and not just as an expansion in inverse powers of the charge. That  will give us full control of the computation, allowing to successfully trace  any  transition between different regimes (see Section~\ref{sec:primary_phonons}).

Intuitively we expect  the radial component of $\phi$ to be a good parameter to control the validity of the semiclassical approximation. The smallness of the size of its quantum fluctuation  relative to its expectation value is a
necessary condition for the semi-classicality of a state
\footnote{For an  illustrative example based on the spinning top see~\cite{Monin:2016jmo}.}. Fluctuations comparable to the expectation value, and thus consistent with the vanishing of $\phi$ (at least somewhere), signal the breakdown  of the semiclassical approximantion.

We will thus  study the large $n$ behavior of the following matrix elements 
\be
\Phi(\t;\ell,n,p) =\prescript{}{A}\la n; \ell, \ell | : \p_\tau^p\hat {\bar \phi} (\tau, \vec n) \p_\tau^p \hat \phi (\tau, \vec n) : | n; \ell, \ell \ra_A .
\label{eq:OrderParameter}
\ee
for arbitrary integer $p$, where $|n,\ell,\ell\rangle_A$ is the primary state found in (\ref{eq:one-phononStateA}).

Rewriting the fields in terms of ladder operators \REf{eq:aaCylinderFieldComplex} and \REf{eq:aaCylinderFieldComplexConj} yields
\begin{equation}
  \Phi(\tau; \ell, n, p) = \alpha_0^2 \sum_{k,k' = 0}^{\ell} \langle \psi_k | \sum_{\substack{ \ell',m' \\ \ell'',m''} } (-1)^p \omega_{\ell'}^p \omega_{\ell''}^p a^\dagger_{\ell',m'} a_{\ell'',m''}Y^*_{\ell',m'} Y_{\ell'',m''} 
\f{e^{\l (\omega_{\ell'}-\omega_{\ell''} \r ) \tau}}{\sqrt{4 \omega_{\ell'} \omega_{\ell''} }} | \psi_{k'} \rangle,
\end{equation}
where we introduced the following notation
\be \label{eq:PrimaryPsiK}
| \psi_k \ra = \g_{k,\ell} (a_{00}^\dagger)^{n-k-1} ( a_{1,1}^\dagger )^{k} a_{\ell-k, \ell-k}^\dagger | 0 \ra.
\ee

Using that for $k, k' \neq \ell-1,\ell$ we have
\begin{align}
 & \la 0 | a_{00}^{n-k-1} a_{11}^{k}a_{\ell-k,\ell-k} a^\dagger_{\ell',m'} a_{\ell'',m''} ( a_{00}^\dagger )^{n-k'-1} ( a_{11}^\dagger )^{k'} a_{\ell-k',\ell-k'}^\dagger | 0 \ra \nonumber \\
 & {} = (n-k-1)! k! \left[ (n-k-1) \delta_{\ell' 0} + k \delta_{\ell' 1} + \delta_{\ell', \ell-k}\right]  \delta_{\ell' \ell''} \delta_{\ell' m'} \delta_{\ell' m''} \delta_{k k'},
\end{align}
and neglecting the terms with $k,k' = \ell-1,\ell$, which are subleading, we get
\begin{equation} \label{eq:OrderParameterSum}
  \Phi (\tau; \ell, n, p) = (-1)^p \alpha_0^2 \sum_{k=0}^{\ell-2} \frac{(2\ell)!(n-k-1)!}{2^{k+1} k! (2\ell-2k)!} \l [ (n-k-1) |Y_{00}|^2 \omega^{2p-1}_0  + k |Y_{11}|^2 \omega_1^{2p-1}+|Y_{\ell-k,\ell-k}|^2 \omega_{\ell-k}^{2p-1}  \r ] .
\end{equation}
We then  use
\begin{equation}
  | Y_{\ell\ell}(\varphi,\theta) | = \frac{1}{2^\ell \ell !}\sqrt{\frac{(2\ell+1)!}{4\pi} } \sin^{\ell} \theta\,,
\end{equation}
which means  $|Y_{\ell-k,\ell-k}|^2$ is maximal at $\theta=\pi/2$. Approximating factorials by Stirling's formula, we finally find 
\begin{align}
\Phi(\pi/2;\ell,n,p) = \f {n}{4^{p+1}\pi} \Bigg[ & Q_0(\ell,p) + \frac{Q_1(\ell,p)}{n} + \frac{Q_2(\ell,p)}{n^2} + \dots  \nonumber \\
& + \f{\ell^{\xi}}{n}\l (P_0(\ell,p) + \f{P_1(\ell,p)}{n}+\f{P_2(\ell,p)}{n^2} + \dots \r ) \Bigg] ,
\label{eq:OrderParameterResult}
\end{align}
where $P_k(\ell)$ and $Q_k(\ell)$ are $n$-independent functions which at large $\ell$ scale as $\ell^k$, and $\x = 2p-\f{1}{2}$.

It can be concluded that for the case at hand $\kappa=1$. In other words, the semiclassical expansion  can be trusted as long as $\ell \ll n$. We expect that for a wide class of observables, even for theories with interaction, computations around the non-trivial saddle can be organized in a systematic series in powers of $\ell /n$. Another instance is examined in Appendix \ref{sec:observable2_estimate}. However, not all quantities have this type of expansion as we now discuss. 

\subsection{Primary states \label{sec:primary_phonons}}
\begin{figure}[h]
\begin{center}
\includegraphics[width=10cm]{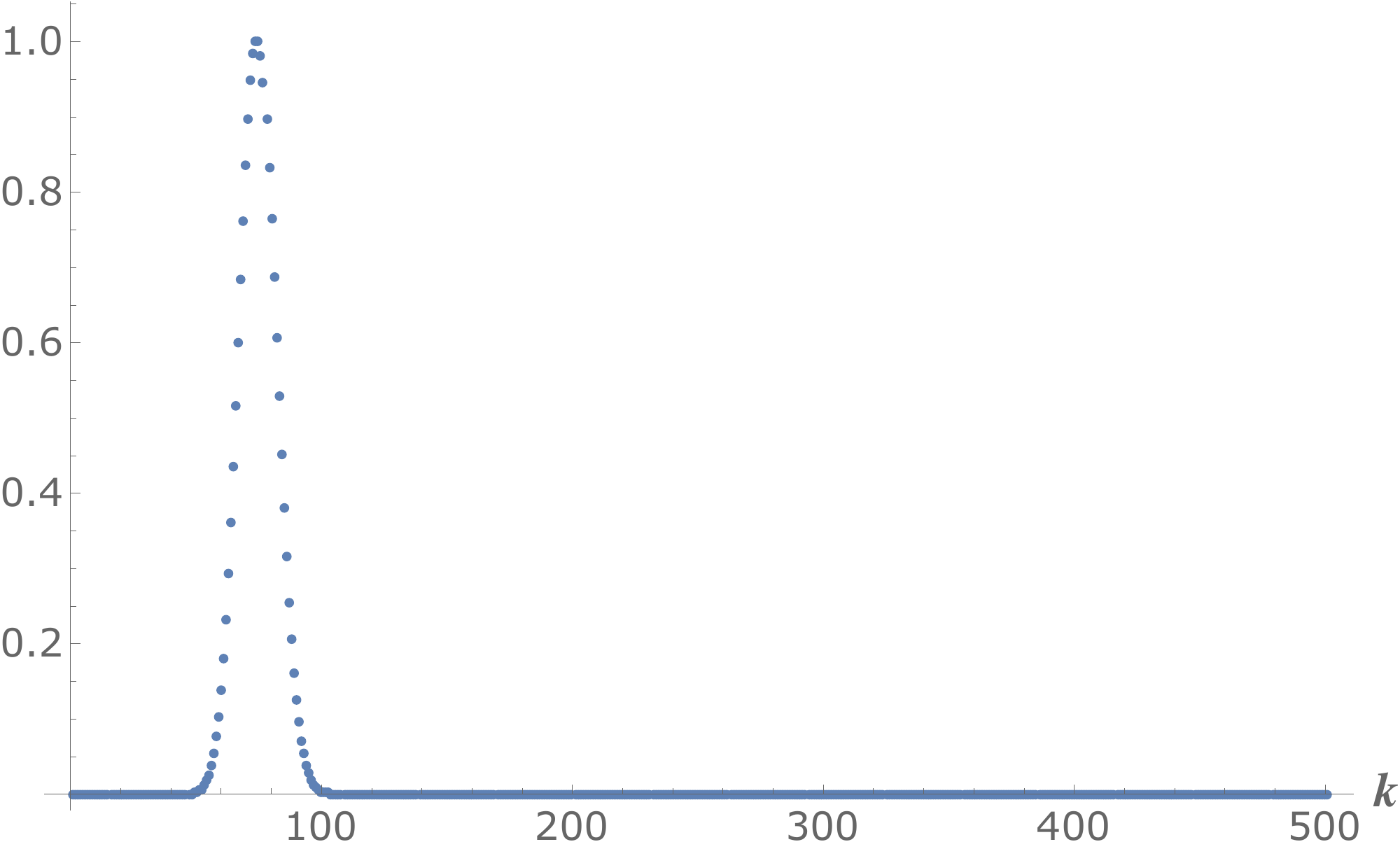}
\caption{\label{fig:relNorm} Normalized $\la \psi_k |\psi_k \ra$ as function of $k$ for $n=10^3$ and $\ell =500$.}
\end{center}
\end{figure}

Let us consider $1/n$ corrections to the operator whose leading term is given by \REf{eq:LOSpinlOperator} and whose associated state   is given in exact form  by \REf{eq:one-phononStateA} \footnote{The spin $\ell$ is bounded by $2 \leq \ell < n$.}. By the notation \REf{eq:PrimaryPsiK} we can write the state succinctly as
\be \label{eq:primSumPsiK}
	|n; \ell, \ell\rangle_A = \alpha_0 \sum_{k=0}^\ell |\psi_k\rangle\,.
\ee
The vectors $|\psi_k\rangle$ are mutually orthogonal, but they are not normalized. Comparing their relative norms we find
\be
\label{eq:expansionParamNorm}
\f{\la \psi_k | \psi _k \ra}{\la \psi_{k-1} | \psi _{k-1} \ra} =\f{(\ell-k+1)(2\ell-2k+1)}{(n-k)k}\sim \frac{\ell^2}{nk}\,,
\ee
where in the last equation we used $k\leq \ell\ll n$. This equation implies the norms $\la \psi_k |\psi_k \ra\propto (\ell^2/n)^k/k!$ approximate  the coefficients in the expansion of the exponential $\exp(\ell^2/n)$. We then have two regimes depending on whether $\ell^2/n\ll 1$ or $\ell^2/n\gsim 1$. In the first case  the  succession $\la \psi_k |\psi_k \ra$ is peaked at $k=0$. Instead, for 
$\ell^2/n\gsim 1$ the succession is peaked at
\be \label{eq:primaryKMax}
  k_{max} = \f{2\ell^2}{n},
\ee
and has a width  of order $\sqrt{k_{max}} = \ell / \sqrt{n}$ (see Figure~\ref{fig:relNorm}). Thus, the primary state \REf{eq:one-phononStateA} is dominated by the sum of $| \psi _k \ra$ roughly in the range $ k_{\max}-\sqrt{k_{\max}}\lsim k\lsim k_{\max}-\sqrt{k_{\max}}$.

This result seems to suggest  that, for primary states,  the  $1/n$ expansion  (\ref{eq:one-phononStateA}), or equivalently \REf{eq:primSumPsiK},
 breaks down at $\ell \sim \sqrt{n}$. However, the expressions for (primary) operators are coordinate dependent. What we have shown  here is that, when  expressed in terms of creation-annihilation operators $a_{\ell,m}$, $a^\dagger_{\ell,m}$, primary operators are written as power series in $\ell/\sqrt{n}$. There may exist other coordinates that partially resum the series leading to a manifest expansion in powers of $\ell/n$. The mere fact that the expectation value \REf{eq:OrderParameter}, which is coordinate-independent, is presented as a power series in $\ell/n$, speaks in favor of that possibility.

Unfortunately, those coordinates are certainly not the creation-annihilation operators corresponding to phonons $A_{\ell,m}$, $A^\dagger_{\ell,m}$. Indeed, rewriting the first two terms in (\ref{eq:one-phononStateA}) using the leading order relation \REf{eq:twoFockSpacesRel}, gives
\be
| n, \ell, \ell \ra _ A \underset{\ell \gg 1}{=} \a_0 \sqrt{(n-1)!}  \l ( A_{\ell\ell}^\dagger - \f{\sqrt{2}\ell}{\sqrt{n}}A_{\ell-1,\ell-1}^\dagger
A_{1,1}^\dagger \r)  |n\rangle .
\ee
One may hope that the second term in parenthesis is cancelled by NLO corrections \REf{eq:NLOFockA}, however, it is straightforward to show that it is not the case. We can show using \REf{eq:gauntL1} that the only potentially relevant term in \REf{eq:NLOFockA}
\be
\frac{(-1)^{\ell-1}\sqrt{\pi}C_{-\ell,m_1,m_2}^{\ell ,\ell_1,\ell_2}}{8\sqrt{2\omega_\ell \omega_{\ell_1}\omega_{\ell_2} n}}
(2+3\ell +\ell_1+\ell_2) A^\dagger_{\ell_1,m_1}A^\dagger_{\ell_2,m_2},
\ee
scales as $O(\ell^0)/\sqrt{n}$, for $\ell_1=m_1=\ell-1$, $\ell_2 =m_2= 1$, so it cannot cancel the term scaling as $\ell/\sqrt{n}$.

Our conclusion of this section is that the semiclassical expansion can be trusted for spins as large as the $U(1)$ charge, $\ell \sim n$, as long as we are dealing with coordinate-independent quantities. On the other hand, if we want to identify primary states, using creation-annihilation operators corresponding to phonons, perturbative expansion breaks down much earlier, for $\ell \sim \sqrt{n}$. We expect that for spins in the window $\sqrt{n}<\ell \ll n$ there should exist different semiclassical backgrounds, expanding around which would allow to describe primary states perturbatively\footnote{Expanding the summand in \REf{eq:PrimaryNormalization} for large $\ell$ and $n$ and computing the sum via saddle-point approximation leads to
\be
\sum_{k=0}^\infty \f{1}{k!} \l ( \f{2\ell^2}{n}\r )^k \exp \l ( -\f{k^2}{\ell} \r ) = \exp \l \{ \f{2\ell^2}{n} \l [ 1-\f{2\ell}{n} + O \l ( \f{\ell^2}{n^2} \r )\r ] \r \},
\ee
which suggests that this result can be obtained perturbatively in a double scaling limit $n \gg 1$, $\ell \gg 1$, $\ell / n =\text{fixed}.$ 

}.

\section{3-pt function \label{sec:3ptfunction}}

\begin{figure}[h]
  \centering
  \includegraphics{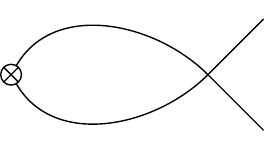}
  \caption{\label{fig:phiphiRenormalization} One-loop renormalization of $\bar\phi \phi$.}
\end{figure}

In the next two sections we will further explore the semiclassical methodology described in Section~\ref{sec:saddle}. Focussing on the Wilson-Fisher fixed point  in $4-\varepsilon$ dimensions, corresponding to the theory in eq. \REf{eq:phi4}, we will derive new results by studying $3$- and $4$-point functions involving two operators with large charge $n$  at next to leading order in $\varepsilon$ (or equivalently in $n^{-1}$). More precisely, we will compute correlators of the class presented in Eq.~\REf{eq:N+2Correlators} involving one or two additional operators $\mc O_i$, i.e. $N$ equals 1 or 2. For simplicity we will focus on insertions of just one specific type of neutral operators
\be
\mc O (x) = (\bar \phi \phi)^k (x).
\ee
We start from the $3$-point function of $\bar\phi\phi$, which, up to the normalization, is fully determined by the scaling dimensions and a single fusion coefficient. The scaling dimension of $\phi^n$ is given by \REf{eq:phinDimension}, while that of $\bar \phi \phi$ can be easily computed using standard perturbation theory through Feynman diagrams as we will see shortly. As a result the only parameter to compute is the fusion coefficient $\lambda_{\bar \phi \phi}$, which appears in the 3-pt function of canonically (re-)normalized operators 
$[{\cal O}_i]$ as\footnote{Canonical normalization corresponds to $\l\la 0| [{\cal O}](x)[{\cal O}](y)|0\r\ra=(x-y)^{-2\Delta_{\cal O}}$.}
\be
\l \la [\bar \phi^n] (x_f) [\bar \phi \phi] (x) [ \phi^n](x_i) \r \ra=\f{\lambda_{\bar \phi \phi}}{(x_f - x_i)^{2\D_{\phi^n}-\D_\mc O} (x-x_i)^{\D_\mc O} (x_f-x)^{\D_\mc O}}.
\ee
On the cylinder, using (\ref{eq:PlaneCylinderOperators}), one can more simply write
\be
 {\lambda_{\bar \phi \phi}}=\lim_{\substack{\tau_f\to \infty \\ \tau_i\to -\infty}}\frac{\la 0| [ \widehat{{\bar\phi}^n}](\tau_f,\vec n_f) [\widehat{\bar \phi \phi}] (\tau, \vec n) [ \widehat{ \phi^n}](\tau_i,\vec n_i) | 0 \ra}{\la 0| [ \widehat{{\bar\phi}^n}](\tau_f,\vec n_f)  [ \widehat{ \phi^n}](\tau_i,\vec n_i) | 0 \ra}
 \equiv \la n|[\widehat{\bar \phi \phi}](\tau,\vec n)|n\ra .
\label{lambdacyl}
\ee
For the theory and the operators at hand, renormalization is multiplicative, so that canonically normalized and bare operators are related by $[{\cal O}_i]={\cal O}_i/Z_i$, with $Z_i$ generally UV divergent. For instance, the 2-point function of $\bar \phi \phi$ is given by 
\be
\la (\bar \phi \phi)(x) (\bar \phi \phi)(y) \ra =\f{Z^2_{\bar \phi \phi}}{(x-y)^{2\D_{\bar \phi \phi}}},
\ee
where at one loop order, i.e. just the diagram in Fig.~\ref{fig:phiphiRenormalization}, 
 \be
Z_{\bar \phi \phi} = \Omega^{-1}_{d-1} (d-2)^{-1} \l [1-\f{\lambda}{8\pi^2} \, \f{1}{4-d} \r ] \l [ 1- \f{\lambda}{16\pi^2} \l ( 1+\g+\log \pi \r ) \r ],
\label{eq:NormRenorm}
\ee
which implies the scaling dimension is
\be
\Delta_{\bar\phi \phi}\equiv (d-2)+\g_{\bar \phi \phi} = (d-2)+ \f{\lambda}{8\pi^2}.
\label{eq:phiphiAnomalousDimension}
\ee

For large $n$, \REf{lambdacyl} can be computed semiclassically by
 expanding around the saddle point (\ref{eq:SaddleSolution}). Equation (\ref{eq:aroundSaddle}) yields in this case
\be
\lambda_{\bar \phi \phi} 
=Z^{-1}_{\bar \phi \phi} \, 
\f{\dst \int \mc D r \mc D \pi  \, (\widehat{\bar\phi\phi}) (\tau, \vec n) e^{-\hat S[r,\pi]}}
{\dst \int \mc D r \mc D \pi  \, e^{-\hat S[r,\pi]}},
\label{eq:SaddleFusion}
\ee
where the path integrals have the boundary conditions specified by  \REf{eq:cylinderPI}.

\paragraph{Leading order:} the computation boils down to evaluating the integrands on the saddle, leading to
\be
\label{eq:fusionPhiPhi}
\lambda_{\bar \phi \phi} = f^2 \Omega_3 = \f{n}{\mu_*},
\ee
where we used \REf{eq:ChargeRelationfnmu} and the leading order result $Z_{\bar \phi \phi}^{-1} = 2 \Omega_{3}$. For small 
$\lambda n$ we have $\mu_*=1$, see \REf{eq:mu_d_phi4}, and the result, $\lambda_{\bar \phi \phi}=n$, coincides with the tree level computation using Feynman diagrams.
In this section the symbol $\mu$ refers to $\mu_4(\lambda n, d)$ while $\mu_*$ refers to $\mu_4(\lambda_* n,4)$. Notice this is the chemical potential of the 4D theory evaluated at the critical coupling  of the theory in  $d=4-\varepsilon$,  given in \REf{eq:phi4FixedPoint}.

\begin{figure}[H]
  \centering
  \includegraphics[width=7cm]{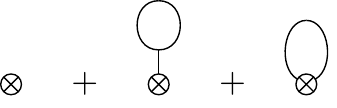}
  \caption{\label{fig:lambdaNLO} Topology of diagrams entering $\langle n | ( \bar \phi \phi ) | n \rangle $ at NLO }
\end{figure}

\paragraph{Next to leading order.} The result is independent of the choice of $(\tau, \vec n)$ in (\ref{eq:SaddleFusion}), therefore, we make the convenient choice $(\tau, \vec n)=(0,\hat n_d)$, with
\be
\hat n_{d}=(\underbrace{0,0,\dots,0,1}_{d}).
\label{eq:direction}
\ee
By expanding around the saddle, the expectation value of the bare operator is then
\be
\l \la n | \l ( \bar \phi \phi \r )(0, \hat n_{d}) | n \r \ra = \f{1}{2} \l \la n | f^2 + 2 f r (0, \hat n_{d}) + r^2(0, \hat n_{d}) | n \r \ra,
\label{eq:3ptFunc1}
\ee
which at NLO, i.e. 1-loop, gives
\be
\l \la n | \l ( \bar \phi \phi \r )(0, \hat n_{d}) | n \r \ra =
\f{f^2}{2} - \l \la r(0, \hat n_{d}) \int d\tau d \Omega_{d-1}\l [ r (\p\pi)^2 - i \m r^2 \dot \pi +\f{\lambda f^2 r^3}{4} \r ] \r \ra +
\f{1}{2}\l \la  r^2(0, \hat n_{d})\r \ra,
\label{eq:3ptFunc2}
\ee
where within the $\l\la\dots \r\ra$ the fields $r, \pi$ are free fields \REf{eq:rpiDYExpansion} propagating according to the quadratic action expanded around the background. The resulting  Feynman diagrams are depicted
 in Fig.~\ref{fig:lambdaNLO}. 
The first step is to find the propagator of $(r,\pi)$. In matrix form this can be written as
\be
D(\tau-\tau',\vec n\cdot\vec n') = \sum_\ell F^{(\ell	)}(\tau-\tau') C^{(d/2-1)}_\ell(\vec n\cdot\vec n'),
\ee
where $C^{(d/2-1)}_{\ell}(\cos \t)$ are Gegenbauer polynomials and $F^{(\ell)}(\tau)$ is a $2\times 2$ matrix whose exact expression is given in Appendix \ref{sec:propagatorCylinder}. The details of the computation can be found in Appendix~\ref{app:lambdaComputation}. The result is
\be
\lambda_{\bar \phi \phi}
=
\f{n}{\m_*}+\f{2(3\m_*^2+1)}{\l [ 2(3\m_*^2-1) \r]^{3/2}}-\f{3-2\m_*^2+3\m_*^4}{2(3\m_*^2-1)}
+\sum_{\ell=1}^\infty \l [ S_\ell(\m_*)  - c_{-1}(\m_*)  \ell -c_{0}(\m_*) -\f{c_{1}(\m_*)}{\ell} \r ],
\label{eq:FusionFinal}
\ee
with
\begin{equation}
  S_\ell(\mu) \equiv S_\ell(\mu,1,4), 
\end{equation}
where
\be
S_\ell(\mu,m,d)= 
\f{2\ell+d-2}{\omega_B^2(0)} \, \f{\omega_B(\ell) \omega_A(\ell) (3\m^2+m^2)-J_\ell(\m^2-m^2)}{\omega_B(\ell) \omega_A(\ell) 
\l [\omega_B(\ell) + \omega_A(\ell)\r ]} \, C_\ell^{(d/2-1)}(1),
\label{eq:Sl_definition}
\ee
while  the  coefficients $c_{-1,0,1}(\m)$ are defined by the asymptotic behavior of the summand
\be
S_\ell(\mu) \underset{\ell\to \infty}{\equiv} c_{-1}(\m) \ell + c_{0}(\m) +  \f{c_{1}(\m)}{ \ell} + \dots,
\ee
so as to render the sum in \REf{eq:FusionFinal} finite. Their exact values are
\be
c_{-1}(\m) = c_0 (\m) = \f{\m^2+1}{3\m^2-1}, ~~ c_1 (\m) = -\f{\m^4+2\m^2-3}{2(3\m^2-1)}.
\ee

As discussed in the Appendix, the transcendental terms proportional to the Euler constant $\gamma$, and to $\ln \pi$, which normally appear in 1-loop expressions, cancel out once we fix $\lambda = \lambda_*$.

For small $\lambda_* n$, expanding $\m_*$ in a power series in $\lambda_* n$
\be
\m_* = 1+\f{\lambda_* n}{16\pi^2}-\f{3}{2}\l( \f{\lambda_* n}{16\pi^2} \r )^2+O(\lambda_*^3n^3),
\ee
we get
\be
\lambda_{\bar \phi \phi} \underset{\lambda_* n \to 0}{=} n \l [ 1- \f{\lambda_* n}{16\pi^2}+\f{5}{2}\l( \f{\lambda_* n}{16\pi^2} \r )^2 + O (\lambda_* n)^3 \r ]+ \l [ 6\z^2(3)-\f{13}{2} \r ]\l( \f{\lambda_* n}{16\pi^2} \r )^2 + O (\lambda_* n)^3 + O (n^{-1}).
\ee
While for large $\lambda_* n$, and therefore  $\m_*\gg 1$,  the sum in \REf{eq:FusionFinal}  approximately satisfies 
\be
\sum_{\ell=1}^\infty \l [ S_\ell(\m_*,1,4)  - c_{-1}(\m_*)  \ell -c_{0}(\m_*) -\f{c_{1}(\m_*)}{\ell} \r ] \underset{\mu \to \infty}{=} \f{1}{6} \m_*^2 \log \m_*.
\ee
Combining that with the leading contribution $n/\m_*$ and using the relation $\m_*^3=\lambda_* n / 4 \Omega_{3}$, which applies in the large $\lambda_* n$ regime,  we get
\bea
\lambda_{\bar \phi \phi} &\underset{\lambda n \to \infty}{=}& \f{8\pi^2}{\lambda_*} \l ( \f{\lambda_* n}{8\pi^2} \r )^{2/3} \l ( 1+\f{\lambda_*}{144\pi^2} \log \f{\lambda_* n}{8\pi^2}\r ) \nn \\
& \simeq &
\f{8\pi^2}{\lambda_*} \l ( \f{\lambda_* n}{8\pi^2} \r )^{\f{2}{3}+\f{\lambda_*}{144\pi^2}} = 
\f{5}{2\eps} \l ( \f{2\eps n}{5} \r )^{\f{2}{3}+\f{\eps}{45}} \sim n^{\f{\D_{\bar \phi \phi}}{d-1}},
\eea
where $\D_{\bar \phi \phi}$ is given by  eq.~\REf{eq:phiphiAnomalousDimension} with $\lambda\to \lambda_*$.
The   scaling with $n$ is precisely as predicted by the large charge EFT description~\cite{Monin:2016jmo}.

\section{4-pt function \label{sec:4ptfunction}}

Focussing again on the Wilson-Fisher fixed point in $d=4-\varepsilon$  we will now study, by the same methodology,   the four point function with two insertions of $(\bar \phi \phi)$. 

Let us recall that a general 4-point correlator in  a CFT can be written using $s$- and  $t$-channel representations
\bea
\la \mc O_4 (x_4) \mc O_3 (x_3) \mc O_2 (x_2) \mc O_1 (x_1) \ra 
& = & 
\f {g_{12,34}(z, \bar z)} {x_{12}^{\D_1+\D_2} x_{34}^{\D_3+\D_4}} \l ( \f {x_{24}} {x_{14}} \r ) ^{\D_1-\D_2} \l ( \f {x_{14}} {x_{13}} \r ) ^{\D_3-\D_4} 
\label{eq:tChannel4ptFunction} \\
& = &
\f {g_{32,14}(1-z, 1-\bar z)} {x_{32}^{\D_3+\D_2} x_{14}^{\D_1+\D_4}} \l ( \f {x_{24}} {x_{34}} \r ) ^{\D_3-\D_2} \l ( \f {x_{34}} {x_{13}} \r ) ^{\D_1-\D_4}
\eea
where $z$ and $\bar z$ are defined by the conformal ratios according to 
\be
u=\bar z z=\f {x_{12}^2 x_{34}^2} {x_{13}^2x_{24}^2},~~ v=(1-z)(1-\bar z)=\f {x_{14}^2 x_{23}^2} {x_{13}^2x_{24}^2}.
\label{eq:Definitionuvzz}
\ee
Modulo kinematic factors, the relevant information is encapsulated in the $g_{ij,kl}(z,\bar z)$. 

For Euclidean signature, the two variables $z\equiv e^{\tau +i\theta}$ and $\bar z\equiv e^{\tau -i\theta}$ are related by complex conjugation. Using conformal transformations to map $x_1\to 0$, $x_4\to \infty$ and
\be
x_3=\hat n\equiv ({0,0,\dots,0,1})\qquad\qquad x_2=\hat n(\theta) e^\tau\equiv ({0,0,\dots,\sin\theta,\cos\theta})e^\tau,
\ee
we can rewrite 
\bea
g_s(z,\bar z)&\equiv &g_{12,34}(z,\bar z) = |z|^{\D_1} \la {\cal O}_4 |\hat {\cal O}_3 (0,\hat n)\hat {\cal O}_2(\tau,\hat n(\theta))| {\cal O}_1 \ra,
\label{eq:CBsChannelGen}\\
g_t(z,\bar z)&\equiv  &g_{32,14}(1-z, 1-\bar z) = \f{| 1-z |^{\D_2+\D_3}}{|z|^{\D_2}}  \la {\cal O}_4 |\hat {\cal O}_3 (0,\hat n)\hat {\cal O}_2(\tau,\hat n(\theta))| {\cal O}_1 \ra.
\label{eq:CBtChannelGen}
\eea
The $g_{ij,kl}(z,\bar z)$ can be decomposed as a sum over the primary operators that appear in the operator product expansion (OPE) of $ij$ and $kl$
\be
g_{ij,kl}(z,\bar z) = \sum_{\alpha} \lambda_{ij\alpha} \bar \lambda_{kl\alpha} g_{\D_\alpha,\ell_\alpha}^{\D_{ji},\D_{kl}}(z, \bar z), ~~ \D_{ij}=\D_i-\D_j,
\label{eq:s-t-channelCB}
\ee
where $\alpha$ labels the primaries while $\Delta_\alpha$, $\ell_\alpha$ and $\lambda_{ij\alpha}$ respectively represent their dimensions, spins and fusion coefficients. The 
conformal blocks $g_{\D,\ell}^{\D_{ji},\D_{kl}}(z, \bar z)$ are completely fixed functions:  their functional form is fixed  by the conformal group and their normalization   by \REf{eq:s-t-channelCB}. Their explicit expressions in $d=2,4$ can be found in~\cite{Dolan:2003hv}. 
What matters for our discussion is that in any dimension  they admit a power series expansion   in $|z|$ ~\cite{Dolan:2003hv,Hogervorst:2013sma}
\be
g_{\D,\ell} ^{\D_{21},\D_{34}} (z,\bar z) =|z| ^{\D} \sum_{k=0}^\infty |z| ^ k \sum^{\ell+k}_{j=j_0(\ell,k)} A_{k,j}^{\D_{21},\D_{34}} (\D,\ell) C_{j}^{(d/2-1)}(\cos \t), ~~ z=|z| e^{i \t}
\label{eq:ConfBlockStatesDescendantsGegenbauer}
\ee
with $j_0(\ell,k)=\max \l ( \ell-k, k-\ell \mod 2\r )$, where the term proportional to $|z|^k C_j(\cos\t)$ corresponds to the level $k$ descendant with spin $j$. The dimension and spin of the intermediate primaries is directly read from this expansion. The $A_{k,j}^{\D_{21},\D_{34}} (\D,\ell)$ are calculable coefficients, in particular $A_{0,0}^{\D_{21},\D_{34}} (\D,0)=1$.

We will here study the specific correlator
\be
\l \la [\bar \phi^n] (x_4) [\bar \phi \phi] (x_3) [\bar \phi \phi] (x_2) [ \phi^n](x_1) \r \ra,
\ee
so that equations (\ref{eq:CBsChannelGen}) and (\ref{eq:CBtChannelGen}) reduce to
\bea
g_s(z,\bar z) &\equiv& g_{\phi^n, \bar \phi\phi; \bar \phi\phi, \bar \phi^n}(z,\bar z) = Z^{-2}_{\bar \phi \phi} 
|z|^{\D_{\phi^n}}\f{\la n | (\widehat {{\bar \phi}  \phi}) (0,\hat n) (\widehat {{\bar \phi} \phi}) (\tau, \hat n(\theta)) | n \ra}{\la n | n \ra},
\label{eq:CBsChannelChargen}\\
g_t(z,\bar z) &\equiv &g_{\bar \phi\phi, \bar \phi\phi; \phi^n, \bar \phi^n}(1-z,1-\bar z) = Z^{-2}_{\bar \phi \phi} 
\f{|1-z|^{2\D_{\bar \phi \phi}}}{|z|^{\D_{\bar \phi \phi}}}
\f{\la n | (\widehat {{\bar \phi}  \phi}) (0,\hat n) (\widehat {{\bar \phi}  \phi}) (\tau, \hat n(\theta)) | n \ra}{\la n | n \ra}\,.\nonumber\\
\label{eq:CBtChannelChargen}
\eea
In the regime $\Delta_{\phi^n}\gg \Delta_{\bar \phi\phi}$, the $s$-channel is controlled by the  ``Heavy-Light'' OPE, while the $t$-channel is controlled by the ``Heavy-Heavy''
and the ``Light-Light'' OPEs.

\subsection{Leading order \label{sec:LO4ptFunction}}

As before, the leading order contribution corresponds to evaluating the path integral on the saddle and gives
\be
\f{\la n | (\hat {\bar \phi} \hat \phi) (0,\hat n_d) (\hat {\bar \phi} \hat \phi) (\tau, \vec n) | n \ra}{\la n | n \ra} = \f{f^4}{4}.
\label{leading4}
\ee
The implications of this result, when considering  the $s$- and $t$-channels are  as follows.

\paragraph{s-channel.}

From \REf{eq:CBsChannelChargen} and \REf{eq:NormRenorm} we obtain
\be
g_s(z,\bar z) = \l (f^2 \Omega_3 \r )^2 |z|^{\D_{\phi^n}}\, .
\ee
Therefore, the only operator appearing in the $\phi^n\times \bar \phi \phi$ OPE is $ \phi ^n (x)$ itself with the fusion coefficient \REf{eq:fusionPhiPhi}. Moreover, we see that at this order the descendants of $\phi^n$ do not contribute. This is to be expected, because the contribution of descendants is suppressed by powers of the ratio $\f{\D_{\bar \phi \phi}}{\D_{\phi^n}}$, and thus by an inverse power of $n$,  just as a consequence of conformal symmetry (see also~\cite{Jafferis:2017zna}). For instance, the first descendant term in the conformal block has  coefficient
\be \label{eq:DescendantCoefficient}
A_{1,1}^{\D_{21},\D_{34}} (\D,0) = \f{(\D_{21}+\D)(\D_{34}+\D)}{4\D},
\ee
which, for the case at hand, equals
\be
 \f{\D_{\bar \phi \phi}^2}{4\D_{\phi^n}}, 
\ee
and is suppressed in the limit $n\gg 1$.

\paragraph{t-channel.} 

From eqs.~(\ref{eq:CBtChannelChargen},\ref{leading4}) we obtain 
\be
g_{t}(y, \bar y) = \l ( \f{n}{\m} \r )^2 \f{| y |^{2\D_{\bar \phi \phi}}}{|1-y|^{\D_{\bar \phi \phi}}}.
\label{eq:LOepsilontChannel}
\ee
Expanding in powers of $y$
\bea
g_{t}(y, \bar y) 
=
\l ( \f{n}{\m} \r )^2 \, |y|^4 \l [ 1+ |y| C_{1}^{(1)}(\cos\t) + |y|^2 C_{2}^{(1)}(\cos\t) + |y|^3 C_{3}^{(1)}(\cos\t) + \dots \r ],
\eea
and comparing with the expansion in conformal blocks, \REf{eq:s-t-channelCB}, we deduce that in this channel there appears a tower of primary operators labelled  by their  spin $\ell$ and by an integer $k$,  with dimension
\be
\D_{(k,\ell)} = 4+2k +2\ell, \qquad\ell, k =0,1,2, \dots,
\label{eq:t-ChannelSpectrumLO}
\ee
and with fusion coefficients satisfying
\be
\lambda^{n,n}_{(k,\ell)} \bar \lambda^{\bar \phi \phi,\bar \phi \phi}_{(k,\ell)}=
\frac{f^4}{4}(-1)^k \frac{ (k!)^2 (k+2\ell)!(k+2\ell+1)! }{ (2k)! (2k+4\ell+1)! }.
\ee
At weak coupling these operators correspond to 
\be
\mc O_{(k,\ell)} (x) = \l ( \bar \phi \phi \, \p^{2k} \p_{\{\m_1} \dots \p_{\m_{2\ell}\}} \bar \phi \phi \r ) (x),
\label{eq:t-ChannelOperatorsLO}
\ee
where $\{ \}$ indicates the traceless symmetric component.

\subsection{NLO \label{sec:NLO4ptFunction}}

\begin{figure}[H]
  \centering
  \includegraphics[width=10cm]{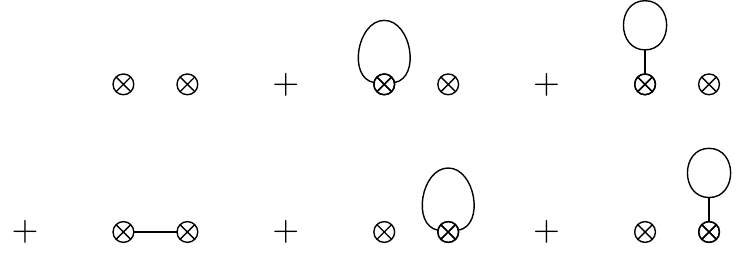}
  \caption{\label{fig:4ptFunction} Topology of diagrams entering $\langle n | ( \bar \phi \phi )  ( \bar \phi \phi )  | n \rangle $ }
\end{figure}

At next to leading order we must consider, in full analogy with \REf{eq:3ptFunc1},
\be
\la n | (\bar \phi \phi) (0,\hat n)  (\bar \phi \phi) (\tau,\hat n(\theta)) | n \ra 
=
\la n | 
\l [ \f{f^2}{2} + f r (0, \hat n) + \f{r^2(0, \hat n) }{2} \r ]  
\l [ \f{f^2}{2} + f r (\tau,\hat n(\theta)) + \f{r^2(\tau, \hat n(\theta)) }{2} \r ] 
| n \ra\,,
\label{eq:4ptFunction1}
\ee
from which both connected and disconnected diagrams arise at NLO, see Fig. \ref{fig:4ptFunction}. 

Disconnected diagrams  just correspond to factorized 3-point functions, which we computed before. Therefore, what is left to compute is the one-phonon exchange connected diagram, which leads to
\be
Z_{\bar \phi \phi}^{-2}\f{\la n | (\hat {\bar \phi} \hat \phi) (0,\hat n_d) (\hat {\bar \phi} \hat \phi) (\tau, \vec n) | n \ra}{\la n | n \ra} = 
\lambda_{\bar \phi \phi}^2 \l [ 1 +\f{4\m\Omega_3}{n}D_{rr}(z,\bar z) \r ],
\ee
with  $\lambda_{\bar \phi \phi}$ the fusion coefficient in \REf{eq:FusionFinal} and with the propagator for the radial mode given by (see Appendix \ref{sec:propagatorCylinder})
\be
D_{rr}(z, \bar z) = \sum_{\ell=0}^\infty \f{\ell+1}{\Omega_{3}} \, \l ( |z|^{\omega_A(\ell)} \, \f{J_\ell - \omega_A^2(\ell) } {2\omega_A(\ell)} 
+ |z|^{\omega_B(\ell)} \, \f{\omega_B^2(\ell)-J_\ell } {2\omega_B(\ell)}\r ) \f{C_\ell^{(1)}
(\cos \t)}{\omega_B^2(\ell)-\omega_A^2(\ell)}.
\label{eq:4ptPropagatorrr}
\ee
Having secured the  four-point function at NLO, we can now turn our attention to the spectrum of operators appearing in the different channels.

\paragraph{s-channel.} The analysis is straightforward. Indeed using \REf{eq:CBsChannelChargen} and \REf{eq:4ptPropagatorrr} we see that the four-point function 

\be
g_{s}(z,\bar z)= \lambda_{\bar \phi \phi}^2 |z|^{\D_{\phi^n}} \l [ 1 +\f{4\m}{n}
\sum_{\ell=0}^\infty \, \l ( |z|^{\omega_A(\ell)} \, \f{J_\ell - \omega_A^2(\ell) } {2\omega_A(\ell)} 
+ |z|^{\omega_B(\ell)} \, \f{\omega_B^2(\ell)-J_\ell } {2\omega_B(\ell)}\r ) \f{(\ell+1) C_\ell^{(1)}
(\cos \t)}{\omega_B^2(\ell)-\omega_A^2(\ell)} \r ]
\label{eq:4ptPropagatorrrS-channel}
\ee
is already in the form \REf{eq:ConfBlockStatesDescendantsGegenbauer}. Therefore, we can identify the  primary operators by simply looking at the powers of $|z|$ in the expansion.  These are in one-to one correspondence with  the $A$- and $B$-type single phonon states found in Section~\REf{sec:SpectrumFluctuations} and result in two separated towers of primaries with dimension
\be
\D_A=\D_{\phi^n}+\omega_A(\ell), ~~\ell \geq 2\qquad \qquad \D_B= \D_{\phi^n}+\omega_{B}(\ell), ~~ \ell \geq  0\, .
\label{eq:onephononSc-channel}
\ee
Notice that the tower of $A$-type primaries starts at $\ell=2$. Indeed, the $\ell=1$ \mbox{$A$-phonon} does appear in \REf{eq:4ptPropagatorrrS-channel} but it  corresponds to the  descendant  $\partial_\mu \phi^n$. Instead $\ell=0$ corresponds to  the ``Goldstone mode", which 
controls the global fluctuations of the phase of $\phi$ and, as such, is not excited by neutral operators like $\bar \phi \phi$. The corresponding fusion coefficients can be read off from the coefficients in front of~$|z|^\D$
\be  \lambda^{\ell}_{\bar \phi \phi,A} = \lambda_{\bar \phi \phi} \sqrt{ \f{4\m}{n} (\ell+1)  \f{J_\ell - \omega_A^2(\ell) } {2\omega_A(\ell)} } , ~~ \ell\geq 1,\ee
\be  \lambda^{\ell}_{\bar \phi \phi,B} = \lambda_{\bar \phi \phi} \sqrt{ \f{4\m}{n} (\ell+1)  \f{\omega_B^2(\ell) - J_\ell } {2\omega_B(\ell)} } , ~~ \ell\geq 0.\ee
We see that these are $n$ suppressed by $\sim \sqrt {\mu/n}$  with respect to  $\lambda_{\bar \phi \phi}$. It should also be noted that these operators enter the OPE without their descendants, similarly to $\phi^n$ at leading order.

\paragraph{t-channel.} The analysis is somewhat more complicated. The reason is that the corresponding expression of the four-point function
\bea
g_{t}(1-z,1 - \bar z) & = & \lambda_{\bar \phi \phi}^2 \f{|1-z|^{2\D_{\bar \phi \phi}}}{|z|^{\D_{\bar \phi \phi}}} \label{eq:4ptPropagatorrrt-channel} \\
&&
 \l [ 1 +\f{4\m}{n}
\sum_{\ell=0}^\infty \, \l ( |z|^{\omega_A(\ell)} \, \f{J_\ell - \omega_A^2(\ell) } {2\omega_A(\ell)} 
+ |z|^{\omega_B(\ell)} \, \f{\omega_B^2(\ell)-J_\ell } {2\omega_B(\ell)}\r ) \f{(\ell+1) C_\ell^{(1)}
(\cos \t)}{\omega_B^2(\ell)-\omega_A^2(\ell)} \r ] \nn
\eea
is written as a power series in $|z|$, and not in $|1-z|$. In order to get the latter, we have to analytically continue the four-point function to the region $z=1$. That would allow to analyze the spectrum of operators appearing in the $t$-channel at next to leading order.

Unfortunately, we do not know how to perform the analytic continuation in closed form\footnote{In other words, we do not possess the propagator in closed form for $z\sim \bar z\sim 1$.}, and instead, we will illustrate the principle with an example. For that, let us consider a simplified situation, $z=\bar z \in \mathbb{R}$, in other words $\t=0$. Introducing the following notation for the summand in \REf{eq:4ptPropagatorrrt-channel}
\be
G(z;\ell)=2\l ( z^{\omega_A(\ell)} \, \f{J_\ell - \omega_A^2(\ell) } {2\omega_A(\ell)} 
+ z^{\omega_B(\ell)} \, \f{\omega_B^2(\ell)-J_\ell } {2\omega_B(\ell)}\r ) \f{(\ell+1) C_\ell^{(1)}
(1)}{\omega_B^2(\ell)-\omega_A^2(\ell)},
\ee
and using its asymptotic behavior (see also \REf{eq:SpectrumLargeL})
\be
G(z;\ell) \underset{\substack{\ell\to \infty \\ z\to 1}}{=} z^{\ell} \l ( \ell+1-\f{3}{2}\f{\m^2-1}{\ell}-(1-z)\ell+\dots\r )
\ee
we can find leading asymptotic of the four point function for $z\to 1$
\be
g_{t}(1-z,1 - z) \underset{z\to 1}{=} \lambda_{\bar \phi \phi}^2 (1-z)^{2\D_{\bar \phi \phi}}
\l \{ 1 +\f{2\m}{n} \l [ \f{1}{(1-z)^2}+\f{\D_{\bar \phi \phi}-1}{1-z}+\f{3}{2} (\m^2-1)\log (1-z) \r ] + \mc R(z) \r \}
\label{eq:4ptt-Channelzto1}
\ee
with the remainder
\be
\mc R(z) = \f{2\m}{n} G(z;0)+\f{2\m}{n} \sum_{\ell=1}^\infty \l [G(z;\ell) - z^{\ell} \l ( \ell+1-\f{3}{2}\f{\m^2-1}{\ell} \r ) \r ] \underset{z\to 1}{=} O((1-z) \log(1-z))
\ee
a  less singular function.

Singular terms in \REf{eq:4ptt-Channelzto1} correspond to different operators. The term proportional to $(1-z)^{-2}$ corresponds to an operator with scaling dimension $\D=2$, which is nothing else but $\bar \phi \phi$ (its anomalous dimension is invisible at this order).\footnote{In general we expect not only $\bar \phi \phi$ but its spin-$\ell$ analogues of the form
\be
\bar \phi \p_{\m_1} \dots \p_{\m_\ell} \phi,
\label{eq:NLOtChannelphiphi}
\ee
to appear in the  $\bar \phi \phi\times \bar \phi \phi$ OPE.}
The second term corresponds to its descendant, whose coefficient is fixed by the conformal symmetry (compare with \REf{eq:DescendantCoefficient}).
Lastly, the term with $\log(1-z)$ can be exponentiated, leading to a modified prefactor
\be
g_{t}(1-z,1 - z) \underset{z\to 1}{\supset} \lambda_{\bar \phi \phi}^2 (1-z)^{2\D_{\bar \phi \phi}} \, (1-z)^{\f{3\m}{n} (\m^2-1)}.
\ee
The resulting exponent should correspond to the scaling dimension of $\D_{(\bar \phi \phi)^2}$ (see \REf{eq:t-ChannelSpectrumLO} and \REf{eq:t-ChannelOperatorsLO}) at  NLO. Indeed, using
\REf{eq:ChargeRelationfnmu}, \REf{eq:muRelationf} and \REf{eq:phiphiAnomalousDimension} we can write
\be
\D_{(\bar \phi \phi)^2} = 2\D_{\bar \phi \phi}+\f{3\m}{n} (\m^2-1) = 4+O(\varepsilon^2),
\label{eq:NLODimesnionSpin0} 
\ee
which  coincides with the computation using Feynman diagrams, see Appendix~\ref{app:phiphiDimensionFeynman}.
The last result can also be directly  derived using the general relation (see e.g. \cite{Baume:2014rla}) $\Delta_{(\bar \phi \phi)^2}= d+\beta'(\lambda_*)$, between the dimension of the interaction term $(\bar\phi\phi)^2$ and $\beta'\equiv \partial_\lambda \beta$. Using \REf{eq:phi4FixedPoint} and $\beta(\lambda)=-\varepsilon \lambda + \frac{5 \lambda^2}{16\pi^2}-\frac{15\lambda^3}{(16\pi^2)^2}+O(\lambda^4)$ immediately gives $\Delta_{(\bar \phi \phi)^2}= 4 + O(\varepsilon^2)$.

We conclude  this section with  two comments. First, it is straightforward to extend the computation presented above to the case when the two `light' operators are $( \bar \phi \phi ) ^k$. \REf{eq:4ptt-Channelzto1} is just minimally modified to  \footnote{The fusion coefficient $\lambda_{(\bar \phi \phi)^k}$ can be computed by repeating the steps of Section \ref{sec:3ptfunction}.}
\be
g_{t}(1-z,1 - z) \underset{z\to 1}{=} \lambda_{(\bar \phi \phi)^k}^2 (1-z)^{2\D_{(\bar \phi \phi)^k}}
\l \{ 1 +\f{2\m k^2}{n} \l [ \f{1}{(1-z)^2}+\f{\D_{(\bar \phi \phi)^k}-1}{1-z}+\f{3}{2} (\m^2-1)\log (1-z) \r ] + \dots \r \},
\ee
which implies that the two leading contributions are associated to $(\bar \phi \phi)^{2k}$ and $(\bar \phi \phi)^{2k-1}$. Moreover, by exponentiating the term with $\log(1-z)$  we obtain, at 1-loop accuracy,  a relation between scaling dimensions
\be
\D_{(\bar \phi \phi)^{2k}} - 2 \D_{(\bar \phi \phi)^{k}} = k^2 \l ( \D_{(\bar \phi \phi)^{2}} - 2 \D_{(\bar \phi \phi)} \r ),
\ee
which can be checked perturbatively using the results of appendix \ref{app:phiphiDimensionFeynman}. This provides an additional cross-check.

The second comment concerns the computation of similar correlators in a general CFT using the universal EFT superfluid description, as done in \cite{Hellerman:2015nra,Monin:2016jmo}.
 Even though the EFT description can be trusted only for sufficiently large separations between the two `light' operators, we can try and use the results of~\cite{Monin:2016jmo} for the 4-point function to formally analyze what operators appear in t-channel. Repeating almost verbatim (albeit unjustifiably) the computation leading to \REf{eq:t-ChannelSpectrumLO} we conclude that the spectrum of operators in this case is given by
\be
\D = \d_1+\d_2+2k+\ell,
\label{eq:EFTtChannelSpectrum}
\ee
which for large $\ell \gg 1$ coincides with the predictions of the analytic bootstrap~\cite{Komargodski:2012ek,Fitzpatrick:2012yx}. This fact indicates there should be away to frame the statement, which is purely within the reach of EFT. But we do not know how.

\section{Summary}

Perhaps the  most synthetic way to state the result of \cite{Hellerman:2015nra}  is by saying
that in an euclidean CFT the insertion of a large charge operator  produces `around' the insertion point a state
that is  equivalent to a conformal superfluid. The equivalence is made fully evident by exploiting the mapping  of the theory to the cylinder. A consequence of this result is that, while the lowest dimension operator of given  charge corresponds to the superfluid ground state at fixed charge density,  the operators of higher dimension must be in one-to-one correspondence with the superfluid excitations. The latter range from states with finite spin involving a finite number of phonons to states involving vortices whose spin scales with the charge \cite{vortices1}. This result is remarkable but, as stated in its generality, it is a bit abstract, not very tangible.

In this paper, considering weakly coupled theories, where the operator spectrum can also be constructed using standard perturbation theory, we made that operator correspondence more tangible. In practice we considered Wilson-Fisher $U(1)$ invariant fixed points and focussed on charge-$n$ operators, like $\phi^n$, with $n\gg 1$. By constructing the operators we have shown how their spectrum automatically exhibits the structure of the Fock space of phonon excitations in the superfluid. As our question mainly concerned counting and structure,  we obtained the above result by focussing  on the simplest case of free field theory.  In that case the spectrum can be fully and exactly worked out in terms of elementary fields and their derivatives, and yet, for  large $n$, one can still describe it in terms of superfluid excitations in a systematic $1/n$ expansion. 

By conformal invariance, operator multiplets are fully classified by their primary operator. We have identified polynomials in fields and derivatives  that play the role of  building blocks in the systematic construction of the primaries. These building blocks are themselves primaries and are labelled by spin $\ell$ and by a discrete label taking two values,  $\mathcal A$ and $\mathcal B$. At a given spin $\ell$, a building block  is fully determined by its highest weight element, carrying $J_3=\ell$. For instance the highest weight $\mathcal A$-blocks
are given by products involving only $\phi$ (no $\bar \phi$) and $\partial_-$ derivatives of the form\footnote{This expression is equivalent to (\ref{eq:define_C_atomic}) with an abuse of notation, since here we do not distinguish between the full field $\phi$ or its creation part.}
\be
\mathcal A_{\ell,\ell}\equiv \sum_{k=0}^{\ell-1} \alpha_k \phi^{\ell -k-1}(\partial_-\phi)^k(\partial_-^{\ell-k} \phi)
\ee
with the $\alpha_k$ some given coefficients. The lower $J_3$ elements of the block $\mathcal A_{\ell, m}$ ($m=-\ell,-\ell+1,\dots, \ell$) are trivially obtained by acting  with the lowering spin operator $J_-$. The $\mathcal A$ blocks are only defined for $\ell\geq 2$. Indeed the $\mathcal A_{00}$-block would be trivially proportional to the identity operator, while the 
$\ell=1$ block would be proportional to $\partial_-\phi$, and as such it would be a descendant not commuting with the special conformal generators. $\mathcal B$-blocks have instead the form
\be
\mathcal B_{\ell,\ell}\equiv \sum_{k=0}^\ell \beta_k \phi^{\ell -k+1}(\partial_-\phi)^k(\partial_-^{\ell-k} \bar \phi)
\ee
with suitable $\beta_k$. Unlike the $\mathcal A$ blocks, the $\mathcal B$-blocks are defined for all $\ell\geq 0$. In particular $\mathcal B_{0,0}= \phi\bar \phi$.
Notice that the blocks have charge $\ell$, thus their charge equals their spin. By a combinatoric argument, we have then proven that all primaries of spin bounded by the charge are obtained by taking products of $\mathcal A$ and $\mathcal B$ blocks, corresponding to expressions of the form
\be
\phi^{n-n_\mathcal A-n_\mathcal B} \left (\prod_{\alpha} \mathcal A_{\ell_\alpha, m_\alpha}\right ) \left (\prod_{\beta} \mathcal B_{\tilde \ell_\beta, \tilde m_\beta}\right ), ~~~ n_\mathcal A=\sum_\alpha \ell_\alpha, ~~~ n_\mathcal B=\sum_\beta \tilde \ell_\beta.
\ee
These expressions are well defined as long as $n-n_\mathcal A-n_\mathcal B\geq 0$. In particular, they are well defined in the limit $n\to \infty$ while keeping the total number of derivatives and powers of $\bar \phi$ finite. This result for the operator spectrum is precisely in one-to-one correspondence with the Fock space of hydrodynamic modes around the superfluid solution. The latter consists of $A$ and $B$-type modes of all possible spins. However $A$-modes of spin $\ell =0$ and $\ell =1$ should not be considered when constructing operators of fixed charge. The former excitations do not correspond to Fock states, and simply give rise to operators that interpolate between subspaces of given charge. The latter are seen to correspond to descendants.

Our building blocks allow to explicitly construct all the primary operators whose number of derivatives  is bounded by their charge.
The spin of these states  then satisfy $\ell\leq n$, and we naturally expect a new regime to arise  for $\ell >n$. On the other hand, as illustrated in section \ref{sec:primary_phonons}, the structure  of the building blocks displays also a subtle change of regime at the smaller value  $\ell\sim \sqrt n$. In the superfluid description the change of regime consists in the fact that for $\ell\ll\sqrt n$, the primary states are approximated by states with a fixed number of phonons, while for $\ell\lsim\sqrt n$ the primary building blocks involve a significant mixture of states with different numbers of phonons. This indicates that at $\ell \gsim \sqrt n$ the interactions among phonons become  important and that states cannot be described  as small fluctuations around  a homogeneous superfluid solution. In other words it appears the superfluid description breaks down for $\ell \gsim \sqrt n$.  However, and remarkably, leaving aside the explicit expression of the primaries, it happens that for $\sqrt{n} < \ell < n $ the  superfluid description
can still be used for computing coordinate (basis) independent quantities, like the expectation value of  $\bar \phi \phi$. This fact strongly suggests that in the window $\sqrt{n} < \ell < n $ there exists another hydrodynamic saddle point allowing a more convenient description of the primaries. We have not investigated that, but this is clearly an issue worth further study.

As we already said, in this paper we haven't addressed at all what happens for even higher spins, $\ell >n$. However,  our combinatoric argument shows that, even in this case, the counting of primary operators (with maximal spin given the number of derivatives) coincides with the counting of superfluid phonon excitations, with the constraint that  each phonon's spin be less than the charge. At the moment we cannot say anything in favor of the existence of yet another hydrodynamic saddle point, rendering computations perturbative in this case, but these puzzling facts clearly warrant further investigation. Indeed, in the interacting case the regime of large spin and large charge is expected to be universally described by vortex dynamics as discussed in \cite{vortices1}. The search for different saddles with respect to which expand at large spin  in the the free and interacting case seems, unavoidably, the next thing to study.

\subsection*{Acknowledgements}

We would like to thank Gabriel Cuomo, Brian Henning and Matt Walters for useful discussions. The work of G.B. and R.R. is partially supported by the Swiss National Science Foundation under contract 200020-188671 and through the National Center of Competence in
Research SwissMAP.


\newpage
\appendices

\section{Coefficient $\mathcal Y_{\ell m}^{\mu_1\dots\mu_\ell}$\label{sec:appYlmCoeff}}

We provide here a few explicit formulas regarding the coefficient defined in (\ref{eq:defineYlmCoeff}). First, we write the spherical harmonics in the basis (\ref{eq:x+-0})
\bea
Y_{\ell m} 
& = &
\label{eq:SphericalCartesian}
[-\mathrm{sign}(m)]^m \sqrt{\f{(2\ell+1)(\ell+m)!(\ell-m)!}{2^{|m|} \, 4\pi }}\sum_{\substack{\a_++\a_-+\a_0=\ell \\ \a_+-\a_--=m}} 
\f{n_+^{\a_+} n_0^{\a_0} n_-^{\a_-} }{(-2)^{\mathrm{min}(\a_+,\a_-)}\a_+!\a_0!\a_-!} \\
& = &
[-\mathrm{sign}(m)]^m \sqrt{\f{(2\ell+1)(\ell+m)!(\ell-m)!}{2^{|m|} \, 4\pi }}\sum_{k ~ \mathrm{step}~2} ^{\ell-|m|}
\f{n_+^{\f{\ell+m-k}{2}} n_0^{k} n_-^{\f{\ell-m-k}{2}} }{(-2)^{\f{\ell-|m|-k}{2}}\l ( \f{\ell+m-k}{2} \r )! k ! \l ( \f{\ell-m-k}{2}\r )!}, \nn
\eea
where the sum over $k$ is taken in steps of 2, starting form $\ell-|m|~\mathrm{mod}~2$.

After the integration we get the coefficients
\begin{align}
  \mathcal Y_{\ell m}^{ \overbrace{+ \ldots +}^{\alpha_+} \overbrace{0\ldots0}^{\alpha_0}\overbrace{-\ldots-}^{\alpha_-}} = 
  ~& \delta_{\alpha_+ + \alpha_0+\alpha_- , \ell} \delta_{\alpha_- - \alpha_+, m}
  \frac{[-\mathrm{sign}(m)]^m \sqrt{\pi} \, (2\ell+1) \sqrt{(\ell+m)!(\ell-m)!}} {\ell !} \nonumber \\
  & \times \sum_{k \text{ step } 2}^{\ell-|m|} \f{(-1)^\f{\ell-k-|m|}{2}}{2^{\f{3}{2}\ell - k - \f{\alpha_0}{2}}}
  \f{\G\l ( \f{k+\alpha_0+1}{2}\r )}{\G \l ( \ell+\f{3}{2}\r )} 
  \f{\l ( \ell-\f{k+\alpha_0}{2}\r )!}{\l ( \f{\ell+m-k}{2}\r )! k! \l ( \f{\ell-m-k}{2}\r )!} . \label{eq:YlmCoeffExplicit}
\end{align}


\section{Explicit expressions for spin 2 and 3 primaries\label{app:spin23}}
In this appendix we give explicit expression for spin-$\ell$ primary operators associated to the primary states (\ref{eq:one-phononStateA}) using operator-state correspondence (\ref{eq:aVsDerivatives}). As discussed in section \ref{sec:oppStateCorr}, it is assumed products of (derivatives of) $\phi$ are normal-ordered and evaluated at the origin.

The counting of primaries given in (\ref{eq:primaryCount}) indicates there is one primary of spin $\ell=0,2,3$ but none of spin $1$.

\paragraph{Spin $0$} is trivial, for we have only one state given by (\ref{eq:chargenspin0}).

\paragraph{Spin $1$}
Explicitly, we have three states
\be
\l ( a_{00}^\dagger \r ) ^{n-1}a_{1,m}^\dagger | 0 \ra,
\ee
which correspond to
\bea
\l ( a_{00}^\dagger \r ) ^{n-1} a^\dagger _{1,1} | 0 \ra & = & -(4 \pi)^{n/2} \phi^{n-1} \p_- \phi | 0 \ra, \\
\l ( a_{00}^\dagger \r ) ^{n-1} a^\dagger _{1,0} | 0 \ra & = & (4 \pi)^{n/2} \phi^{n-1} \p_0 \phi | 0 \ra,  \\
\l ( a_{00}^\dagger \r ) ^{n-1} a^\dagger _{1,-1} | 0 \ra & = & (4 \pi)^{n/2} \phi^{n-1} \p_+ \phi | 0 \ra.
\eea
It is straightforward to show using \REf{eq:K0aabb}, \REf{eq:K-aabb} and \REf{eq:K+aabb} that those states, hence operators, are descendants, as we would expect since these operators can be written as derivatives of~$\phi^n$.

\paragraph{Spin $2$}

We can write two spin-2 operators by combining $n$ fields $\phi$ and two derivatives in a traceless and symmetric way
\be
O^{(2,1)}_{\m \n} = \phi^{n-1}\l ( \p_\m\p_\n \phi - \f{\d_{\m \n}}{3}\p^2 \phi  \r ), ~~ 
O^{(2,2)}_{\m \n} = \phi^{n-2}\l ( \p_\m \phi \p_\n \phi - \f{\d_{\m \n}}{3}( \p \phi )^2  \r ).
\ee
One linear combination of these is the spin-two primary.

To give examples of primary states with non-maximal $J_3$ eigenvalue, let us repeat the method of section \ref{sec:vacuum_primaries_construction} in this simple case. We consider a state
\be
| n ; 2, 0 \ra_{A} = \a_1 \l ( a_{00}^\dagger \r ) ^{n-1} a^\dagger _{2,0} | 0 \ra+\a_2 \l ( a_{00}^\dagger \r ) ^{n-2} \l ( a^\dagger _{1,0} \r )^2 | 0 \ra+
\b_2 \l ( a_{00}^\dagger \r ) ^{n-2} a^\dagger _{1,-1} a^\dagger _{1,1}  | 0 \ra.
\ee
Acting with $K_\pm$ and $K_0$ we see that this state is primary provided $\a_2=\b_2=-\a_1$. It follows from (\ref{eq:aVsDerivatives}) that
\bea
| n ; 2, 0 \ra_{A}
& = & 
(4 \pi)^{n/2}\a_1 \l [ \f{1}{3} \phi^{n-1}\l ( \p_0^2 \phi - \p_+\p_- \phi  \r ) -  \phi^{n-2}\l ( \p_0 \phi \p_0 \phi - \p_+ \phi \p_- \phi  \r ) \r ]| 0 \ra \\
& = & 
(4 \pi)^{n/2}\f{\a_1}{2} \l [ \f{1}{3} \phi^{n-1}\l ( 3\p_0^2 \phi - \p^2 \phi  \r ) -  \phi^{n-2}\l ( 3 \p_0 \phi \p_0 \phi - (\p \phi)^2   \r ) \r ]| 0 \ra \\
& = & 
(4 \pi)^{n/2}\f{3\a_1}{2} \l [ \f{1}{3} \phi^{n-1}\l ( \p_0^2 \phi - \f{1}{3}\p^2 \phi  \r ) -  \phi^{n-2}\l ( \p_0 \phi \p_0 \phi - \f{1}{3}(\p \phi)^2   \r ) \r ]| 0 \ra \\
& = & 
(4 \pi)^{n/2}\f{3\a_1}{2} \l ( \f{1}{3}O^{(2,1)}_{00} - O^{(2,2)}_{00} \r) | 0 \ra.
\eea
Hence, we conclude the operator $O_{\mu\nu}^{(2,1)} - 3 O_{\mu\nu}^{(2,2)}$ is primary.

\paragraph{Spin $3$}

In this case we have an anzatz
\bea
| n ; 3, 0 \ra_{A} 
& = & 
\a_1 \l ( a_{00}^\dagger \r ) ^{n-1} a^\dagger _{3,0} | 0 \ra \\
& + &\a_2 \l ( a_{00}^\dagger \r ) ^{n-2} a^\dagger _{2,0} a^\dagger _{1,0} | 0 \ra+\b_2 \l ( a_{00}^\dagger \r ) ^{n-2} a^\dagger _{2,1} a^\dagger _{1,-1} | 0 \ra+\g_2 \l ( a_{00}^\dagger \r ) ^{n-2} a^\dagger _{2,-1} a^\dagger _{1,1} | 0 \ra \nn \\
& + &
\a_3 \l ( a_{00}^\dagger \r ) ^{n-3} \l ( a^\dagger _{1,0} \r )^3 | 0 \ra + \b_3 \l ( a_{00}^\dagger \r ) ^{n-3} a^\dagger _{1,-1} a^\dagger _{1,1}a^\dagger _{1,0} | 0 \ra. \nn 
\eea
As before, acting with $K$ and imposing that the state be primary we get
\bea
| n ; 3, 0 \ra_{A}
& = & 
(4 \pi)^{n/2}\a_1 \l[ \l ( a_{00}^\dagger \r ) ^{n-1} a^\dagger _{3,0}  \r. \\
& - &3 \l ( a_{00}^\dagger \r ) ^{n-2} a^\dagger _{2,0} a^\dagger _{1,0} -\sqrt{2} \l ( a_{00}^\dagger \r ) ^{n-2} a^\dagger _{2,1} a^\dagger _{1,-1} -\sqrt{2}\l ( a_{00}^\dagger \r ) ^{n-2} a^\dagger _{2,-1} a^\dagger _{1,1}  \nn \\
& + &
\l. 
2 \l ( a_{00}^\dagger \r ) ^{n-3} \l ( a^\dagger _{1,0} \r )^3  + 6 \l ( a_{00}^\dagger \r ) ^{n-3} a^\dagger _{1,-1} a^\dagger _{1,1}a^\dagger _{1,0} \r ] | 0 \ra, \nn 
\eea
which corresponds to
\be
| n ; 3, 0 \ra_{A}  = (4 \pi)^{n/2}\a_1 \l ( \f{1}{6} O^{(3,1)}_{000}-\f{5}{6} O^{(3,2)}_{000}+5 O^{(3,3)}_{000}\r )  | 0 \ra=
(4 \pi)^{n/2} \f{ \a_1}{6} \l (O^{(3,1)}_{000}-5 O^{(3,2)}_{000}+30 O^{(3,3)}_{000}\r )  | 0 \ra,
\ee
with
\bea
O^{(3,1)}_{\m \n \lambda} & = & \phi^{n-1} \p_\m \p_\n \p_\lambda \phi - \f{1}{5} \l ( \d_{\m \n} \p_\lambda+ \d_{\m \lambda} \p_\n + \d_{\lambda \n} \p_\m \r ) \partial^2 \phi, \\
O^{(3,2)}_{\m \n \lambda} & = & \phi^{n-2} \p_\m \p_\n \phi \p_\lambda \phi+\phi^{n-2} \p_\m \phi \p_\n \p_\lambda \phi + \phi^{n-2} \p_\n \phi \p_\m \p_\lambda \phi  \\ 
&&
- \f{\d_{\m \n}}{5} \l ( \p _ \lambda\phi \partial^2 \phi + \p_\lambda (\partial \phi )^2 \r )
- \f{\d_{\m \lambda}}{5} \l ( \p _ \n \phi \partial^2 \phi + \p_\n (\partial \phi )^2 \r )
- \f{\d_{\lambda \n}}{5} \l ( \p _ \m \phi \partial^2 \phi + \p_\m (\partial \phi )^2 \r ), \nn \\
O^{(3,3)}_{\m \n \lambda} & = & \phi^{n-3} \p_\m\phi \p_\n\phi \p_\lambda \phi - \f{1}{5} \l ( \d_{\m \n} \p_\lambda \phi+ \d_{\m \lambda} \p_\n \phi+ \d_{\lambda \n} \p_\m \phi \r ) (\partial \phi )^2.
\eea
the three spin-3 operators. We conclude operator $ O^{(3,1)}_{\mu\nu\lambda}-5 O^{(3,2)}_{\mu\nu\lambda}+30 O^{(3,3)}_{\mu\nu\lambda} $ is primary.


\section{Counting primaries with $\ell>n$\label{app:counting_primaries}}

Here we give several examples of the formula (\ref{eq:primaryCount}) presented in main text.

First, let us consider the case of charge 2. We detail the partitions mentioned in the argument of the main text. We do this for the examples of spin 4 and 5:
\begin{itemize}

\item
$\mathrm{Prim}(4,2) =1$
\be
\ba{l | l || l | l || c}
p(4,2)& p(3,2) & p^*(4,2) & p^*(3,2) & \mathrm{Prim}(4,2) \\
\hline
(4) & (3) & (1,1,1,1) & (1,1,1) & \times\\ 
(3,1) & (2,1) & (2,1,1) & (2,1) & \times \\
(2,2) & & \mathbf{(2,2)} & & \checkmark
\ea
\ee

\item
$\mathrm{Prim}(5,2) =0$
\be
\ba{l | l || l | l || c}
p(5,2)& p(4,2) & p^*(5,2) & p^*(4,2) & \mathrm{Prim}(5,2) \\
\hline
(5) & (4) & (1,1,1,1,1) & (1,1,1,1)& \times \\ 
(4,1) & (3,1) & (2,1,1,1) & (2,1,1) & \times \\
(3,2) &(2,2) & (2,2,1) & (2,2) & \times
\ea
\ee

\end{itemize}
In general, we find there is one primary operator for even spins and none for odd spins.

Let us now give a more involved example with charge 3 and spin 8, resulting in $\mathrm{Prim}(8,3) =2$.
\be
\ba{l | l || l | l || c}
p(8,3)& p(7,3) & p^*(8,3) & p^*(7,3) & \mathrm{Prim}(8,3) \\
\hline
(8) & (7) & (1,1,1,1,1,1,1,1) & (1,1,1,1,1,1,1) & \times \\
(7,1) & (6,1)  & (2,1,1,1,1,1,1) & (2,1,1,1,1,1)& \times \\ 
(6,2) & (5,2)  & (2,2,1,1,1,1) & (2,2,1,1,1) & \times \\
(6,1,1) & (5,1,1) & (3,1,1,1,1,1) & (3,1,1,1,1)& \times \\
(5,3) & (4,3) & (2,2,2,1,1) & (2,2,2,1)& \times \\
(5,2,1) & (4,2,1) & (3,2,1,1,1) & (3,2,1,1)  & \times \\
(4,4) & & \mathbf{(2,2,2,2)} &  & \checkmark \\
(4,3,1) &(3,3,1) & (3,2,2,1) & (3,2,2)  & \times \\
(4,2,2) & (3,2,2) & (3,3,1,1) & (3,3,1) & \times \\
(3,3,2) &  &\mathbf{(3,3,2)} & & \checkmark
\ea
\ee

For arbitrary spin and charge $n=3$ an explicit expression is given by
\begin{equation}
\mathrm{Prim}(\ell,3) = \begin{cases} \left \lfloor \frac{\ell}{6} \right \rfloor     ,    & \text{if } \ell = 6p + 1 \text{ for some } p\in \mathbb N,     \\
		       \left \lfloor \frac{\ell}{6} \right \rfloor + 1 ,                   & \text{if } \ell \neq 6p + 1 \text{ for all } p\in \mathbb N .
         \end{cases}
\end{equation}

\item
In general, the number of primaries can be found from
\be
\sum_{\ell=0}^\infty \mathrm{Prim}(\ell,n) x^\ell = \prod_{k=2}^n \f{1}{(1-x^k)}.
\ee

\section{NLO Fock states on non-trivial background\label{sec:appNLOFockStates}}

Here we give the next to leading order result for the annihilation operators $B_{\ell m}$ over non-trivial background computed in section \ref{sec:twoFock}. The result uses the Gaunt coefficients defined in \REf{eq:gaunt}.

For $\ell=0$:
\begin{equation} \label{eq:NLOFockB0}
\begin{split}
  B_{00} ~=~ & \frac{a_{00} b_{00}}{\sqrt{n}} + \frac{6 n b_{00} b_{00}^\dagger -3 a_{00}^2 b_{00}^2 + (a_{00}^\dagger)^2(b_{00}^\dagger)^2 }{8 n^{3/2}} \\
           & + \sum_{\substack{\ell > 0 \\ \text{all } m }} \frac{1}{8(1+2\ell)n^{3/2}}  \Big( 
              (-1)^m 4 n (1+\ell) a_{\ell m}b_{\ell ,-m}
                     - (-1)^m (3+2\ell)a_{00}^2 b_{\ell m} b_{\ell,-m} \\[-15pt]
           & \hspace{110pt} - (-1)^m (1+2\ell)(a_{00}^\dagger)^2 a_{\ell m} a_{\ell,-m}
                     - (1+4\ell ) 2 n a_{\ell m} a_{\ell,m}^\dagger \\
           & \hspace{110pt} - 4 a_{00}^2 b_{\ell m} a_{\ell m}^\dagger
                     + (-1)^m (-1+2\ell ) a_{00}^2 a_{\ell m}^\dagger a_{\ell,-m}^\dagger \\
           & \hspace{110pt} - (-1)^m 4 n \ell a_{\ell m}^\dagger b_{\ell,-m}^\dagger
                     + 2 n (3+4\ell) b_{\ell m} b_{\ell m}^\dagger  \\
           & \hspace{110pt} + (-1)^m (1+2\ell) (a_{00}^\dagger)^2 b_{\ell m}^\dagger b_{\ell,-m}^\dagger \Big)\,.
\end{split}
\end{equation}

For $\ell>0$:

\begin{equation} \label{eq:NLOFockB}
\begin{split}
	B_{\ell m} =\  & \frac{a_{00} b_{\ell m}}{\sqrt{n}} +
		\frac{1}{4(1+2\ell)n^{3/2}} \Big(  (3+4\ell) \big( nb_{00}^\dagger -b_{00}a_{00}^2 \big)b_{\ell m} 
										+ 2 n b_{00} a_{\ell m} \\
					&  \hspace{124pt} + (-1)^{m} \big( (3+2\ell) n b_{00} + (1+2\ell) b_{00}^\dagger (a_{00}^\dagger)^2 \big)b_{\ell,-m}^\dagger \\
					&  \hspace{124pt} - (-1)^m \big( 2\ell n b_{00}^\dagger + 2(1+\ell) b_{00} a_{00}^2\big) a_{\ell,-m}^\dagger \Big) \\
			& + \!\!\!\! \sum_{\substack{\ell_1,\ell_2 > 0 \\ \text{all } m_1,m_2 }}  \!\!\!\! \frac{(-1)^m\sqrt{\pi}C_{-m,m_1,m_2}^{\ell,\ell_1,\ell_2}}{8\sqrt{2\omega_\ell\omega_{\ell_1}\omega_{\ell_2}}n^{3/2}}
			\Big( -(3+3\ell+\ell_1+\ell_2) a_{00}^2 b_{\ell_1,m_1}b_{\ell_2,m_2} \\[-20pt]
			& \hspace{145pt} +  2 (2+\ell+3\ell_1-\ell_2) n b_{\ell_1,m_1} a_{\ell_2,m_2} \\
			& \hspace{145pt} + 2 (-1)^{m_2}(3+\ell+3\ell_1+\ell_2) n b_{\ell_1,m_1}b_{\ell_2,-m_2}^\dagger \\
			& \hspace{145pt} - (1-\ell+\ell_1+\ell_2) (a_{00}^\dagger)^2 a_{\ell_1,m_1}a_{\ell_2,m_2} \\
			& \hspace{145pt} + 2 (-1)^{m_1}(\ell+\ell_1-\ell_2) (a_{00}^\dagger)^2 b_{\ell_1,-m_1}^\dagger a_{\ell_2,m_2} \\
			& \hspace{145pt} + (-1)^{m_1+m_2}(1+\ell+\ell_1+\ell_2)(a_{00}^\dagger)^2 b_{\ell_1,-m_1}^\dagger b_{\ell_2,-m_2}^\dagger \\
			& \hspace{145pt} - 2 (-1)^{m_2} (2+3\ell+\ell_1-\ell_2) a_{00}^2 b_{\ell_1,m_1} a_{\ell_2,-m_2}^\dagger \\
			& \hspace{145pt} - 2(-1)^{m_2} (1-\ell+\ell_1+3\ell_2) n a_{\ell_1,m_1}a_{\ell_2,-m_2}^\dagger \\
			& \hspace{145pt} + 2 (-1)^{m_1+m_2} (\ell+\ell_1-3\ell_2) n b_{\ell_1,-m_1}^\dagger a_{\ell_2,-m_2}^\dagger \\
			& \hspace{145pt} - (-1)^{m_1+m_2}(1+3\ell-\ell_1-\ell_2)a_{00}^2 a_{\ell_1,-m_1}^\dagger a_{\ell_2,-m_2}^\dagger \Big) \,.
\end{split}
\end{equation}

\section{Some asymptotics of Gaunt coefficients \label{app:Gaunt}}

In the main text we are interested in the expansion at large $\ell$ of $A_{\ell\ell}^\dagger$. Thus we provide here some formulas for the asymptotics of relevant Gaunt coefficients.
We use special cases of $3j$ symbols \cite{dlmf:3j}
\begin{equation}
  \begin{pmatrix} \ell & \ell_1 & \ell_2 \\ 0 & 0 & 0 \end{pmatrix} = \begin{cases} 0 & L \text{ odd}, \\
                                                                 (-1)^\frac{L}{2} \sqrt{\frac{(L-2\ell)!(L-2\ell_1)!(L-2\ell_2)!}{(L+1)!}} \frac{\left(\frac{L}{2}\right)!}{\left(\frac{L-2\ell}{2}\right)!\left(\frac{L-2\ell_1}{2}\right)!\left(\frac{L-2\ell_2}{2}\right)!} & L \text{ even},
                                                    \end{cases}
\end{equation}
\begin{equation}
  \begin{pmatrix}
    \ell & \ell_1 & \ell_2 \\ \ell & -\ell -m_2 & m_2
  \end{pmatrix}
  = (-1)^{\ell-\ell_1-m_2} \sqrt{ \frac{(2\ell)! (L-2\ell)! (\ell+\ell_1+m_2)! (\ell_2-m_2)!}{(L+1)! (L-2\ell_1)! (L-2\ell_2)! (-\ell+\ell_1-m_2)!(\ell	_2+m_2)!}}, 
\end{equation}
where $L = \ell+\ell_1+\ell_2$. We can use Stirling formula to estimate these at large spin. If we consider $\ell$ to be large, due to triangle inequality \REf{eq:triangle}, at least one of $\ell_1,\ell_2$ has to be of order $\ell$.

If we assume $\ell_1\sim \ell_2\sim \ell$, we have
\begin{equation} \label{eq:gauntL1L2}
  \frac{ C_{\ell,-\ell-m_2,m_2}^{\ell,\ell_1,\ell_2} }{\sqrt{\omega_\ell \omega_{\ell_1} \omega_{\ell_2}}} \overset{\substack{\ell_1\sim \ell_2\sim \ell \\ \ell\to \infty}}{\longrightarrow} g_1\!\!\left( \frac{\ell_1}{\ell},\frac{\ell_2}{\ell},\frac{m_2}{\ell}\right)^{\ell/2} h_1\!\!\left(\frac{\ell_1}{\ell},\frac{\ell_2}{\ell},\frac{m_2}{\ell}\right) \left( \ell^{-7/4} + \mathcal O(\ell^{-11/4})\right) ,
\end{equation}
where
\begin{equation}
  g_1 (x,y,z) = \frac{ 4 (-1)^{1-x+y-2z} (x+y-1)^{x+y-1} (x+z+1)^{x+z+1} (y-z)^{y-z} }{ (x-y+1)^{x-y+1} (y-x+1)^{y-x+1} (x+y+1)^{x+y+1} (x-z-1)^{x-z-1} (y+z)^{y+z} } ,
\end{equation}
whose absolute value is bounded by $1$ and for each pair $x,y$ there is one unique $z$ such that $|g_1(x,y,z)| = 1$, namely $z = \frac{x^2-y^2-1}{2}$, and
\begin{equation}
  h_1 (x,y,z) = \frac{2}{\pi^{5/4}} \frac{ (y-z)^{1/4} (1+x+z)^{1/4} }{ (x-y+1)^{1/2} (y-x+1)^{1/2} (x+y+1) (x-z-1)^{1/4} (y+z)^{1/4} } .
\end{equation}
Hence, for each $\ell_1,\ell_2$ there is only one $m_1,m_2$ for which the coefficient is not exponentially suppressed, and for that choice \REf{eq:gauntL1L2} is of order $\ell^{-7/4}$.

On the other hand if we assume $(\ell_1-\ell) \sim \ell_2 \sim 1$ (the case $\ell_2\sim \ell, \ell_1\sim 1$ will of course give similar result)
\begin{equation} \label{eq:gauntL1}
  \frac{ C_{\ell,-\ell-m_2,m_2}^{\ell,\ell_1,\ell_2} }{\sqrt{\omega_\ell \omega_{\ell_1} \omega_{\ell_2}}} \overset{\substack{\ell_1\sim \ell \\ \ell\to \infty}}{\longrightarrow}
  (-1)^\ell h_2\!\left(\ell_1-\ell,\ell_2,m_2 \right) \ell^{\frac{\ell-\ell_1+m_2}{2}} \left( \ell^{-1} + \mathcal O(\ell^{-2})\right) ,
\end{equation}
with
\begin{equation}
  h_2 (x,y,z) = \frac{ (-1)^\frac{x+y-2z}{2} 2^\frac{x-2y+z-1}{2} (y-x)! \sqrt{(y-z)!} }{ \sqrt{\pi} \left(\frac{x+y}{2}\right)! \left(\frac{y-x}{2}\right)! \sqrt{(y+z)!(-x-z)!} } .
\end{equation}
Hence \REf{eq:gauntL1} is least suppressed in case $m_2 = \ell_1 - \ell$ which is its maximum allowed value since $|m_1| = |-\ell-m_2| \leq \ell_1$, and in that case \REf{eq:gauntL1} is of order $\ell^{-1}$.

\section{Norm of an excited state\label{sec:observable2_estimate}}

To furnish one more example of the perturbative expansion of quantities involving spinning charged states discussed in \ref{sec:observable1_estimate}, we now consider the computation of $\langle n | A_{\ell\ell} A_{\ell\ell}^\dagger |n\rangle$. This is equivalent to the norm of the state $A_{\ell\ell}^\dagger | n \rangle$. Writing the state as a power series in $n^{-1/2}$
\begin{equation}
  A_{\ell\ell}^\dagger | n \rangle = |\Psi_0 \rangle + \frac{1}{\sqrt{n}} |\Psi_1\rangle + \frac{1}{n} |\Psi_2 \rangle + \dots
\end{equation}
we have from \REf{eq:NLOFockA}
\begin{equation} \label{eq:A_ll_n_terms}
\begin{split}
 & |\Psi_0\rangle = \frac{a_{00}a_{\ell\ell}^\dagger}{\sqrt{n}} | n \rangle \\
 & |\Psi_1\rangle = \Bigg(\frac{(1+4 \ell) b_{00}^\dagger a_{\ell\ell}^\dagger }{4(1+2\ell)} + \!\!\!\! \sum_{\substack{\ell_1,\ell_2 > 0 \\ \text{all } m_1,m_2}}  \!\!\!\! \frac{(-1)^m\sqrt{\pi}C_{\ell,m_1,m_2}^{\ell,\ell_1,\ell_2}}{8\sqrt{2\omega_\ell\omega_{\ell_1}\omega_{\ell_2}}n} \Big( 
	      - (2+3\ell+\ell_1+\ell_2) a_{00}^2 a_{\ell_1,-m_1}^\dagger a_{\ell_2,-m_2}^\dagger \\
   & \hspace{240pt} + 2 (1+\ell-\ell_1+3\ell_2)n  b_{\ell_1,-m_1}^\dagger a_{\ell_2,-m_2}^\dagger \\
   & \hspace{240pt} + (\ell-\ell_1-\ell_2)(a_{00}^\dagger)^2 b_{\ell_1,-m_1}^\dagger b_{\ell_2,-m_2}^\dagger \Big) \Bigg) |n\rangle 
\end{split}
\end{equation}
since many terms vanish when applied to $|n\rangle$. We have not computed $|\Psi_2\rangle$ as this would require the NNLO expression for $A_{\ell\ell}^\dagger$. It is easy to see that $\langle \Psi_0 | \Psi_0 \rangle = 1$ and $\langle \Psi_0 | \Psi_1 \rangle = 0$. Thus order $n^{-1}$ correction to the norm is given by
\begin{equation}
  || A_{\ell\ell}^\dagger | n \rangle ||^2 = 1 + \frac{1}{n} \big( \langle \Psi_1 | \Psi_1 \rangle + \langle \Psi_0 | \Psi_2 \rangle + \langle \Psi_2 | \Psi_0 \rangle \big) .
\end{equation}
We cannot directly evaluate the last two terms, but we can analyze the term $\langle \Psi_1 | \Psi_1 \rangle$. This will be given by an infinite sum over spins such as $\ell_1,\ell_2$, which we have no reason to expect will converge. Hence $\langle \Psi_0 | \Psi_2\rangle$ has to be an infinite sum as well, such that its divergent part cancels with that of $\langle \Psi_1 | \Psi_1 \rangle$. The order of magnitude of the spins for which the cancellation starts taking effect can only be the only characteristic spin of the problem : $\ell$. We can thus estimate that the tails of both sums will cancel when summed spins are greater than $\ell$. In other words, the behavior of both sums can be approximated, barring some unexpected cancellations, by estimating the behavior of the sum in $\langle \Psi_1|\Psi_1\rangle$ with a cutoff of order $\ell$. We observe all terms of $|\Psi_1\rangle$ in \REf{eq:A_ll_n_terms} are orthogonal to each other, so we must estimate the norm of these individual terms in the limit of large summed spins $\ell_1,\ell_2$.

First, we consider $\ell_1 \sim \ell_2 \sim \ell$. We estimate the contribution of such terms to the norm as
\begin{equation} \label{eq:divergentSumL1L2}
  \sum_{\substack{ \ell_1 \sim \ell_2 \sim \ell \\ m_1, m_2}} \frac{ \left| C_{\ell,m_1, m_2}^{\ell, \ell_1, \ell_2} \right|^2 }{ \omega_\ell \omega_{\ell_1} \omega_{\ell_2}} (\ell^2)
  \sim  \sum_{ \ell_1 \sim \ell_2 \sim \ell} \left( \ell^{-7/4} \right)^2 (\ell^2)
  \sim \ell^2 \times \ell^{-3/2}
  \sim \ell^{1/2} ,
\end{equation}
Let us briefly explain this estimation. We start with a single sum because of the orthogonality of terms in \REf{eq:A_ll_n_terms}, and the summand is the square of the coefficient in that equation. Then, as observed in the appendix \ref{app:Gaunt}, in this regime there is only one choice of $m_1,m_2$ which give a non-suppressed term (\ref{eq:gauntL1L2}). Finally, the double sum over $\ell_1,\ell_2$ yields an additional $\ell^2$ factor.

Secondly, we consider the case $ \ell_1 - \ell \sim \ell_2 \sim 1$. Again, there is only one choice of $m_1,m_2$ that yields the dominant term (\ref{eq:gauntL1}). Neglecting other terms, we estimate the contribution to the norm as
\begin{equation}
  \sum_{\substack{ \ell_1 \sim \ell, \ell_2\sim 1 \\ m_1, m_2}} \frac{ \left| C_{\ell,m_1, m_2}^{\ell, \ell_1, \ell_2} \right|^2 }{\ell \ell_1 \ell_2} (\ell^2)
  \sim  \sum_{ \ell_1 \sim \ell, \ell_2\sim 1} \left( \ell^{-1} \right)^2 (\ell^2)
  \sim \ell \times 1
  \sim \ell ,
\end{equation}
where in the second estimation the sum yields a single $\ell$ factor since only $\ell_1$ is summed up to order $\ell$.
We notice this contribution is dominating that of \REf{eq:divergentSumL1L2}. Evidently the case $ \ell_2 - \ell \sim \ell_1 \sim 1$ gives an equal contribution.

 Therefore, the series expansion of the norm is estimated schematically as
 \begin{equation}
   || A_{\ell\ell}^\dagger |n\rangle ||^2 \sim 1 + \frac{\ell+ \dots}{n} + \mathcal O(n^{-2}).
 \end{equation}
where the dots represent terms which are subdominant at large $\ell$. 

We see the result is again expressed as a series in $\frac{\ell}{n}$.

\section{Propagator on the cylinder\label{sec:propagatorCylinder}}

In this section we show how to 
construct propagators corresponding to fluctuations of fields $r,\pi$ in \REf{eq:rpiDYExpansion}.
From time translation and rotation symmetry we know the propagator can be written as
\begin{equation}
  \langle  x(\tau_1,\vec n_1) y(\tau_2, \vec n_2) \rangle = D_{x y}(\tau_1-\tau_2,\vec n_1 \cdot \vec n_2) ,
\end{equation}
where $x,y \in \{ r,\pi\}$ are fields and $\langle \dots \rangle$ is the $\tau$-ordered Wick contraction. The quadratic Lagrangian yields a matrix equation similar to \REf{eq:EOMFluctuationsMatrix} for the propagator (note that it is not diagonal due to mixing between $\dot \pi$ and $r$)
\be
-\l(
\ba{cc}
\p_\tau^2+\D_{\mathbb S^{d-1}} -M^2 & 2i \m \p_\tau \\
-2i \m \p_\tau & \p_\tau^2+\D_{\mathbb S^{d-1}}
\ea
\r)
\l(
\ba{cc}
D_{rr} & D_{r\pi} \\
D_{\pi r} & D_{\pi \pi}
\ea
\r) = \d(\tau_1-\tau_2)\d^{(\mathbb S^{d-1})}(\vec n_1 \cdot \vec n_2 ).
\ee
Expanding in spherical harmonics (in this case only with $\vec m = \vec 0$, which corresponds to Gegenbauer polynomials)
\be
D(\tau,\vec n_1 \cdot \vec n_2) = \sum_\ell F^{(\ell)}(\tau) C^{(d/2-1)}_\ell(\cos (\vec n_1 \cdot \vec n_2)),
\label{eq:PropagatorGeneralP}
\ee
we obtain
\be 
\label{eq:ModePropagator}
-N_\ell \Omega_{d-2}
\l(
\ba{cc}
\p_\tau^2-J_\ell -M^2 & 2i \m \p_\tau \\
-2i \m \p_\tau & \p_\tau^2-J_\ell
\ea
\r)
\l(
\ba{cc}
F^{(\ell)}_{rr} & F^{(\ell)}_{r\pi} \\
F^{(\ell)}_{\pi r} & F^{(\ell)}_{\pi \pi}
\ea
\r) = C_\ell^{(d/2-1)}(1)\d(\tau),
\ee
where the normalization factor of Gegenbauer polynomials is given by
\be
N_\ell \int_{-1}^1 C^{(d/2-1)}_{\ell}(x) C^{(d/2-1)}_{\ell}(x) (1-x^2)^{\f{d-3}{2}} d x =\f{2^{4-d} \, \pi \, \G(\ell+d-2)}{(2\ell+d-2) \, \ell! \, \G^2\l (\f{d}{2}-1\r )}.
\ee
We can look for solutions of this equations for $\tau <0$ and $\tau >0$, which will be given by expressions similar to \REf{eq:rpiDYExpansion}, and then find the propagator by matching this solutions at $\tau =0$ with a specific discontinuity of derivatives. Alternatively, we can Fourier transform \REf{eq:ModePropagator} and use (see~\cite{dlmf:Gegenbauers}, table 18.6.1)
\be
C^{(d/2-1)}_{\ell}(1) = \f{\G(\ell+d-2)}{\ell! \, \G(d-2)},
\label{eq:Gegenbauer1}
\ee
to obtain
\be
F^{(\ell)}(\tau) = \f{2\ell+d-2}{(d-2)\Omega_{d-1}}\int \f{d \omega}{2\pi} e^{-i \omega \tau}  \f{M^{(\ell)}(\omega)}
{(\omega^2+\omega_B^2(\ell))(\omega^2+\omega_A^2(\ell))},
\label{eq:momentumPropagator}
\ee
with
\be
M^{(\ell)} (\omega)=\l(
\ba{cc}
\omega^2+J_\ell & 2 \m \omega \\
-2 \m \omega & \omega^2+J_\ell +M^2
\ea
\r).
\ee
For $\ell\neq 0$ integration in \REf{eq:momentumPropagator} can be easily done using Cauchy's theorem, resulting in
\be
F^{(\ell)}(\tau)=\f{2\ell+d-2}{(d-2)\Omega_{d-1}} \, \l ( \f{M^{(\ell)}(-i\omega_A(\ell))e^{-\omega_A (\ell)\tau}}{2\omega_A(\ell)} - 
\f{M^{(\ell)}(-i\omega_B(\ell))e^{-\omega_B(\ell) \tau}}{2\omega_B(\ell)} \r ) \f{1}{\omega_B^2(\ell)-\omega_A^2(\ell)}
\label{eq:propagatorJneq0tau>0}
\ee
for $\tau>0$, and
\be
F^{(\ell)}(\tau)=\f{2\ell+d-2}{(d-2)\Omega_{d-1}} \, \l ( \f{M^{(\ell)}(i\omega_A(\ell))e^{\omega_A(\ell) \tau}}{2\omega_A(\ell)} - 
\f{M^{(\ell)}(i\omega_B(\ell))e^{\omega_B(\ell) \tau}}{2\omega_B(\ell)} \r ) \f{1}{\omega_B^2(\ell)-\omega_A^2(\ell)}
\label{eq:propagatorJneq0tau<0}
\ee
for $\tau<0$. The same result can obviously be obtained directly from \REf{eq:rpiDYExpansion}. Indeed, say for $\tau_1<\tau_2$ computing non-zero spin contribution to time ordered correlator we get
\be
\la n | r(\tau_2) r(\tau_1) | n\ra_\ell = \l ( \f{J_\ell-\omega_A(\ell)^2}{2\omega_A(\ell)}e^{-\omega_A(\ell)|\tau_2-\tau_1|} + \f{\omega_B(\ell)^2-J_\ell}{2\omega_B(\ell)}e^{-\omega_B(\ell)|\tau_2-\tau_1|}\r ) \f{1}{\omega_B^2(\ell)-\omega_A^2(\ell)} \sum_{\vec m} Y_{\ell\vec m} Y^*_{\ell\vec m},
\ee
which upon using \REf{eq:Gegenbauer1} and (see~\cite{Frye:2012jj})
\be
\sum_{\vec m} Y_{\ell\vec m} (\vec n_1)Y^*_{\ell\vec m} (\vec n_2)= \f{2\ell+d-2}{(d-2) \Omega_{d-1}}C_\ell^{(d/2-1)} (\vec n_1 \cdot \vec n_2)
\ee
reproduces \REf{eq:propagatorJneq0tau>0}. Similarly, we can compute $r\pi$ and $\pi \pi$ components of the propagator.

Dealing with $\ell=0$ modes is somewhat more subtle. The difficulty is that apart from the gapped mode corresponding to 
$(B_{0\vec0}, B_{0\vec0}^\dagger)$ there is also the gapless mode $\hat \pi, p_\pi$, for which
\be
	p_\pi | n \ra = 0,
\ee
and which does not have the Fock space structure. It does not present a problem for $\la r r \ra$, indeed using \REf{eq:rpiDYExpansion} we get
\be
\la n | r(\tau_2) r(\tau_1) | n \ra_0 = \f{1}{2\omega_B(0)\Omega_{d-1}} e^{-\omega_B(0) |\tau_2-\tau_1|},
\ee
which is consistent with \REf{eq:propagatorJneq0tau>0} and \REf{eq:propagatorJneq0tau<0}. On the other hand considering correlators linear in $\pi$ is problematic. However, that is not an issue, for in all instances the field $\pi$ appears only in the exponent\footnote{Bear in mind that $\hat \pi$ is defined on a compact space (circle), since charge is quantized. As such, the corresponding canonical momentum $ p_\pi$ is defined only on the space of periodic functions. Otherwise $p_\pi$ is not Hermitian. Indeed, the following relation holds
\be
\int_0^{2\pi} d\hat \pi \psi_2^*(\hat \pi) \l [-i \p_{\hat \pi} \psi_1(\hat \pi) \r ] = \int_0^{2\pi} d\hat \pi (\hat \pi) \l [-i \p_{\hat \pi}\psi_2\r ]^* \psi_1(\hat \pi),
\ee
only if $\psi_2(\hat \pi) \psi_1(\hat \pi)\Big |_{0}^{2\pi}=0$, i.e. for periodic functions $\psi_i(\hat\pi)$.},
hence, we need only to worry about correlators involving $e^{i \pi(\tau)/f}$. For instance, using Baker-Campbell-Hausdorff formula we obtain 
(for $\tau<0$)
\be
\la e^{-i\pi(0)/f} e^{i \pi(\tau)/f} \ra _0 =\exp \l [ -\f{1-\f{4\m^2}{\omega_B^2(0)}}{2\Omega_{d-1}f^2} \tau \r ] 
\exp \l [ \f{1}{f^2}\f{4\m^2}{\omega_B^2(0)} \f{e^{\omega_B(0)\tau}-1}{2\omega_B(0)\Omega_{d-1}}\r ].
\ee 
Comparing with the naive expectation
\be
\la e^{-i\pi(0)/f} e^{i \pi(\tau)/f} \ra _0 = 1+ \f{D^{(0)}_{\pi\pi}(|\tau|)-D^{(0)}_{\pi\pi}(0)}{f^2}+O(f^{-4}),
\ee 
it is consistent to define (compare with \REf{eq:propagatorJneq0tau>0} and \REf{eq:propagatorJneq0tau<0})
\be
F^{(0)}_{\pi \pi}(\tau) = -\f{1-\f{4\m^2}{\omega_B^2(0)}}{2\Omega_{d-1}} |\tau| + \f{4\m^2}{\omega_B^2(0)} \f{e^{-\omega_B(0)|\tau|}}{2\omega_B(0)\Omega_{d-1}}+\const .
\ee
Similarly, computing $\la e^{-i\pi(0)/f} r(0) e^{i \pi(\tau)} \ra _0$ allows to define
\be
F^{(0)}_{r \pi}(\tau) = \mathrm{sign}(\tau)\f{i\m}{\omega_B^2(0)} \f{e^{-\omega_B(0)|\tau|}}{\Omega_{d-1}}+\const .
\ee

\section{3-pt function computation \label{app:lambdaComputation}}

We start from
\be
\l \la n | \l ( \bar \phi \phi \r )(0, \hat n_{d}) | n \r \ra = \f{1}{2} \l \la n | f^2 + 2 f r (0, \hat n_{d}) + r^2(0, \hat n_{d}) | n \r \ra,
\ee
and then expanding around the saddle, we get for the expectation value of bare fields
\be
\l \la n | \l ( \bar \phi \phi \r )(0, \hat n_{d}) | n \r \ra =
\f{f^2}{2} - \l \la r(0, \hat n_{d}) \int d\tau d \Omega_{d-1}\l [ r (\p\pi)^2 - i \mu r^2 \dot \pi +\f{\lambda f^2 r^3}{4} \r ] \r \ra +
\f{1}{2}\la  r^2(0, \hat n_{d}) \ra.
\ee
Here as in section \ref{sec:3ptfunction}, the symbol $\mu$ refers to $\mu_4(\lambda n,d)$. One must keep in mind that $m$, $f$, $J_\ell$, $\omega_{A,B}(\ell)$ are functions of $d$, $\mu$ and $n$. The symbol $\mu_*$ will refer to $\mu_4(\lambda_* n,4)$.
Using the propagator \REf{eq:PropagatorGeneralP} we compute contractions for terms without spatial derivatives
\bea
\int \la r_1 r_2^3\ra 
& \to & 
3\Omega_{d-1} \l [ \int d\tau F^{(0)}_{rr} (\tau)\r ] \, \l [ \sum_{\ell=0}^\infty 
F^{(\ell)}_{rr} (0) C_\ell^{(d/2-1)}(1) \r], \\
\int \la r_1 \dot \pi_2 r_2^2\ra 
& \to & 
2\Omega_{d-1} \l [ \int d\tau F^{(0)}_{rr} (\tau)\r ] \, \l [ \sum_{\ell=0}^\infty \dot F^{(\ell)}_{\pi r} (0) C_\ell^{(d/2-1)}(1) \r] \nn \\ 
&&
+\Omega_{d-1}  \l [ \int d\tau \dot F^{(0)}_{\pi r} (\tau)\r ] \, \l [ \sum_{\ell=0}^\infty 
F^{(\ell)}_{rr} (0) C_\ell^{(d/2-1)}(1) \r], \\
\int \la r_1 \dot \pi_2^2 r_2\ra 
& \to & 
2 \Omega_{d-1} \l [ \int d\tau \dot F^{(0)}_{\pi r} (\tau)\r ] \, \l [ \sum_{\ell=0}^\infty \dot F^{(\ell)}_{\pi r} (0) C_\ell^{(d/2-1)}(1) \r] \nn \\ 
&&
-\Omega_{d-1}  \l [ \int d \tau F^{(0)}_{r r} (\tau)\r ] \, \l [ \sum_{\ell=0}^\infty 
\ddot F^{(\ell)}_{\pi \pi} (0) C_\ell^{(d/2-1)}(1) \r], \\
\int \la r^2\ra 
& \to & 
\l [ \sum_{\ell=0}^\infty 
F^{(\ell)}_{rr} (0) C_\ell^{(d/2-1)}(1) \r],
\eea
where indices indicate evaluation point, for example $r_1 = r(\tau_1, \vec n_1)$.
For the last term 
\be
\int g^{ij}_2\la r_1 \p_i \pi_2 \p_j \pi_2 r_2\ra 
\ee
in order to find contraction of two fields at the same point it is necessary to introduce a splitting, compute derivative(s) and then consider the limit. For example 
\be\label{eq:contractionSpatialDerivatives1}
\la r_2 \p_i \pi_2\ra = \lim_{\vec n_2'\to \vec n_2} \p'_i D_{\pi r}(0,\vec n_2'\cdot \vec n_2) = 0,
\ee
which vanishes since $\vec n_2'\cdot \vec n_2$ is maximal for $\vec n_2' =  \vec n_2$.
Similarly, using the chain rule and the same argument we show
\be\label{eq:contractionSpatialDerivatives2}
\la \p_i \pi_2 \p_j \pi_2\ra = \lim_{\vec n_2'\to \vec n_2} \p'_i \p_j D_{\pi \pi}(0,\vec n_2'\cdot \vec n_2) = \left. \frac{d}{d x} D_{\pi\pi}(0,x) \right|_{x= 1}  (\partial_i \vec n_2) \cdot(\partial_j \vec n_2)
\ee
and one can show, for example by choosing specific coordinates on the sphere, that
\begin{equation}
  g_2^{ij} (\partial_i \vec n_2) \cdot(\partial_j \vec n_2) = d-1.
\end{equation}
Using (see~\cite{dlmf:Gegenbauers}, eq. (18.9.19))
\be \label{eq:dGegenbauer}
\f{d}{dx} C_n^{(\lambda)}(x) = 2\lambda C_{n-1}^{(\lambda+1)}(x),
\ee
and (see \REf{eq:Gegenbauer1})
\be
C_{\ell-1}^{(d/2)}(1)= C_\ell^{(d/2-1)}(1) \, \f{J_\ell}{(d-1)(d-2)},
\ee
we obtain
\be
\int g^{ij}_2\la r_1 \p_i \pi_2 \p_j \pi_2 r_2\ra \to 
\Omega_{d-1} J_\ell \l [ \int d \tau F^{(0)}_{r r} (\tau)\r ]
\, \l [ \sum_{\ell=0}^\infty F_{\pi \pi}^{(\ell)}(0)C_\ell^{(d/2-1)}(1) \r].
\ee

For what follows we will need explicit expressions
\bea
F_{rr}^{(0)}(\tau) & = & \f{e^{-\omega_B(0) |\tau|}}{2 \Omega_{d-1}\,\omega_B(0)}, \\
\dot F_{\pi r}^{(0)}(\tau) & = & \f{i \mu \, e^{-\omega_B(0) |\tau|}}{\Omega_{d-1} \, \omega_B(0)}, \\
F_{r r}^{(0)}(0) & = & \f{1}{2 \Omega_{d-1}\,\omega_B(0)}, \\
F_{r r}^{(\ell)}(0) & = & \f{2\ell+d-2}{\Omega_{d-1}(d-2)} \f{\omega_B(\ell)\omega_A(\ell)+J_\ell}{2\omega_B(\ell)\omega_A(\ell) \l[ \omega_B(\ell)+\omega_A(\ell) \r ]}, ~~ {\ell \neq 0}, \\
F_{\pi \pi}^{(\ell)}(0) & = & \f{2\ell+d-2}{\Omega_{d-1}(d-2)} \f{\omega_B(\ell)\omega_A(\ell)+J_\ell+2(\m^2-m^2)}{2\omega_B(\ell)\omega_A(\ell) \l[ \omega_B(\ell)+\omega_A(\ell) \r ]}, ~~ {\ell \neq 0}, \\
\dot F_{\pi r}^{(0)}(0) & = & \f{i \mu}{ \Omega_{d-1} \, \omega_B(0)}, \\
\dot F_{\pi r}^{(\ell)}(0) & = & \f{2\ell+d-2}{\Omega_{d-1}(d-2)} \f{i \mu}{\omega_B(\ell)+ \omega_A(\ell)}, ~~ {\ell \neq 0}, \\
\ddot F_{\pi \pi}^{(0)}(0) & = & \f{2\m^2}{\Omega_{d-1} \, \omega_B(0)}, \\
\ddot F_{\pi \pi}^{(\ell)}(0) & = & \f{2\ell+d-2}{\Omega_{d-1}(d-2)} \f{\omega^2_+(\ell)+\omega_A^2(\ell)+\omega_B(\ell)\omega_A(\ell)-J_\ell-2(\m^2-m^2)}{2 \l [\omega_B(\ell)+ \omega_A(\ell) \r ]}, ~~ {\ell \neq 0},
\eea
and integrals
\bea
\int d\tau F^{(0)}_{rr} (\tau) & = & \f{1}{\Omega_{d-1}\,\omega_B^2(0)}, \\
\int d\tau \dot F^{(0)}_{\pi r} (\tau) & = & \f{2i \m}{\Omega_{d-1}\,\omega_B^2(0)}.
\eea

We obtain
\bea
\l \la n | \bar \phi \phi (0,\hat n_d)  | n \r \ra & = & \frac{n}{2 \mu \Omega_{d-1}}
-\f{2(\m^2-m^2)}{\omega_B^2(0)}\sum_{\ell=0}^\infty F_{rr}^{(\ell)}(0) C_\ell^{(d/2-1)}(1)
-\f{2i\m}{\omega_B^2(0)}\sum_{\ell=0}^\infty \dot F_{\pi r}^{(\ell)}(0) C_\ell^{(d/2-1)}(1) \nn \\
&& {}
+\f{1}{\omega_B^2(0)}\sum_{\ell=0}^\infty \l [ \ddot F_{\pi \pi}^{(\ell)}(0) - J_\ell F_{\pi \pi}^{(\ell)}(0) \r ]C_\ell^{(d/2-1)}(1).
\eea
Which can be further simplified with
\be
\omega_B^2(\ell)\omega_A^2(\ell) = J_\ell^2+2J_\ell (\m^2-m^2),~~ 
\omega_B^2(\ell)+\omega_A^2(\ell) =2 (J_\ell +3\m^2-m^2)
\ee
leading to
\be
\l \la n | (\bar \phi \phi)(0,\hat n_{d})  | n \r \ra = \frac{n}{2 \mu \Omega_{d-1}}
+\sum_{\ell=0}^\infty \f{1}{\omega_B^2(0)} \f{2\ell+d-2}{\Omega_{d-1}(d-2)} C_\ell^{(d/2-1)}(1)
\f{\omega_B(\ell) \omega_A(\ell) (3\m^2+m^2)-J_\ell(\m^2-m^2)}{\omega_B(\ell) \omega_A(\ell) 
\l [\omega_B(\ell) + \omega_A(\ell)\r ]}.
\label{eq:3pt_Full}
\ee
Denoting the summand in \REf{eq:3pt_Full} by 
\be
\f{S_\ell(\m,m,d)}{(d-2)\Omega_{d-1}}
\ee
and considering its asymptotics at $\ell \to \infty$
\be
S_\ell(\m,m,d) \underset{\ell\to \infty}{\equiv} c_{-1}(\m,m,d) \ell^{d-3}+c_{0}(\m,m,d) \ell^{d-4} + c_{1}(\m,m,d) \ell^{d-5} + \dots,
\ee
we get
\bea
\l \la n | (\bar \phi \phi ) (0,\hat n_d) | n \r \ra & = & \frac{n}{2 \Omega_{d-1}\mu_4((\lambda_*+\delta \lambda)n, d) } \\
&& {} +
\f{1}{(d-2)\Omega_{d-1}} \Bigg \{S_0(\m,m,d) +\sum_{\ell=1}^\infty \Big [S_\ell(\m,m,d)  - c_{-1}(\m,m,d)  \ell^{d-3} \nn \\
&& {}
-c_{0}(\m,m,d)  \ell^{d-4} - c_{1}(\m,m,d)  \ell^{d-5} \Big ]  \nn \\
&& {} +
c_{-1}(\m,m,d) \z(3-d)+c_{0}(\m,m,d) \z(4-d) + c_{1}(\m,m,d)\z(5-d) \Bigg \}_{\lambda_*} \nn
\eea
where we put explicit dependence of $\mu_4$ on the 1-loop coupling counterterm, with
\be
\d \lambda = \f{5(\lambda_*)^2}{16 \pi^2} \f{1}{4-d},
\ee
and 1-loop terms don't need any counterterm corrections at this order. The renormalized coupling is denoted by $\lambda_*$ which at this stage is considered as independent of the dimension.
Expanding the first term in $\lambda$ and other terms in $4-d$, keeping only $O(1)$ terms (these are the only ones that are needed at this order), leads to
\bea
\l \la n | (\bar \phi \phi ) (0,\hat n_d) | n \r \ra & = &
\Bigg\{ \f{n}{2\Omega_{d-1}\mu} - \f{5\lambda^2}{16 \pi^2}\f{1}{4-d}\f{n}{2\Omega_{d-1}\mu^2} \, \f{\p \mu}{\p \lambda}
\\
&& {} + 
\f{1}{(d-2)\Omega_{d-1}} \Bigg( R(\m_*) + \f{c_{1P}(\m,m)}{4-d}+c_{1F}(\m_*,1) \Bigg) \Bigg\}_{\lambda_*} \nonumber
\eea
where we introduced
\bea
R(\m)&=& S_0(\m,1,4) +\sum_{\ell=1}^\infty \l [ S_\ell(\m,1,4)  - c_{-1}(\m,1,4)  \ell -c_{0}(\m,1,4) -\f{c_{1}(\m,1,4)}{\ell} \r ]  \nn \\
&& {} + 
c_{-1}(\m,1,4) \z(-1)+c_{0}(\m,1,4) \z(0),
\label{eq:FiniteSum}
\eea
and
\be
c_{1}(\m,m,d)\z(5-d) \underset{d\to 4}{=}\f{c_{1P}(\m,m)}{4-d}+c_{1F}(\m,m),
\ee
with
\bea
\label{eq:Zeta1Pole}
c_{1P}(\m,m) & = & \f{m^2+2m^4+\m^2-3m^2\m^2-\m^4}{2(3\m^2-m^2)}, \\
c_{1F}(\m,m) & = &  \f{12m^4-5\m^2-6\m^4-m^2(18\m^2+5)}{12(3\m^2-m^2)}.
\label{eq:Zeta1Finite}
\eea
In the theory at hand, equation \REf{eq:mu_d_phi4} takes the form
\begin{equation} \label{eq:mu_equation}
\mu^2-m^2 = \frac{n \lambda}{4 \mu \Omega_{d-1}} .
\end{equation}
This implies
\be
\f{\p \m}{\p \lambda} = \f{n}{4 \Omega_{d-1} (3 \m^2-m^2)} ,
\ee
which yields
\bea
\l \la n | (\bar \phi \phi ) (0,\hat n_d) | n \r \ra 
& = &
\Bigg\{ \f{n}{2\m \Omega_{d-1}}-\f{5}{16 \pi^2} \, \f{1}{4-d} \, \f{\lambda^2 n^2}{8 \Omega_{d-1}^2 \m^2 (3\m^2 -m^2)} \\
&& {} + 
\f{1}{(d-2)\Omega_{d-1}} \Bigg( R(\m_*) + \f{c_{1P}(\m,m)}{4-d}+c_{1F}(\m_*,1) \Bigg) \Bigg\}_{\lambda_*} .\nn
\label{eq:3pt_Expanded}
\eea
Taking into account normalization \REf{eq:NormRenorm} and expanding in $\lambda$ we get
\bea
\lambda_{\bar \phi \phi} & = & Z^{-1}_{\bar \phi \phi } \l \la n | (\bar \phi \phi ) (0,\hat n_d) | n \r \ra \\
& = &
\Bigg\{ \f{n (d-2)}{2\m}+ \frac{\lambda n(d-2)}{16 \pi^2 \mu (4-d)} + \frac{\lambda n (d-2)}{32\pi^2 \mu}(1+\gamma+\log\pi)  \nn \\
& & {}
-\f{5}{16 \pi^2} \, \f{1}{4-d} \, \f{\lambda^2 n^2}{8 \Omega_{d-1}^2 \m^2 (3\m^2 -m^2)}+ R(\m_*) + \f{c_{1P}(\m,m)}{4-d}+c_{1F}(\m_*,1) \Bigg\}_{\lambda_*}. \nn
\eea
Using (\ref{eq:mu_equation}) to substitute $\lambda n$, we can gather the three order $\frac{1}{4-d}$ terms, then expand $\mu$, $m$ and $\Omega_{d-1}$ in $4-d$ to show cancellation of divergences and get a finite part
\begin{align}
  & \frac{1}{4-d} \left( \frac{(d-2)\Omega_{d-1} (\mu^2-m^2)}{4 \pi^2 } -\f{5 \Omega_{d-1} (d-2) (\mu^2-m^2)^2}{8 \pi^2 (3\m^2 -m^2)} + c_{1P}(\m,m) \right) \nonumber \\
 & \overset{d\to 4}{\longrightarrow} \ \  \frac{2(\mu_*^2+1)-(\gamma + \log \pi)(\mu_*^4+2\mu_*^2-3)}{4(3\mu_*^2-1)} .
\end{align}
Some $\gamma+\log\pi$ appeared from 
\be \label{eq:OmegaDerivative}
\frac{1}{\Omega_3} \left.\frac{\partial \Omega_{d-1}}{\partial d}\right|_{d=4} = \f{1}{2} \l ( \g +\log \pi -1\r ) .
\ee
We can as well substitute $\lambda n$ in the term
\begin{equation}
  \frac{\lambda n (d-2)}{32\pi^2 \mu}(1+\gamma+\log\pi) \ \overset{d\to 4}{\longrightarrow}\ \frac{\mu_*^2-1}{2}(1+\gamma+\log\pi) .
\end{equation}
We see there is not yet full cancellation of $\gamma+\log\pi$ terms. The reason is the very first term, which is enhanced by $n$, contains $\mu(\lambda_* n,d)$ which we still have to expand in $4-d$, bringing $n(4-d)$ contributions at NLO. To this end, we can express the derivative of $\m$ with respect to $d$ from~\REf{eq:mu_d_phi4}
\be
\left. \frac{\partial \mu}{\partial d}\right|_{d=4} =\left. \f{\m}{3\m^2-1} \l [ 1-\frac{1}{\Omega_3} (\m^2-1) \frac{\partial \Omega_{d-1}}{\partial d} \r ] \right|_{d=4}
\ee
and use (\ref{eq:OmegaDerivative}). This yields
\begin{equation}
  \frac{n(d-2)}{2 \mu} = \frac{n}{\mu_*} - \frac{(4-d)n \left( \mu_*^2-1 \right) (2+\gamma+\log\pi) }{2 \mu_*\left( 3\mu_*^2-1 \right)} .
\end{equation}
Now that $\frac{1}{4-d}$ poles have cancelled and everything has been expanded to relevant order in $4-d$, we can take the theory at the fixed point (\ref{eq:phi4FixedPoint}). This yields for the previous equation
\begin{equation}
  \frac{n(d-2)}{2 \mu} = \frac{n}{\mu_*} - \frac{5 \left( \mu_*^2-1 \right)^2 (2+\gamma+\log\pi) }{4 ( 3\mu_*^2-1 )} .
\end{equation}
Putting everything together, we notice all $\gamma+\log\pi$ terms cancel, and we get the final result
\begin{equation}
  \lambda_{\bar \phi \phi} = \frac{n}{\mu_*}- \frac{2 \mu_*^4-7\mu_*^2 +3}{2 (3\mu_*^2-1)} + R(\mu_*) + c_{1F}(\mu_*,1) .
\end{equation}

Plugging explicit expressions from~\REf{eq:FiniteSum} and~\REf{eq:Zeta1Finite} results in
\be
\lambda_{\bar \phi \phi}
=
\f{n}{\mu_*}+\f{2(3\mu_*^2+1)}{\l [ 2(3\mu_*^2-1) \r]^{3/2}}-\f{3\mu_*^4-2\mu_*^2+3}{2(3\mu_*^2-1)}
+\sum_{\ell=1}^\infty \l [ S_\ell(\mu_*)  - c_{-1}(\mu_*)  \ell -c_{0}(\mu_*) -\f{c_{1}(\mu_*)}{\ell} \r ],
\ee
where we noted
\begin{equation}
  S_\ell(\mu) = S_\ell(\mu,1,4) , \qquad c_i(\mu) = c_i(\mu,1,4) .
\end{equation}

\section{Feynman diagram computation of $\protect\Delta_{(\protect\bar \protect\phi \protect\phi)^k}$}\label{app:phiphiDimensionFeynman}

We compute diagrammatically the one-loop anomalous dimension of $(\bar \phi \phi)^k$ in theory \REf{eq:phi4}. We consider the MS renormalization of operators in the following momentum-space correlator\footnote{To be more precise, the operator $(\bar\phi\phi)^k$ mixes with other operators \cite{Brown:1979pq}. However, since  this operator is a primary of the critical theory, the mixing cancels in that case. This means we can neglect the mixing when diagramatically computing the anomalous dimension off-criticality.}:
\begin{equation}
  \langle (\bar \phi \phi)^k \bar\phi(p) \cdots \phi(p)\cdots \rangle = Z_{(\bar\phi\phi)^k} Z_\phi^{2k} \langle \left[ (\bar \phi \phi)^k \right] [\bar\phi](p)\cdots [\phi](p) \cdots \rangle \,,
\end{equation}
where there are $k$ insertions of field $\phi(p)$ and $\bar\phi(p)$.
The field renormalization factor $Z_\phi$ has no one-loop contribution so we consider it equal to $1$. The bare $(\bar\phi\phi)^k$ operator is normalized :
\begin{equation}
  \btf{x}
    \vertex [bigcross] (x) {};
    \vertex [above left=.6cm of x] (a);
    \vertex [above right=.6cm of x] (c);
    \vertex [below left=.6cm of x] (b);
    \vertex [below right=.6cm of x] (d);
  \etf ~=~ 1\,.
\end{equation}
We do not draw exterior lines in the diagrams.

There are three diagrams at one-loop level, of which we compute the divergent part:
\begin{equation}
  \btf{x}
    \vertex [bigcross] (x) {};
    \vertex at ($(x)$) (xb);
    \vertex [right=1cm of x] (y);
    \vertex [above left=.6cm of x] (a);
    \vertex [below left=.6cm of x] (b);
    \vertex [above right=.6cm of y] (c);
    \vertex [below right=.6cm of y] (d);
    \diagram*[inline=(y)] {
      (xb) -- [out=45, in=135, with arrow=0.5,looseness=1.3] (y),
      (xb) -- [out=-45, in=-135, with arrow=0.5, looseness=1.3] (y),
      (y) -- [with arrow=0.6] (c),
      (y) -- [with arrow=0.6] (d),
    };
  \etf ~=~ 
  \btf{x}
    \vertex (x);
    \vertex [bigcross, right=1cm of x] (y) {};
    \vertex at ($(y)$) (yb);
    \vertex [above left=.6cm of x] (a);
    \vertex [below left=.6cm of x] (b);
    \vertex [above right=.6cm of y] (c);
    \vertex [below right=.6cm of y] (d);
    \diagram*[inline=(y)] {
      (a) -- [fermion] (x),
      (b) -- [fermion] (x),
      (x) -- [out=45, in=135, with arrow=0.5,looseness=1.3] (yb),
      (x) -- [out=-45, in=-135, with arrow=0.5, looseness=1.3] (yb),
    };
  \etf ~=~ \frac{k(k-1)}{4}(-\lambda) \frac{1}{8\pi^2 \varepsilon} + O(\varepsilon^0)
   \qquad
  \btf{q}
    \vertex (q);
    \vertex [above=0.5cm of q] (x);
    \vertex [bigcross] at (x) (xb) {};
    \vertex [below=0.5cm of q] (y);
    \vertex [above left = .6cm of x] (a);
    \vertex [above right = .6cm of x] (c);
    \vertex [below left = .6cm of y] (b);
    \vertex [below right = .6cm of y] (d);
    \diagram*[inline=(q)] {
      (y) -- [out=135, in=-135, with arrow=0.5,looseness=1.3] (x),
      (x) -- [out=-45, in=45, with arrow=0.5,looseness=1.3] (y),
      (b) -- [fermion] (y),
      (y) -- [with arrow=0.6] (d),
    };
  \etf ~=~ k^2(-\lambda)\frac{1}{8\pi^2 \varepsilon} + O(\varepsilon^0)\,.
\end{equation}
Summing all diagrams yields
\begin{equation}
  \langle (\bar \phi \phi)^k \bar\phi(p) \cdots \phi(p)\cdots \rangle = 1 - \frac{k(3k-1)\lambda}{16 \pi^2 \varepsilon} + O(\lambda\varepsilon^0,\lambda^2)\,,
\end{equation}
from which we get
\begin{equation}
  Z_{(\bar\phi\phi)^k} = 1-\frac{k(3k-1)\lambda}{16 \pi^2 \varepsilon} + O(\lambda^2)\,.
\end{equation}
The one-loop anomalous dimension is then given by
\begin{equation}
  \gamma_{(\bar\phi\phi)^k} = -\lambda \varepsilon \frac{\partial \log Z_{(\bar\phi\phi)^2}}{\partial \lambda}  = \frac{k(3k-1)\lambda}{16\pi^2}+ O(\lambda^2)\,.
\end{equation}
At the Wilson-Fisher fixed point (\ref{eq:phi4FixedPoint}) the dimension is
\begin{equation}
  \Delta_{(\bar\phi\phi)^k} = 2 k \left( \frac{d}{2}-1\right) + \gamma_{(\bar\phi\phi)^k} = 2k + \frac{3k(k-2)}{5} \varepsilon + O(\varepsilon^2)\,.
\end{equation}

\newpage

\bibliographystyle{utphys}
\bibliography{Identifying_Large_Charge_Operators}{}

\end{document}